\documentclass[a4paper,12pt]{article}

\usepackage{latexsym}
\usepackage{ }
\usepackage{graphicx, subfigure} 
\usepackage{cite} 
\usepackage{tabularx}
\usepackage{array}
\usepackage{graphics}     
\usepackage{amsmath}
\usepackage{amssymb}
\usepackage{longtable} 
\usepackage{verbatim} 
\usepackage[table]{xcolor}
\usepackage{booktabs}
\usepackage{hyperref} 
\usepackage[utf8]{inputenc}
\usepackage{multirow}
\usepackage{soul}

\definecolor{light-gray}{gray}{0.90}
\definecolor{dgreen}{rgb}{0.0, 0.5, 0.0}

\definecolor{forestgreen}{rgb}{0.13, 0.55, 0.13}
\definecolor{emerald}{rgb}{0.31, 0.78, 0.47}
\definecolor{mintgreen}{rgb}{0.60, 1.00, 0.60}
\definecolor{olivegreen}{rgb}{0.42, 0.56, 0.14}
\definecolor{seagreen}{rgb}{0.18, 0.55, 0.34}
\definecolor{limegreen}{rgb}{0.20, 0.80, 0.20}
\definecolor{springgreen}{rgb}{0.00, 1.00, 0.50}

\renewcommand{\arraystretch}{1.2} 
 
\newcommand{\cen}[1]{\multicolumn{1}{c}{#1}}
\newcommand{\ra}[1]{\renewcommand{\arraystretch}{#1}}
\newcommand{\rb}[1]{\renewcommand{\tabcolsep}{#1}}

\begin{document}

\hfill {\tt MITP/25-055,CERN-TH-2025-164}  

\def\thefootnote{\fnsymbol{footnote}}
 
\begin{center}

\vspace{3.cm}

{\Large\bf {Data-driven analyses and model-independent fits for present $b\to s \ell \ell$ results}}

\setlength{\textwidth}{10cm}
                    
\vspace{2.cm}
{\large\bf  
T.~Hurth$^{a,}$\footnote{Email: tobias.hurth@cern.ch},
F.~Mahmoudi$^{b,c,d,}$\footnote{Email: nazila@cern.ch}, Y.~Monceaux$^{b,}$\footnote{Email: y.monceaux@ip2i.in2p3.fr},
S.~Neshatpour$^{b,}$\footnote{Email: s.neshatpour@ip2i.in2p3.fr}
}
 
\vspace{1.cm}
{\em $^a$PRISMA Cluster of Excellence and  Institute for Physics (THEP),
Johannes Gutenberg University, D-55099 Mainz, Germany}\\[0.2cm]
{\em $^b$Universit\'e Claude Bernard Lyon 1, CNRS/IN2P3, \\
Institut de Physique des 2 Infinis de Lyon, UMR 5822, F-69622, Villeurbanne, France}\\[0.2cm]
{\em $^c$Theoretical Physics Department, CERN, CH-1211 Geneva 23, Switzerland}\\[0.2cm]
{\em $^d$Institut Universitaire de France (IUF), 75005 Paris, France }\\[0.2cm]

\end{center}

\renewcommand{\thefootnote}{\arabic{footnote}}
\setcounter{footnote}{0}

\vspace{1.cm}
\thispagestyle{empty}
\centerline{\bf ABSTRACT}
\vspace{0.5cm}
We present a critical assessment of the present $B$ anomalies in the exclusive $b \to s \ell\ell$ mode based on the QCD factorisation (QCDf) approach. In particular, we analyse the impact of different local form factor calculations and of the largest bin in the 
low-$q^2$ region. 

We also present a model-independent analysis of the new results of the LHCb and CMS experiments on the $B \to K^* \mu^+\mu^-$ angular observables. In addition, we update the global fit by including all $b \to s$ observables incorporating the new data from CMS and LHCb. In these analyses, we use 10\% or higher guesstimates of the non-factorisable power corrections as additional uncertainties, serving as a placeholder for robust estimates of these contributions.

Updating earlier results, we also analyse the combined LHCb and CMS data on the $B \to K^* \mu^+\mu^-$ angular observables using data-driven approaches to find indications whether these tensions between the QCDf predictions and the present data are due to underestimated subleading hadronic contributions or due to new physics effects.

\newpage

\section{Introduction}
Since the LHCb collaboration recently found the theoretically clean ratios namely $R_K$ and $R_{K^{*}}$ --which test lepton universality-- to be SM-like~\cite{LHCb:2022vje},  the focus is back on the long-standing tensions in the angular observables and branching ratios of exclusive $b \to s$ observables~\cite{LHCb:2013ghj,LHCb:2014cxe,LHCb:2015tgy,LHCb:2015wdu,LHCb:2015svh,LHCb:2020lmf,LHCb:2020gog,LHCb:2021xxq,LHCb:2021zwz}, which have very recently been confirmed again by CMS~\cite{CMS:2024atz} and LHCb~\cite{LHCb:2025update}. However, the theoretical description of these exclusive decays within the QCD factorisation (QCDf) approach~\cite{Beneke:2001at,Beneke:2004dp} is incomplete because non-factorisable power corrections cannot be calculated in general, in other words there is no exact SM prediction within this approach.  Therefore, it is not possible to decide whether the tensions are due to these unknown power corrections or due to new physics effects.

A possible solution to the problem was that the previously measured deviations in the theoretically clean ratios $R_K$ and $R_{K^{*}}$~\cite{LHCb:2017avl,LHCb:2021trn} were shown to be consistent with the deviations in the angular observables and branching ratios. In this way, the discovery of new physics in the ratios would also have indirectly confirmed the new physics interpretation of the deviations in the angular observables and branching ratios. 
In the present post-$R_{K^{(*)}}$ era, one must look for other solutions.

There are estimates on the power corrections based on analyticity and the $z$-expansion beyond the QCDf approach which leads to rather small values for these contributions~\cite{Bobeth:2017vxj,Gubernari:2020eft,Gubernari:2022hxn,Gopal:2024mgb}. But there are also claims that not all contributions are included in these analyses  like rescattering effects of intermediate hadrons~\cite{Ciuchini:2022wbq}.  A complete analysis of all the effects due to the anomalous thresholds does not exist yet~\cite{Mutke:2024tww}.
Moreover, the leading term in the computation in~\cite{Gubernari:2022hxn} is proportional to the local form factors (as is the leading term in QCDf). Therefore, updating local form factor predictions would also lead to a reevaluation of these non-local contributions.
There is also the claim (see Ref.~\cite{Piscopo:2023opf}, below Eq.~2.26) that for example in case of the vacuum-to-B three-particle matrix elements with the gluon and the light spectator quark the dominant region is the one in which the two fields are aligned on different light-cone 
directions;\footnote{Within the SCET context soft functions which {\it live} on both light cones have been recently discussed for exclusive and also for inclusive mode in Refs~\cite{Qin:2022rlk,Bartocci:2024bbf}.} thus, a  local light cone expansion of matrix elements with fields aligned on one light cone direction - as used in Refs.~\cite{Bobeth:2017vxj,Gubernari:2020eft,Gubernari:2022hxn} - might not lead to the complete result (see also \cite{Melikhov:2022wct,Melikhov:2023pet}).

The tensions in the exclusive $b \to s$ decays can also be verified by measurements of the corresponding inclusive decays, as was shown in Refs.~\cite{Huber:2024rbw,Huber:2020vup}. Such measurements are feasible at the BELLE-II experiment~\cite{Belle-II:2018jsg}, and may also be possible at the LHCb experiment, particularly in the high-$q^2$ region~\cite{Huber:2024rbw,Amhis:2021oik}. 

In principle, the new concept of refactorisation within the effective field theory \cite{Liu:2020wbn,Beneke:2022obx,Hurth:2023paz} can also lead to a solution by establishing  factorisation at the subleading level for such complicated decays in the long run. Then,  the non-factorisable power corrections would be calculable in the QCDf approach. 

In the meanwhile, it is possible to work out a model-independent analysis of all $b \to s$  observables, updating previous analyses~\cite{Hurth:2014vma,Hurth:2016fbr,Hurth:2017hxg,Hurth:2017sqw,Arbey:2019duh,Hurth:2020ehu,Hurth:2021nsi,Hurth:2023jwr,Mahmoudi:2024zna} and using a guesstimate of $10\%$ and also higher percentages for {the uncertainty due to} the non-factorisable power corrections. This serves as a placeholder for a robust estimate of these contributions. 
Furthermore, we also consider all observables in the high-$q^2$ region, where the Operator Product Expansion (OPE) and Heavy Quark Effective Theory (HQET) provide the theoretical description~\cite{Grinstein:2004vb,Bobeth:2011nj,Bobeth:2011gi,Bobeth:2011nj}. But quark-hadron duality violations could, in principle, introduce larger uncertainties. These observables in the high-$q^2$ region have less sensitivity to the NP Wilson coefficients.
There have been other groups presenting similar model-independent analyses~\cite{Ciuchini:2022wbq,Greljo:2022jac,Alguero:2023jeh,Ali:2025xkw}.

We might also tackle the problem using data-driven approaches in order to get indications about the nature of the tensions.  In the last years, we have presented various types of data-driven 
approaches~\cite{Chobanova:2017ghn,Neshatpour:2017qvi,Arbey:2018ics,Hurth:2020rzx,Hurth:2021nsi}. In the second part of the paper, we update them and get a combined picture. More recently, data-driven analyses were also presented in 
Refs~\cite{Alguero:2019ptt,Alguero:2023jeh,Bordone:2024hui}.

The paper is organised as follows: 
Section~\ref{sec:CMS_vs_LHCb} compares the recent CMS measurements~\cite{CMS:2024atz} of $B \to K^{*}\mu^+\mu^-$ angular observables with the corresponding LHCb results from 2020~\cite{LHCb:2020lmf} and 2025~\cite{LHCb:2025update}, discussing form factor choices, the treatment of the low-$q^2$ region, and the use of $P_i^{(\prime)}$ versus $S_i$ observables.
Section~\ref{sec:global_fits} presents updated global fits to all $b \to s\ell\ell$ data, examining the impact of the largest low-$q^2$ bins and varying assumptions on form factor choices and non-factorisable power corrections. Section~\ref{sec:data_driven} adopts a data-driven approach to contrast NP and hadronic fits, tests helicity and $q^2$ independence, and assesses the statistical preference between the different fits. Section~\ref{sec:conclusions} summarises our conclusions.

\section{Comparison of CMS and LHCb results}
\label{sec:CMS_vs_LHCb}

We compare the new measurements of angular observables in the $B\to K^* \mu^+\mu^-$ decay
by the CMS collaboration~\cite{CMS:2024atz} with the corresponding ones by the LHCb~collaboration~\cite{LHCb:2020lmf,LHCb:2025update}. 
The recent CMS measurement of the $B \to K^* \mu^+\mu^-$ angular observables is provided only for the $P_i^{(\prime)}$ basis~\cite{Matias:2012xw,Descotes-Genon:2013vna}. Therefore, for a more coherent comparison, we also use the $P_i^{(\prime)}$ observables for the LHCb measurements (unlike in our previous work~\cite{Hurth:2021nsi,Neshatpour:2022pvg,Hurth:2023jwr}, where we used the $S_i$  observables~\cite{Altmannshofer:2008dz}). We discuss the consequences of this choice below.

For the $B \to K^*$ form factors, we use the results from Ref.~\cite{Gubernari:2023puw}, denoted by ``GRvDV23'', as {the} default in our analysis, which is based on a combined fit to lattice QCD data~\cite{Horgan:2013hoa,Horgan:2015vla} and light-cone sum rule (LCSR) calculations~\cite{Gubernari:2018wyi,Gubernari:2020eft} with $B$-meson distribution amplitudes. We also compare these predictions with those based on the ``BSZ15'' form factors from Ref.~\cite{Bharucha:2015bzk}, which we used in our previous work. Those are based on a combined fit to the same lattice results~\cite{Horgan:2013hoa,Horgan:2015vla}
and {a} LCSR calculation with $K^*$-meson distribution amplitudes.

While the local hadronic matrix elements are described by form factors, the non-local, non-factorisable contributions are more involved. These are computed at leading order in $\Lambda/E_{K^*}$ and $\Lambda/m_b$ within the QCDf   framework~\cite{Beneke:2001at,Beneke:2004dp}. However, power corrections beyond this leading order are not calculable in this approach. Although several efforts have been made to estimate these contributions, they remain not fully under control~\cite{Khodjamirian:2010vf,Khodjamirian:2012rm,Bobeth:2017vxj,Asatrian:2019kbk,Gubernari:2020eft,Gubernari:2022hxn}.

For our SM predictions, we make a heuristic estimate of these power corrections by assigning a 10\% uncertainty at the amplitude level to the leading-order {non-factorisable}  QCDf terms. 
We also consider larger guesstimates of $50\%$ and $100\%$ in the global fit, including other modes (see section 3).

\begin{figure}[h!]
\begin{center}
\includegraphics[width=0.41\textwidth]{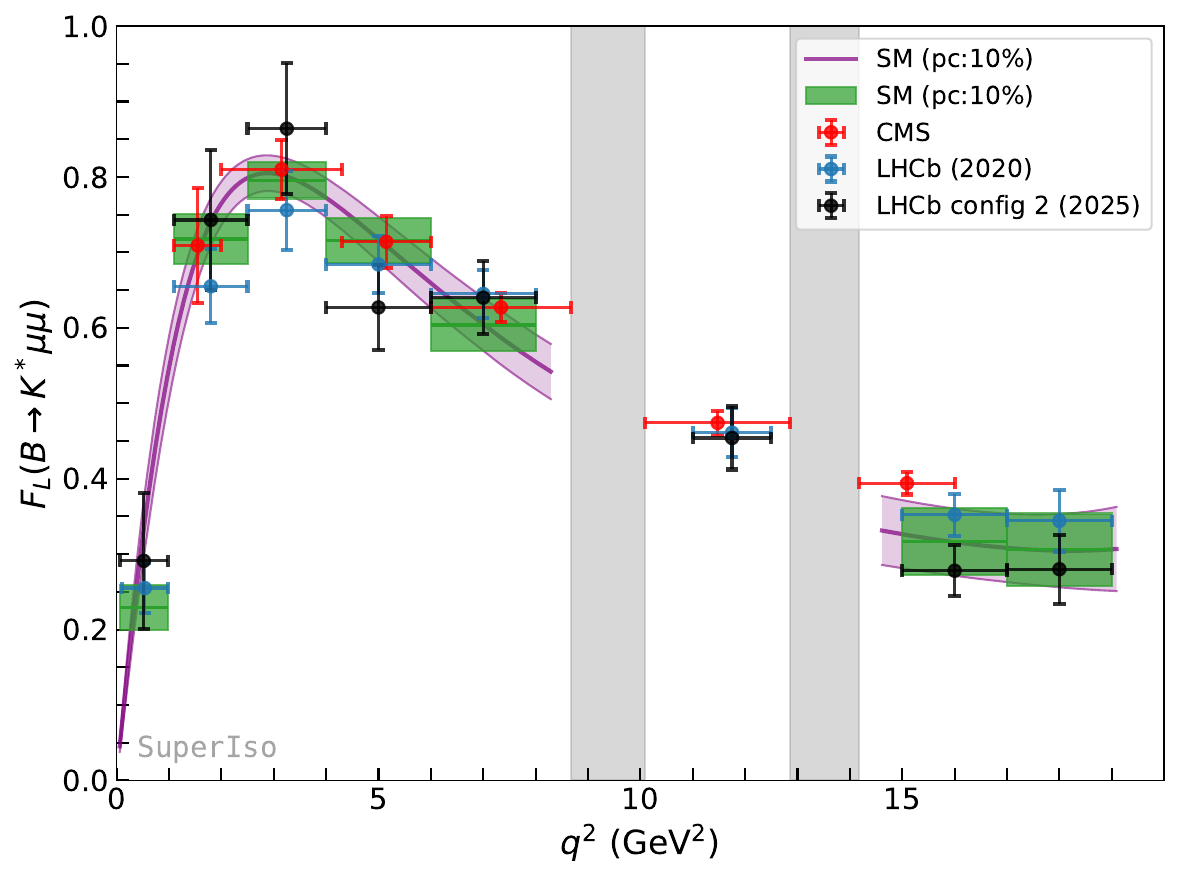}
\includegraphics[width=0.41\textwidth]{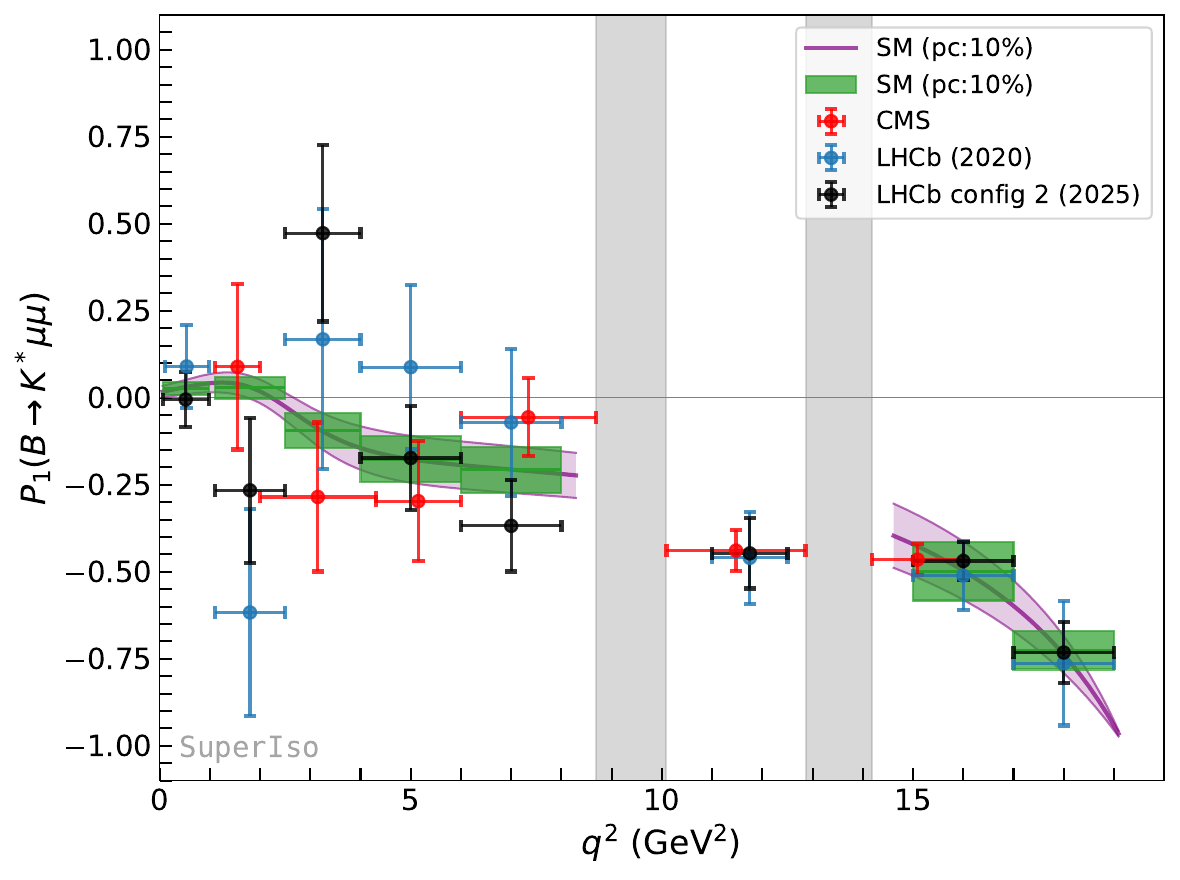}
\includegraphics[width=0.41\textwidth]{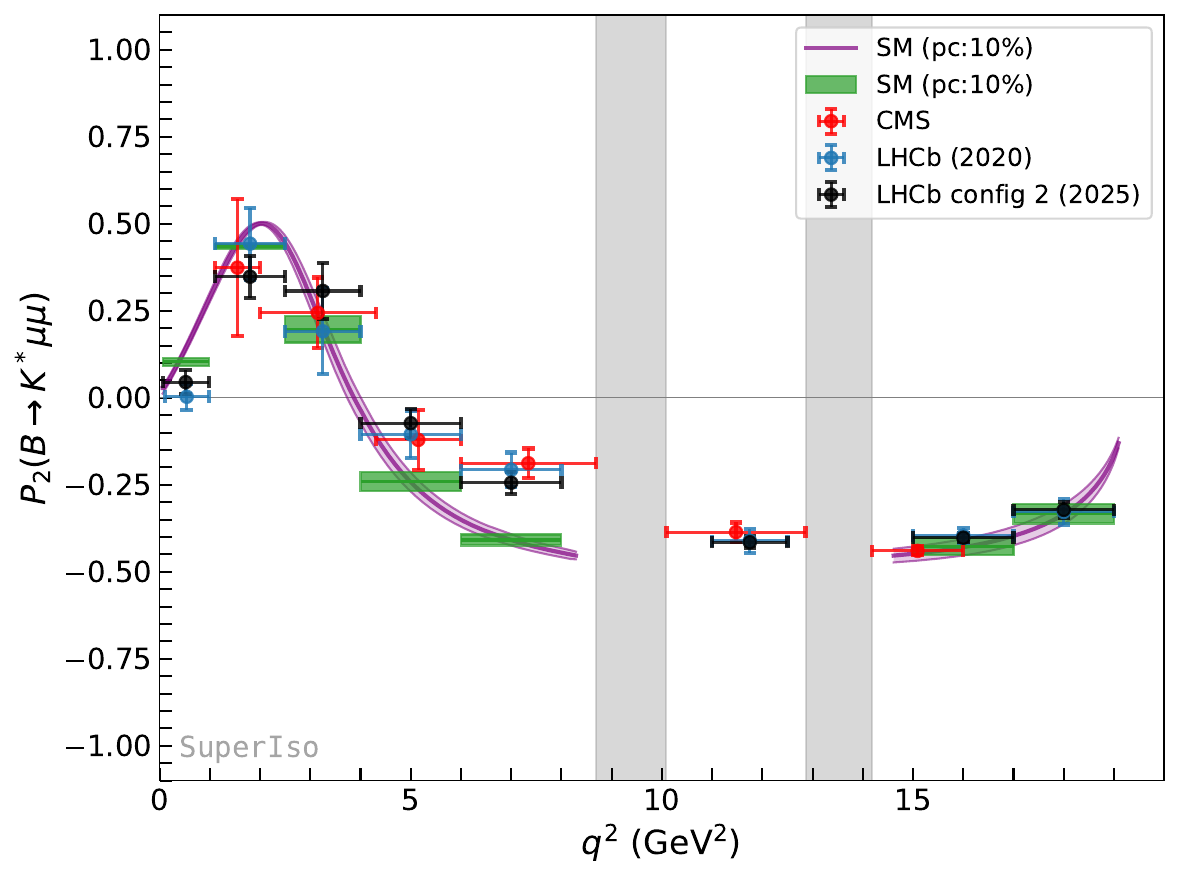}
\includegraphics[width=0.41\textwidth]{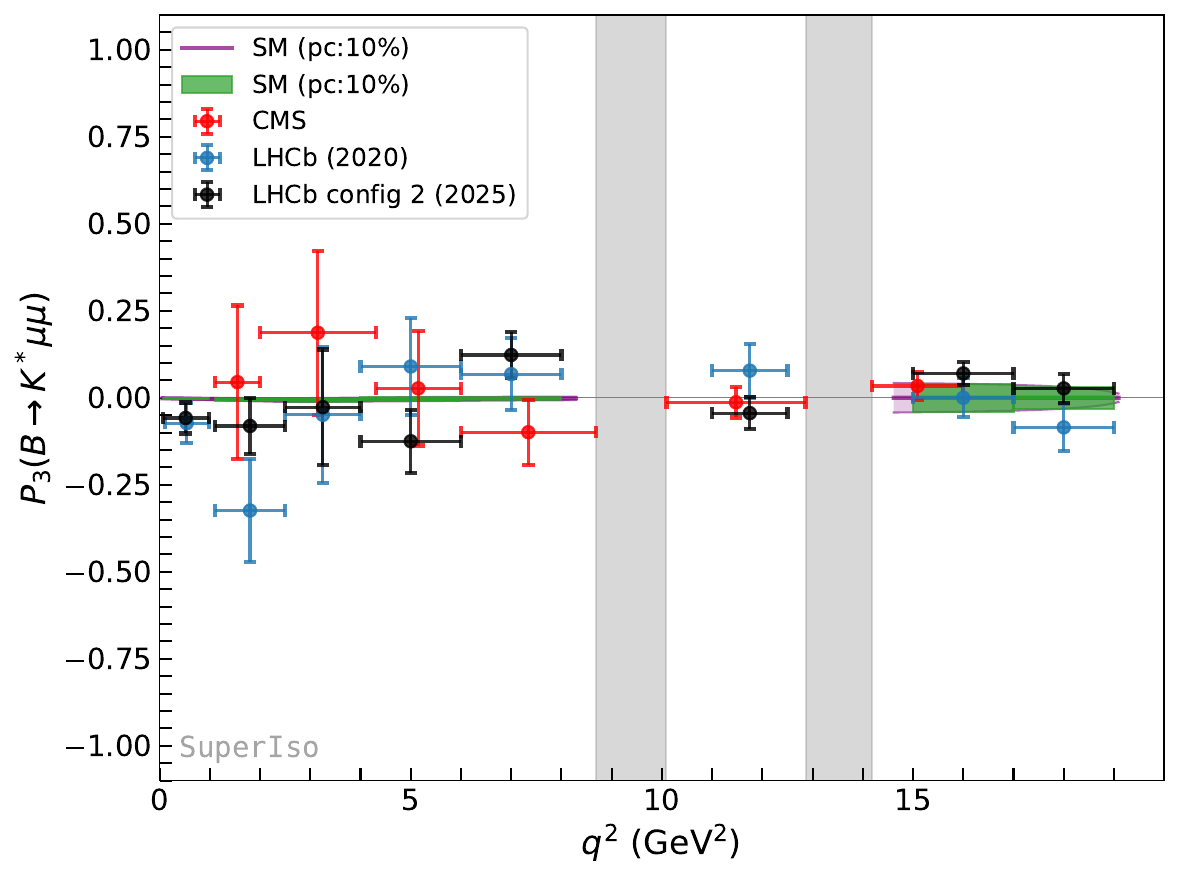}
\includegraphics[width=0.41\textwidth]{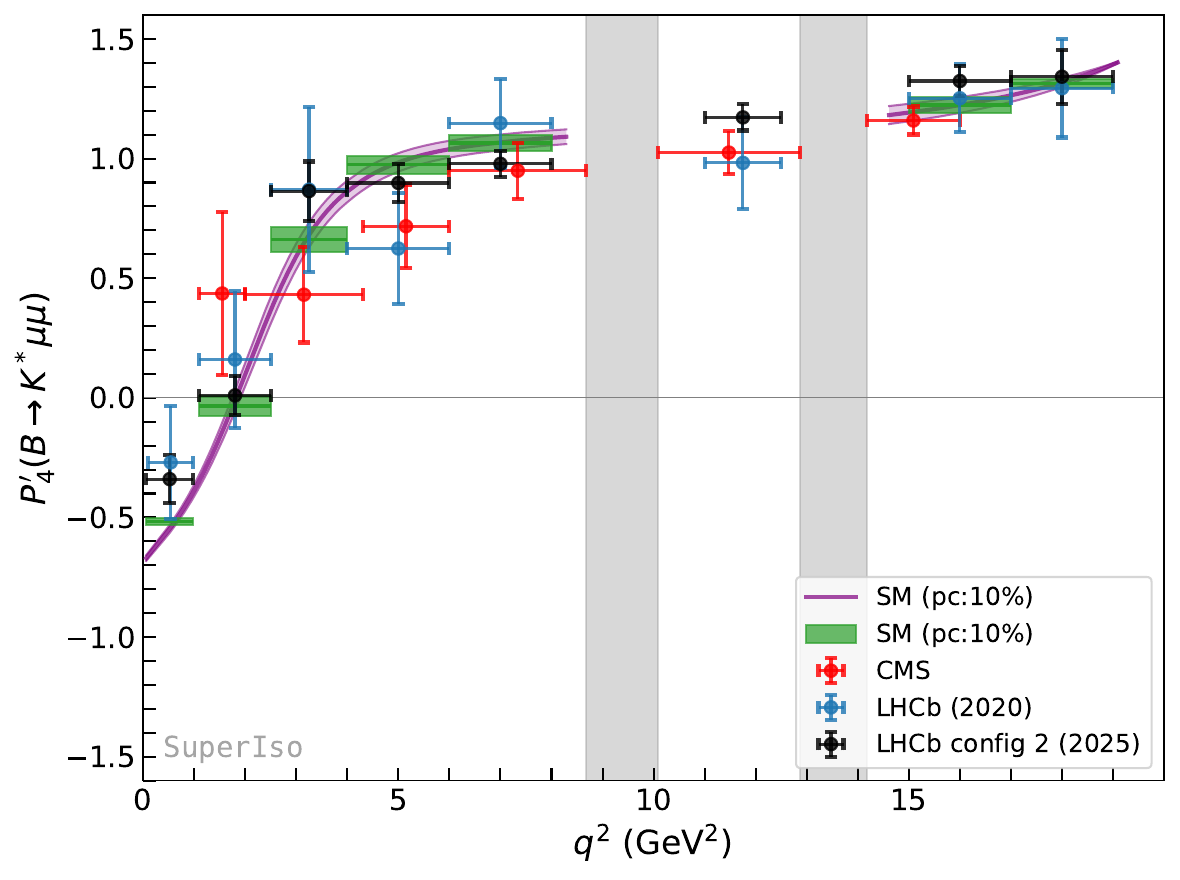}
\includegraphics[width=0.41\textwidth]{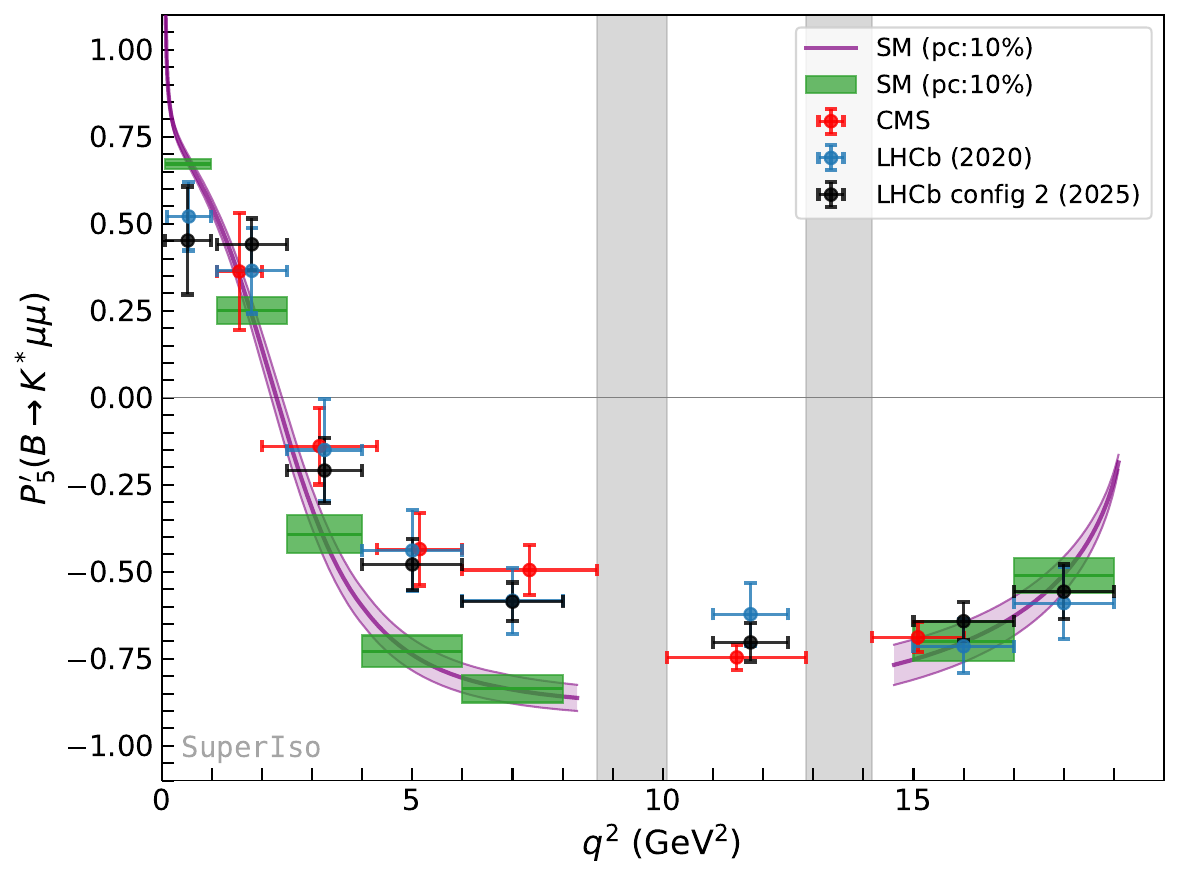}
\includegraphics[width=0.41\textwidth]{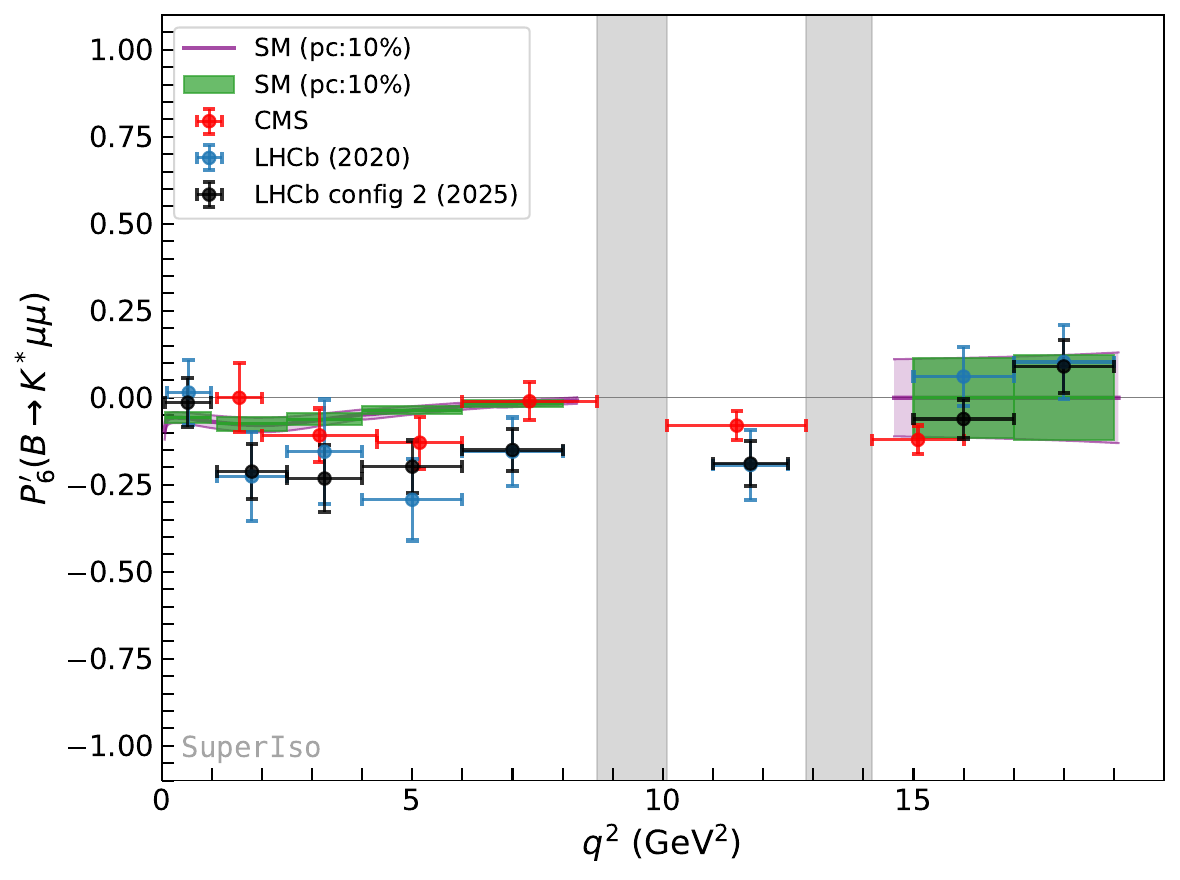}
\includegraphics[width=0.41\textwidth]{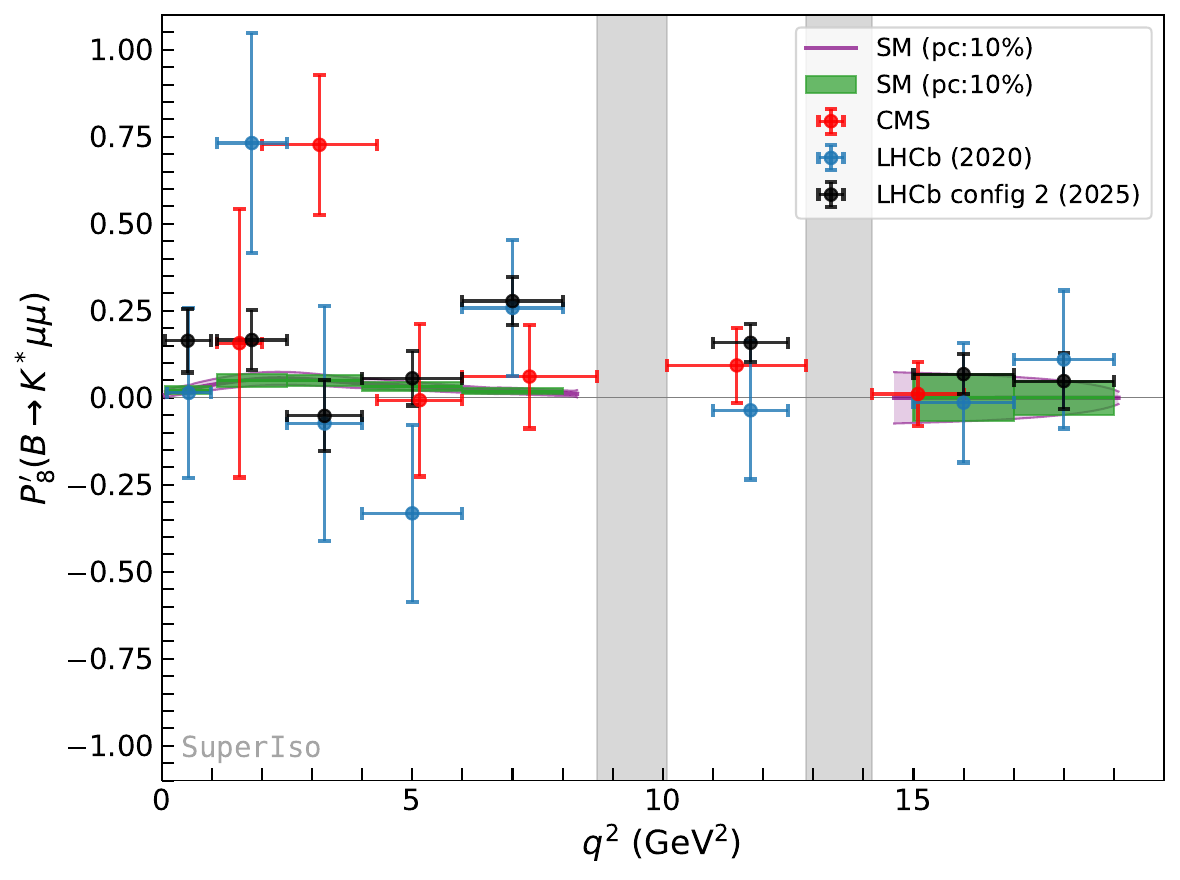}
\vspace{-0.2cm}
\caption{\small
$B^0 \to K^{*0}\mu^+\mu^-$ angular observables with SM predictions using GRvDV23 form factors and assuming 10\% power corrections. The data points correspond to the CMS measurement~\cite{CMS:2024atz} and the LHCb results from 2020~\cite{LHCb:2020lmf} and 2025~\cite{LHCb:2025update}.
\label{fig:B0Kstar0mumu_Pi_CMS_LHCb_SM_pc_10_100_200}}
\end{center}
\end{figure}

Additionally, since the QCDf framework is only valid in the low-$q^2$ region below the charm threshold, $q^2 < 4m_c^2 \approx 7\,\mathrm{GeV}^2$~\cite{Beneke:2001at,Beneke:2004dp}, the predictions for the largest low-$q^2$ bins (i.e., $[6,8]$ for LHCb and $[6,8.68]$ for CMS) should be treated with greater caution because the validity of the QCDf approach is not guaranteed in this region. 
{Therefore, for our default dataset in this paper --unless otherwise stated-- we include the low-$q^2$ region below the $J/\psi$ resonance, excluding the largest low-$q^2$ bins, and we also consider the high-$q^2$ region above the $\psi(2S)$ resonance.}
However, in the global fit we will analyse the impact of the largest low-$q^2$ bins (see section 3).

\subsection{Theory predictions vs measurements}
\label{sec:BKstarmumu_SM_vs_Exp}
In order to illustrate the consistency of the CMS and LHCb data on the various observables within the $B\to K^*\mu^+\mu^-$ decay, we show the measurements and the SM guesstimates for each observable separately in Figure~\ref{fig:B0Kstar0mumu_Pi_CMS_LHCb_SM_pc_10_100_200}.The CMS results correspond to the 140~fb$^{-1}$ dataset~\cite{CMS:2024atz}, while for LHCb we consider both the 2020 measurement based on 4.7~fb$^{-1}$~\cite{LHCb:2020lmf} and the recent 2025 update with 8.4~fb$^{-1}$ data~\cite{LHCb:2025update}. We first compare the CMS data with the LHCb 2020 results, and then separately with the updated LHCb 2025 results to illustrate the evolution in precision and consistency between the experiments.
We consider the SM predictions using GRvDV23-FF with 10\% power corrections both for the low- and high-$q^2$ bins. 
It is interesting that most observables are compatible with the QCDf predictions, including a $10\%$ guesstimate for the non-factorisable power corrections, but the individual deviations, for example in the observables $P_2$ and $P_5{^\prime}$, are consistently reproduced by both CMS and LHCb datasets.

\subsection{Comparison with one- or two-parameter NP fits}\label{sec:BKstarmumu_NPfit}

\begin{figure}[t!]
\begin{center}
\includegraphics[width=0.49\textwidth]{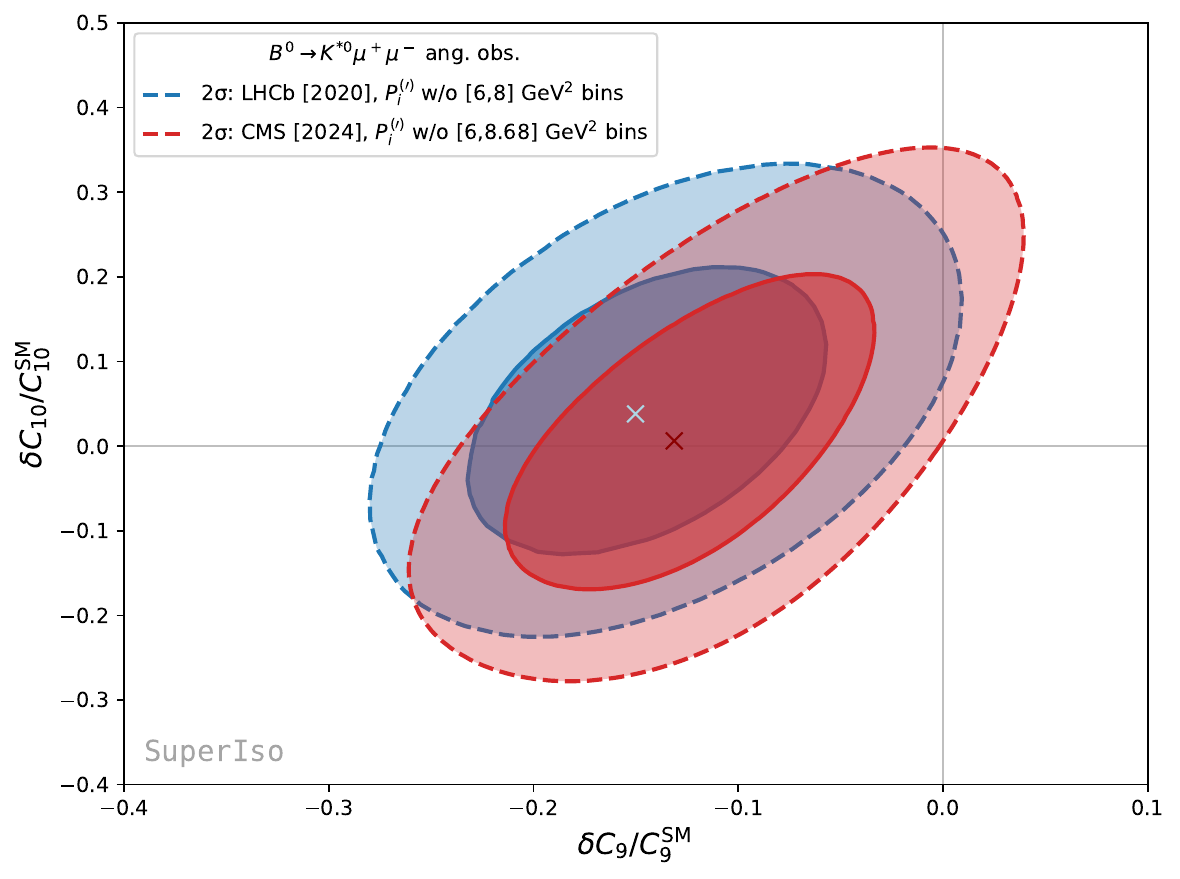}
\includegraphics[width=0.49\textwidth]{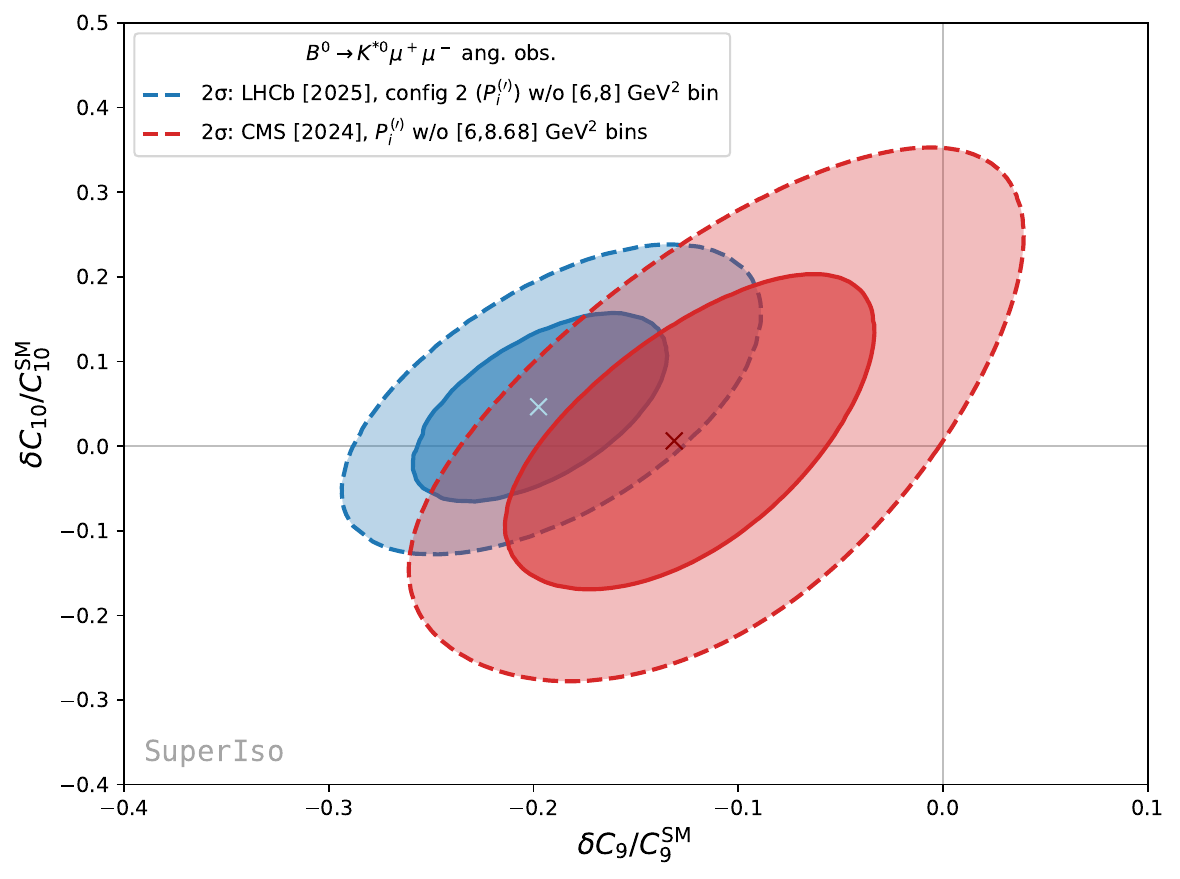}
\vspace{-0.2cm}
\caption{\small
The $1$ and $2\sigma$ confidence level (C.L.) of the $\{C_9, C_{10}\}$ fit to angular $B\to K^*\mu^+\mu^-$ observables (without $[6,8]$ and $[6.,8.68]$ GeV$^2$ bins), using the measurements from CMS~\cite{CMS:2024atz} and separately, two different LHCb results. The left panel uses the earlier LHCb 2020 data~\cite{LHCb:2020lmf}, while the right panel uses the updated LHCb 2025 result~\cite{LHCb:2025update}.
The fits lead to Pull$_\text{SM}$ of $2.0\sigma$ for CMS and 2.3 and 5.1$\sigma$ for LHCb 2020 and 2025 data, respectively.
} 
\label{fig:angObs_CMS_vs_LHCb_choice2}
\end{center}
\end{figure}

In Tables~\ref{tab:Pi_LHCb_CMS_1and2D} and Figure~\ref{fig:angObs_CMS_vs_LHCb_choice2} we compare the NP fit to the $P_i^{(\prime)}$ angular observables of $B\to K^* \mu^+\mu^-$ using the CMS measurement~\cite{CMS:2024atz} and both the earlier and updated LHCb results~\cite{LHCb:2020lmf,LHCb:2025update}.
The one-parameter NP fits show excellent consistency between the CMS measurement and the LHCb 2020 result, which had comparable uncertainties. In contrast, the very recent LHCb results, with their reduced uncertainties, yield higher tensions. It can also be seen that the two-parameter NP fits are compatible with each other at the 1-sigma level. The NP significances are $2.0\sigma$ for CMS and 2.3 and $5.1\sigma$ for the 2020 and 2025 LHCb results, respectively. 
This is a confirmation of the previous LHCb measurements, reaffirming the discrepancy with the QCDf predictions (with the 10\% guesstimate on non-factorisable power corrections).
In the rest of this paper, we will only consider the updated 2025 LHCb data\footnote{The LHCb collaboration provides the data in six different configurations; in this work, unless otherwise stated, we consider configuration 2, which includes the $P_i^{(\prime)}$ observables.}.
\begin{table}[th!]
\begin{center}
\setlength\extrarowheight{3pt}
\hspace*{-0.3cm}
\scalebox{0.8}{
{\color{black}
\begin{tabular}{|l|r|r|c|}
\hline 
\multicolumn{4}{|c|}{	\small Angular observables excluding $q^2\in[6,8.68]$ GeV$^2$ } \\[-4pt]							
\multicolumn{4}{|c|}{\small $P_i^{(\prime)}$ by CMS 2024\quad	 ($\chi^2_{\rm SM}=		39.0	$)} \\ \hline			
& b.f. value & $\chi^2_{\rm min}$ & ${\rm Pull}_{\rm SM}$  \\										
\hline \hline										
$\delta C_{9} $    	& $ 	-0.56	\pm	0.20	 $ & $ 	32.7	 $ & $	2.5	\sigma	 $  \\
\hline										
$\delta C_{10} $    	& $ 	-0.80	\pm	0.50	 $ & $ 	36.6	 $ & $	1.6	\sigma	 $  \\
\hline							          			
\hline										
$\delta C_{\rm LL}$	& $ 	-0.56	\pm	0.24	 $ & $ 	34.6	 $ & $	2.1	\sigma	 $  \\
\hline										
$\delta C_{\rm LR}$	& $ 	-0.39	\pm	0.14	 $ & $ 	33.2	 $ & $	2.4	\sigma	 $  \\
\hline										
$\delta C_{\rm RL}$	& $ 	0.00	\pm	0.13	 $ & $ 	34.4	 $ & $	2.1	\sigma	 $  \\
\hline										
$\delta C_{\rm RR}$	& $ 	0.28	\pm	0.25	 $ & $ 	37.7	 $ & $	1.1	\sigma	 $  \\
\hline										
\hline										
\multirow{2}{*}{$\{ \delta C_{9} ,\, \delta C_{10} \}$}	& $\delta C_{9} = 	-0.55	\pm	0.25	$ & \multirow{2}{*}{$	32.7	$} & \multirow{2}{*}{$	2.0	\sigma	$} \\
& $\delta C_{10} =		0.00	\pm	0.50						$ & & \\
\hline
\end{tabular}
}
}

\scalebox{0.8}{
\begin{tabular}{|l|r|r|c|}
\hline 
\multicolumn{4}{|c|}{	\small Angular observables excluding $q^2\in[6,8.68]$ GeV$^2$} \\[-4pt]							
\multicolumn{4}{|c|}{\small $P_i^{(\prime)}$ by LHCb~2020\quad		 ($\chi^2_{\rm SM}=		64.3	$)} \\ \hline			
& b.f. value & $\chi^2_{\rm min}$ & ${\rm Pull}_{\rm SM}$  \\										
\hline \hline										
$\delta C_{9} $    	& $ 	-0.66	\pm	0.21	 $ & $ 	56.7	 $ & $	2.8	\sigma	 $  \\
\hline										
$\delta C_{10} $    	& $ 	-0.70	\pm	0.50	 $ & $ 	62.1	 $ & $	1.5	\sigma	 $  \\
\hline							          			
\hline										
$\delta C_{\rm LL}$	& $ 	-0.55	\pm	0.24	 $ & $ 	60.1	 $ & $	2.1	\sigma	 $  \\
\hline										
$\delta C_{\rm LR}$	& $ 	-0.49	\pm	0.17	 $ & $ 	57.2	 $ & $	2.7	\sigma	 $  \\
\hline										
$\delta C_{\rm RL}$	& $ 	0.00	\pm	0.17	 $ & $ 	63.5	 $ & $	0.9	\sigma	 $  \\
\hline										
$\delta C_{\rm RR}$	& $ 	0.14	\pm	0.27	 $ & $ 	64.1	 $ & $	0.5	\sigma	 $  \\
\hline										
\hline										
\multirow{2}{*}{$\{ \delta C_{9} ,\, \delta C_{10} \}$}	& $\delta C_{9} = 	-0.63	\pm	0.24	$ & \multirow{2}{*}{$	56.6	$} & \multirow{2}{*}{$	2.3	\sigma	$} \\
& $\delta C_{10} =		-0.10	\pm	0.50						$ & & \\
\hline
\end{tabular}
{\color{black}
\begin{tabular}{|l|r|r|c|}
\hline 
\multicolumn{4}{|c|}{	\small Angular observables excluding $q^2\in[6,8.68]$ GeV$^2$ } \\[-4pt]							
\multicolumn{4}{|c|}{\small  $P_i^{(\prime)}$ (config 2) by LHCb~2025\quad ($\chi^2_{\rm SM}=		103.3	$)} \\ \hline			
& b.f. value & $\chi^2_{\rm min}$ & ${\rm Pull}_{\rm SM}$  \\										
\hline \hline										
$\delta C_{9} $    	& $ 	-0.89	\pm	0.14	 $ & $ 	74.2	 $ & $	5.4	\sigma	 $  \\
\hline										
$\delta C_{10} $    	& $ 	-1.02	\pm	0.30	 $ & $ 	91.5	 $ & $	3.4	\sigma	 $  \\
\hline							          			
\hline										
$\delta C_{\rm LL}$	& $ 	-0.71	\pm	0.17	 $ & $ 	90.8	 $ & $	3.5	\sigma	 $  \\
\hline										
$\delta C_{\rm LR}$	& $ 	-0.62	\pm	0.11	 $ & $ 	76.1	 $ & $	5.2	\sigma	 $  \\
\hline										
$\delta C_{\rm RL}$	& $ 	0.00	\pm	0.13	 $ & $ 	93.6	 $ & $	3.1	\sigma	 $  \\
\hline										
$\delta C_{\rm RR}$	& $ 	0.26	\pm	0.17	 $ & $ 	100.7	 $ & $	1.6	\sigma	 $  \\
\hline										
\hline										
\multirow{2}{*}{$\{ \delta C_{9} ,\, \delta C_{10} \}$}	& $\delta C_{9} = 	-0.84	\pm	0.17	$ & \multirow{2}{*}{$	73.9	$} & \multirow{2}{*}{$	5.1	\sigma	$} \\
& $\delta C_{10} =		-0.17	\pm	0.30						$ & & \\
\hline
\end{tabular}
}
}
\caption{Comparison of NP fit to the $P_i^{(\prime)}$ angular observables for $B\to K^* \mu^+ \mu^-$ with LHCb and CMS, excluding the $[6,8]$ and $[6,8.68]$ GeV$^2$ bins, respectively. 
The results are given assuming 10\% power corrections. The upper panel shows the CMS results, while the lower panel shows the LHCb measurements from 2020 (left) and 2025 (right).
\label{tab:Pi_LHCb_CMS_1and2D}
} 
\end{center} 
\end{table}
%

\begin{figure}[h!]
\begin{center}
\includegraphics[width=0.7\textwidth]{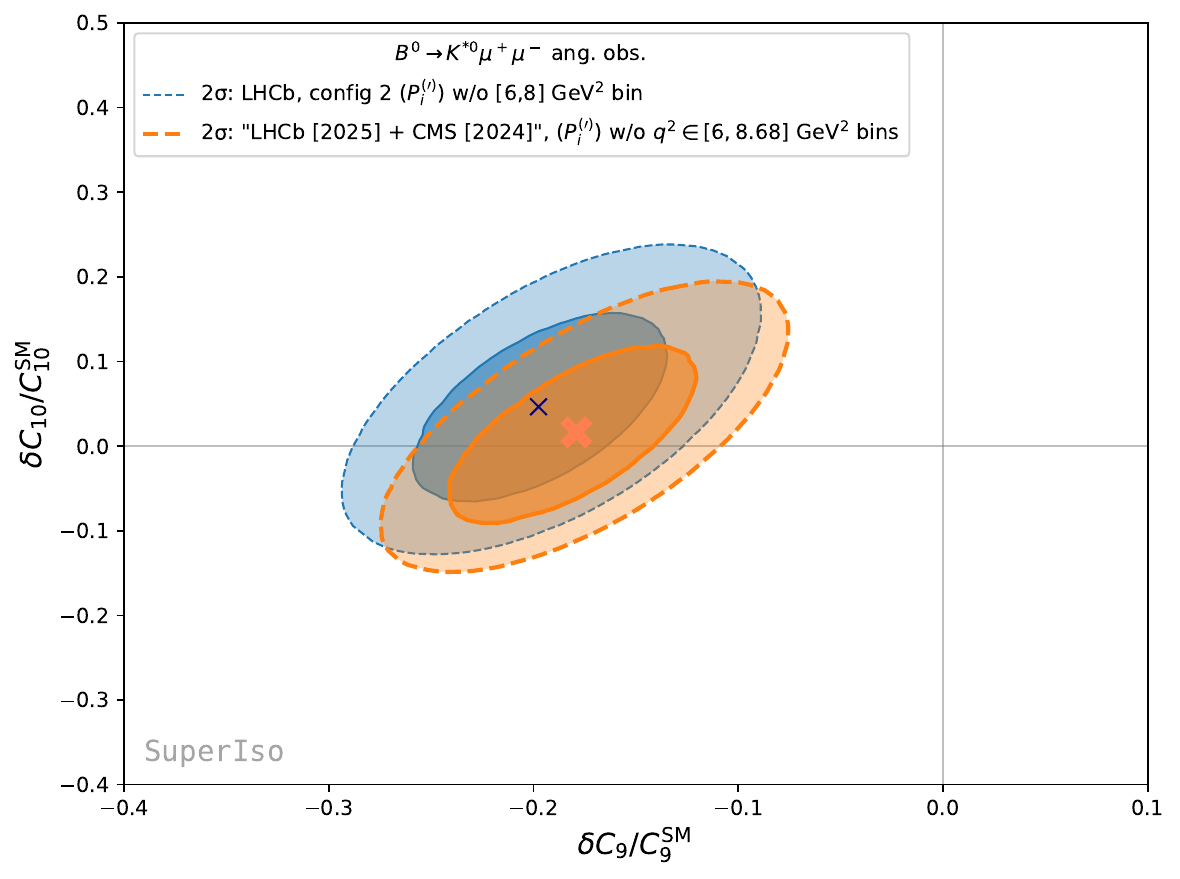}
\vspace{-0.2cm}
\caption{\small
The $1\sigma$ and $2\sigma$ C.L. regions for the $\{C_9, C_{10}\}$ fit to angular observables in $B \to K^* \mu^+ \mu^-$, when a fit to the LHCb measurements~\cite{LHCb:2020lmf} with and without CMS data~\cite{CMS:2024atz} (excluding the $q^2 \in [6, 8.68]~\text{GeV}^2$ bins in both cases) is made, with Pull$_\text{SM}$ of $5.1$ and $4.9\sigma$, respectively.
\label{fig:angObs_LHCb_with_or_without_CMS}}
\end{center}
\end{figure}

The impact of adding the CMS data to the LHCb data is shown explicitly in Figure~\ref{fig:angObs_LHCb_with_or_without_CMS}, where the two-parameter NP fit to the LHC measurements with and without CMS data is shown. The NP significance decreases slightly from 5.1 to 4.9$\sigma$. This reduction arises because the CMS data prefers a smaller NP contribution to $\delta C_9$ than suggested by the updated LHCb measurements.

\begin{figure}[h!]
\begin{center}
\includegraphics[width=0.7\textwidth]{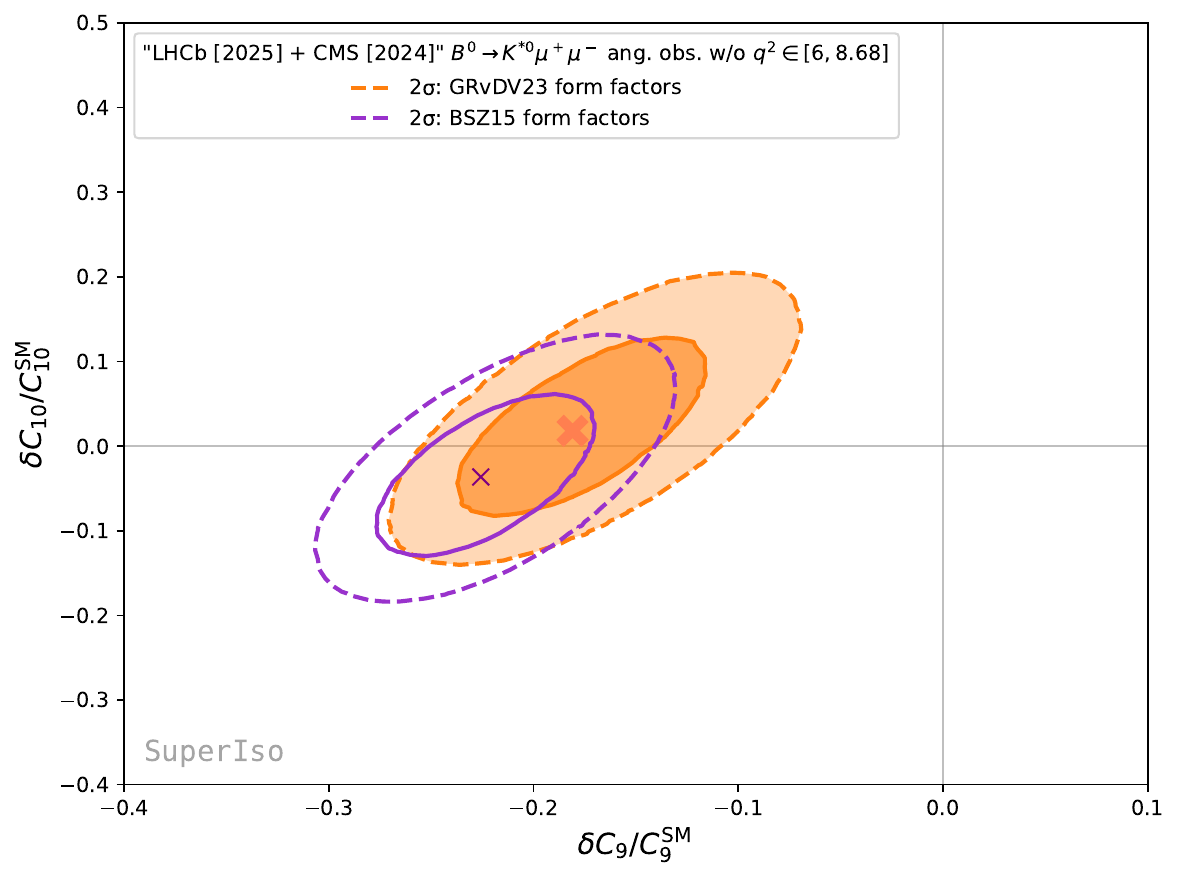}
\vspace{-0.2cm}
\caption{\small
The $1\sigma$ and $2\sigma$ confidence level regions for the $\{C_9, C_{10}\}$ fit to angular observables in $B \to K^* \mu^+ \mu^-$, when in the combined LHCb and CMS fit two different sets of form factors: GRvDV23-FF (orange region) and BSZ15-FF (purple contours) are used, resulting in Pull$_\text{SM}$ of $4.9$ and $6.1\sigma$, respectively.
\label{fig:angObs_LHCb_CMS_Pi_choice_1_2}}
\end{center}
\end{figure}

We note that the choice of local form factors has a major impact, even on the angular observables fit. Figure~\ref{fig:angObs_LHCb_CMS_Pi_choice_1_2} shows the two-parameter fit to the combined CMS and LHCb data when LCSR with $B$ meson amplitudes (GRvDV23) or with $K^*$ meson amplitudes (BSZ15) are considered for the theoretical prediction. In both cases, the LCSR results are combined with the same lattice results (see above for the references). The NP significance is in the former case 4.9$\sigma$, in the latter 6.1$\sigma$.  The authors of Ref.\cite{Gubernari:2022hxn} observe a similar pattern with the older experimental results.

\begin{figure}[t!]
\begin{center}
\includegraphics[width=0.7\textwidth]{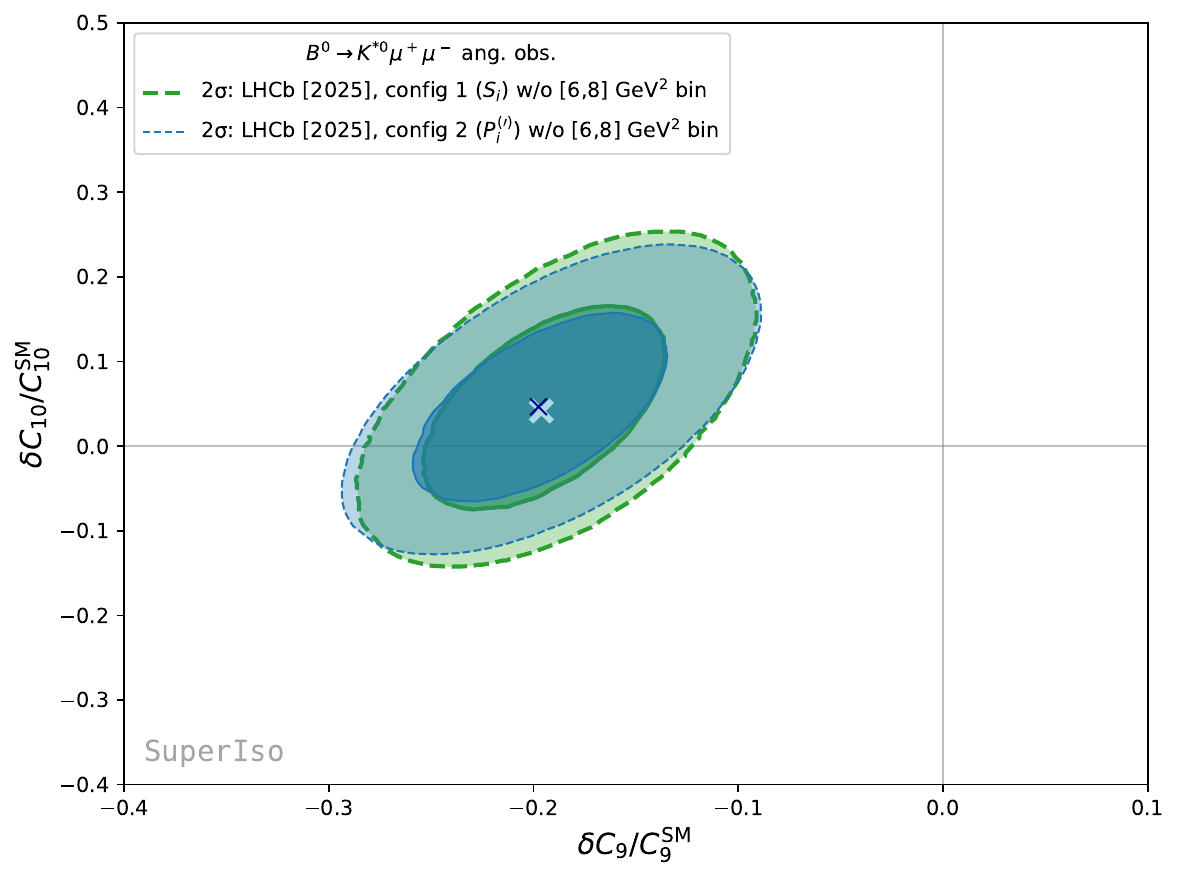}
\vspace{-0.2cm}
\caption{\small
The $1$ and $2\sigma$ C.L. of the $\{C_9, C_{10}\}$ fit to angular $B\to K^*\mu^+\mu^-$ observables, using the $P_i^{(\prime)}$ and $S_i$ measurements from LHCb~\cite{LHCb:2020lmf}, excluding the $[6,8]$ GeV$^2$ bins, resulting in Pull$_\text{SM}$ of $5.1$ and $4.8\sigma$, respectively.
\label{fig:angObs_LHCb_Pi_vs_Si_choice2}}
\end{center}
\end{figure}

Finally, we test, using the LHCb data, whether the choice of the normalised $P_i^{(\prime)}$ observables (config.~2) instead of the $S_i$ observables (config.~1) makes a difference.
Figure~\ref{fig:angObs_LHCb_Pi_vs_Si_choice2} shows that  the best-fit points coincide perfectly, and both configurations yield very similar Pull$_{\rm SM}$ values, with config.~2 giving 5.1$\sigma$ and config.~1 giving 4.8$\sigma$. 

\clearpage 
\section{Global fits}
\label{sec:global_fits} 
In this section, we consider global fits including all $b\to s \ell \ell$ observables. We also explore the impact of the highest low-$q^2$ bins we have left out in the {previous section} because the validity of the QCDf approach in these highest low-$q^2$ bins is questionable. 
Thus, we consider two datasets in our global fit: one including these bins (containing 263 observables), and one excluding them (with 230 observables). The full list of observables for both datasets is provided as ancillary files and also given in Appendix~\ref{sec:ObsList}. These can be used with \texttt{SuperIso}~\cite{Mahmoudi:2007vz,Mahmoudi:2008tp,Mahmoudi:2009zz,Neshatpour:2022fak} to obtain their SM or NP predictions.

Unless otherwise stated, we use the following form factor inputs for the exclusive semileptonic decays involving $B \to V$ and $B \to P$ transitions: For the $B \to K$ form factors, we adopt the combined fit to light-cone sum rule and lattice QCD results from Ref.~\cite{Gubernari:2018wyi} (GKvD18), consistent with our previous analyses~\cite{Hurth:2021nsi,Neshatpour:2022pvg,Hurth:2023jwr}. The LCSR calculation is based on $B$-meson distribution amplitudes, while the lattice input is taken from Ref.~\cite{Bouchard:2013eph}. In Appendix~\ref{app:BR_BKmumu_FFcheck}, we again illustrate the impact of alternative form factor inputs on the SM prediction for $\text{BR}(B \to K \mu^+ \mu^-)$ as done in the last section for the angular observables in the $B \to K^* \ell^+\ell^-$ mode. For $B \to K^*$ and $B_s \to \phi$, we use the form factors provided in Ref.~\cite{Gubernari:2023puw} (GRvDV23), which are based on a combined fit to lattice QCD data~\cite{Horgan:2013hoa,Horgan:2015vla} and light-cone sum rule calculations~\cite{Gubernari:2018wyi,Gubernari:2020eft} with $B$-meson distribution amplitudes. This represents an update to our earlier work, in which we used the BSZ15 results~\cite{Bharucha:2015bzk} with light-meson distribution amplitudes combined with the same lattice data~\cite{Horgan:2013hoa,Horgan:2015vla}. In Appendix~\ref{app:P5prime_B0Kstar0mumu_FFcheck} we show the impact of the different choices in the case of the observables $P_2$ and $P_5^\prime$ for the $B\to K^* \mu^+\mu^-$ decay, while Appendix~\ref{app:BR_Bsphimumu_FFcheck} illustrates their effect on BR($B_s \to \phi\mu^+\mu^-$).
 
Our current choice ensures a consistent treatment across decay modes using form factors derived from LCSR calculations with $B$-meson distribution amplitudes.

In Table~\ref{tab:2025GRvDV_all_vs_allwo68_1D} the results of one-parameter NP fits without (left) and with the highest bins above 6 $ {\rm GeV}^2$ (right) are shown. 
The NP Wilson coefficients $\delta C_7$ and $\delta C_{10}$ get strongly constrained by the radiative penguin
decays $b \to s \gamma$ and the $B_s \to \mu^+\mu^-$ decay, respectively. 
The NP significance in the one-parameter fits to the Wilson coefficients $\delta C_9$ and also $\delta C_{LR}$ changes dramatically when the highest low-$q^2$ bins are included; they change from $5.2\sigma$ to $7.4\sigma$ and from $4.1\sigma$ to $6.1\sigma$ respectively.\footnote{In $C_{XY}$, $X$ refers to the chirality of the quark current and $Y$ to the chirality of the lepton current. Thus, we have
\begin{equation}
    \delta C_{LL} \equiv \delta C_9 = - \delta C_{10}, \nonumber
    \hspace{0.5cm} 
    \delta C_{RL} \equiv \delta C^\prime_{9} = - \delta C^\prime_{10} \, ,\quad
    \delta C_{RR} \equiv \delta C_9^\prime =  \delta C^\prime_{10}, \nonumber
    \hspace{0.5cm} 
    \delta C_{LR} \equiv \delta C_{9} =  \delta C_{10} \, .
\end{equation}}
We emphasise again that these values are obtained assuming a 10\% guesstimate for the power corrections and with the mentioned choices of form factors.
The impact of $\delta C_9$ on key $b \to s \ell\ell$ observables is further illustrated in Appendix~\ref{sec:dC9_effects}.

\begin{figure}[t!]
\begin{center}
\includegraphics[width=0.7\textwidth]{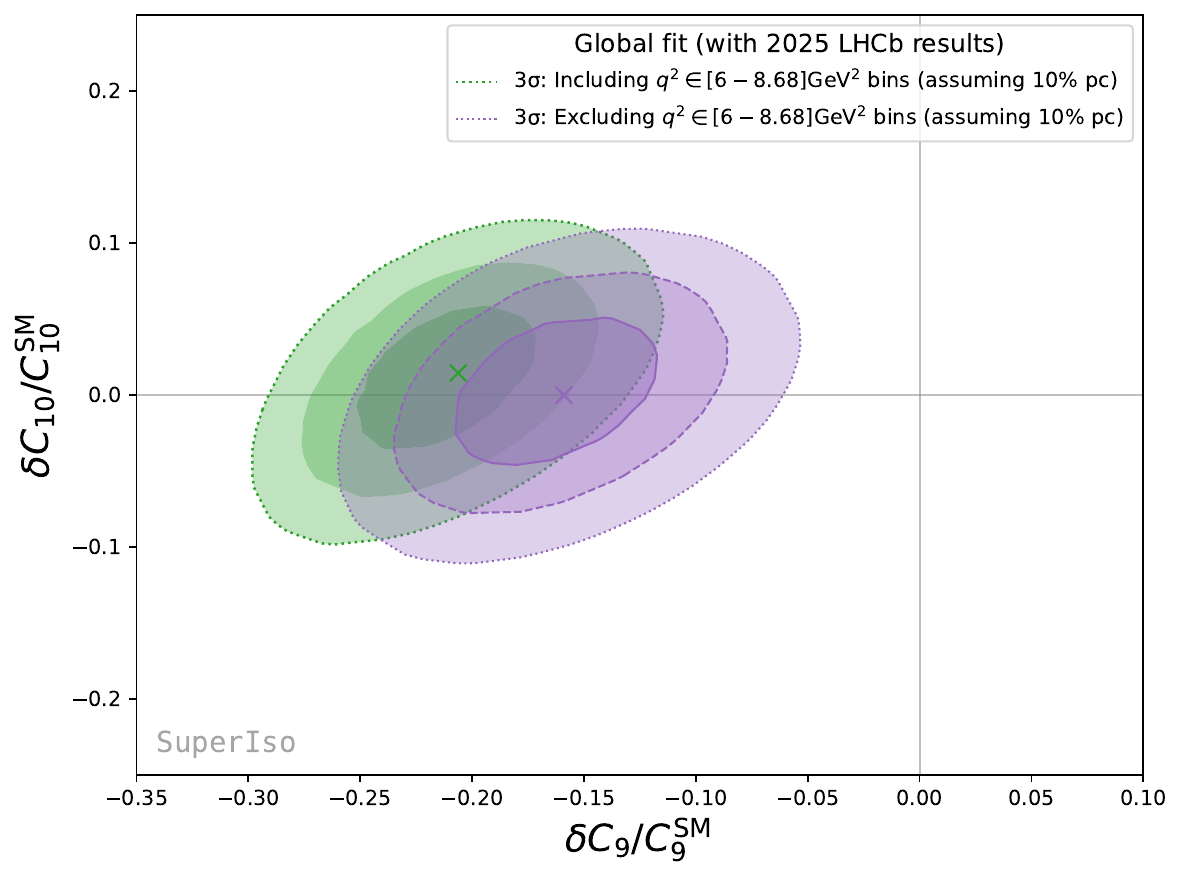}
\vspace{-0.2cm}
\caption{\small
Two-dimensional fit of $\{C_9, C_{10}\}$ to all observables with (green) or without (purple) $q^2 \in [6,8.68]$ GeV$^2$ bins with Pull$_\text{SM}$ of 4.8 and 7.1$\sigma$, respectively. These values are obtained assuming a 10\% guesstimate for the power corrections.
\label{fig:global_fit_with_vs_without_6_8}}
\end{center}
\end{figure}

Analogously, the NP significance in the two-parameter fits, shown in Figure~\ref{fig:global_fit_with_vs_without_6_8},  changes from $4.8\sigma$ to $7.1\sigma$.
One can speculate that this drastic increase in NP significance is due to the fact that these high bins in the low-$q^2$ bins just below the resonances are not well described by the QCDf approach. This circumstance makes it very difficult to establish a conclusion regarding new physics based on these results in the largest low-$q^2$ bins within this framework. 

\begin{table}[t!]
\begin{center}
\setlength\extrarowheight{3pt}
\hspace*{-0.3cm}
\setlength\extrarowheight{3pt}
\hspace*{-0.3cm}
\scalebox{0.85}{
{\color{black}
\begin{tabular}{|l|c|r|c|}
\hline 
\multicolumn{4}{|c|}{	\small All observables excluding $q^2 \in [6,8.68]$ GeV$^2$ bins		} \\[-4pt]							
\multicolumn{4}{|c|}{\small (Config. 2 for $B\to K^*\mu\mu$ by LHCb);\; $\chi^2_{\rm SM}=		329.3	$} \\ \hline			
& b.f. value & $\chi^2_{\rm min}$ & ${\rm Pull}_{\rm SM}$  \\										
\hline \hline										
$\delta C_7$	& $ 	-0.01	\pm	0.01	 $ & $ 	328.0	 $ & $	1.1	\sigma	 $  \\
$\delta C_{Q_{1}} $    	& $ 	-0.02	\pm	0.06	 $ & $ 	329.3	 $ & $	0.0	\sigma	 $  \\
$\delta C_{Q_{2}} $    	& $ 	0.00	\pm	0.01	 $ & $ 	329.3	 $ & $	0.0	\sigma	 $  \\
\hline										
$\delta C_{9} $    	& $ 	-0.69	\pm	0.12	 $ & $ 	302.6	 $ & $	5.2	\sigma	 $  \\
\hline										
$\delta C_{10} $    	& $ 	-0.19	\pm	0.12	 $ & $ 	326.9	 $ & $	1.5	\sigma	 $  \\
\hline							          			
\hline										
$\delta C_{\rm LL}$	& $ 	-0.31	\pm	0.13	 $ & $ 	323.0	 $ & $	2.5	\sigma	 $  \\
\hline										
$\delta C_{\rm LR}$	& $ 	-0.34	\pm	0.08	 $ & $ 	312.7	 $ & $	4.1	\sigma	 $  \\
\hline										
$\delta C_{\rm RL}$	& $ 	0.00	\pm	0.08	 $ & $ 	326.1	 $ & $	1.8	\sigma	 $  \\
\hline										
$\delta C_{\rm RR}$	& $ 	0.11	\pm	0.11	 $ & $ 	328.3	 $ & $	1.0	\sigma	 $  \\
\hline										
\hline										
\multirow{2}{*}{$\{ \delta C_{9} ,\, \delta C_{10} \}$}	& \small $\delta C_{9} = 	-0.69	\pm	0.12	$ & \multirow{2}{*}{$	302.6	$} & \multirow{2}{*}{$	4.8	\sigma	$} \\
&\small $\delta C_{10} =		-0.01	\pm	0.13						$ & & \\
\hline
\end{tabular}
}
{\color{black}
{
\begin{tabular}{|l|c|r|c|}
\hline 
\multicolumn{4}{|c|}{	\small All observables including $q^2 \in [6,8.68]$ GeV$^2$ bins		} \\[-4pt]							
\multicolumn{4}{|c|}{\small (Config. 2 for $B\to K^*\mu\mu$ by LHCb);\; $\chi^2_{\rm SM}=		431.7	$} \\ \hline			
& b.f. value & $\chi^2_{\rm min}$ & ${\rm Pull}_{\rm SM}$  \\										
\hline \hline										
$\delta C_7$	& $ 	-0.02	\pm	0.01	 $ & $ 	429.3	 $ & $	1.5	\sigma	 $  \\
$\delta C_{Q_{1}} $    	& $ 	-0.02	\pm	0.05	 $ & $ 	431.7	 $ & $	0.0	\sigma	 $  \\
$\delta C_{Q_{2}} $    	& $ 	0.00	\pm	0.01	 $ & $ 	431.7	 $ & $	0.0	\sigma	 $  \\
\hline										
$\delta C_{9} $    	& $ 	-0.92	\pm	0.10	 $ & $ 	377.6	 $ & $	7.4	\sigma	 $  \\
\hline										
$\delta C_{10} $    	& $ 	-0.32	\pm	0.12	 $ & $ 	424.3	 $ & $	2.7	\sigma	 $  \\
\hline							          			
\hline										
$\delta C_{\rm LL}$	& $ 	-0.35	\pm	0.13	 $ & $ 	424.0	 $ & $	2.8	\sigma	 $  \\
\hline										
$\delta C_{\rm LR}$	& $ 	-0.47	\pm	0.07	 $ & $ 	394.6	 $ & $	6.1	\sigma	 $  \\
\hline										
$\delta C_{\rm RL}$	& $ 	0.00	\pm	0.09	 $ & $ 	428.1	 $ & $	1.9	\sigma	 $  \\
\hline										
$\delta C_{\rm RR}$	& $ 	0.07	\pm	0.11	 $ & $ 	431.4	 $ & $	0.5	\sigma	 $  \\
\hline										
\hline										
\multirow{2}{*}{$\{ \delta C_{9} ,\, \delta C_{10} \}$}	&\small $\delta C_{9} = 	-0.89	\pm	0.11	$ & \multirow{2}{*}{$	377.4	$} & \multirow{2}{*}{$	7.1	\sigma	$} \\
&\small $\delta C_{10} =		-0.05	\pm	0.13						$ & & \\
\hline
\end{tabular}
}
}
}
\caption{Comparison of NP fit excluding (including) the LHCb [6, 8] and CMS [6, 8.68] GeV$^2$ bins on the left (right). These values are obtained assuming a 10\% guesstimate for the power corrections.
\label{tab:2025GRvDV_all_vs_allwo68_1D}} 
\end{center} 
\end{table}
%
%

For the non-factorisable power corrections in the $B \to K^{(*)} \ell\ell$ and $B_s \to \phi \ell\ell$ decays, we have always assumed a 10\% uncertainty to the leading-order non-factorisable QCDf amplitude as the default value in our model-independent analyses in the past. Here we analyse how much an increase in the guesstimate of the non-local power corrections (by increasing it from 10\% to 50\% and 100\%) affects the NP significance in the global fit. As noted in the introduction, these guesstimates serve only as a placeholder until solid and robust estimates of these power corrections are available.

The results are given in Tables~\ref{tab:allobs_except_608_GRvDV23_50_100_nonFactErr} and clearly indicate that $100\%$  non-factorisable power corrections compared to the leading non-factorisable QCDf amplitude are needed on average, to get the one-parameter fit to the NP Wilson
coefficient $\delta C_9$ down to 2.3$\sigma$. Such large contributions are not currently expected in view of the analysis in Ref.~\cite{Gubernari:2022hxn}, which predicts a very small non-local contribution. If there are corrections or additional contributions, it is not expected that they will induce an order-of-magnitude change. However, the result demonstrates that the deviation between data and QCDf predictions is rather significant.

\begin{table}[h!]
\begin{center}
\setlength\extrarowheight{3pt}
\hspace*{-0.5cm}
\scalebox{0.6}{
{\color{black}
\begin{tabular}{|l |c|r|c|}
\hline 
\multicolumn{4}{|c|}{	\small All observables except $q^2 \in$ [6-8.68] GeV$^2$ bins 	} \\[-4pt]							
\multicolumn{4}{|c|}{\small{\bf 	10\% pc	} \; ($\chi^2_{\rm SM}=		329.3	$)} \\ \hline			
& b.f. value & $\chi^2_{\rm min}$ & ${\rm Pull}_{\rm SM}$  \\										
\hline \hline									
$\delta C_{9} $    	& $ 	-0.69	\pm	0.12	 $ & $ 	302.6	 $ & $	5.2	\sigma	 $  \\
\hline										
$\delta C_{10} $    	& $ 	-0.19	\pm	0.12	 $ & $ 	326.9	 $ & $	1.5	\sigma	 $  \\
\hline							          			
\hline										
$\delta C_{\rm LL}$	& $ 	-0.31	\pm	0.13	 $ & $ 	323.0	 $ & $	2.5	\sigma	 $  \\
\hline										
$\delta C_{\rm LR}$	& $ 	-0.34	\pm	0.08	 $ & $ 	312.7	 $ & $	4.1	\sigma	 $  \\
\hline										
$\delta C_{\rm RL}$	& $ 	0.00	\pm	0.08	 $ & $ 	326.1	 $ & $	1.8	\sigma	 $  \\
\hline										
$\delta C_{\rm RR}$	& $ 	0.11	\pm	0.11	 $ & $ 	328.3	 $ & $	1.0	\sigma	 $  \\
\hline										
\hline										
\multirow{2}{*}{$\{ \delta C_{9} ,\, \delta C_{10} \}$}	& \small $\delta C_{9} = 	-0.69	\pm	0.12	$ & \multirow{2}{*}{$	302.6	$} & \multirow{2}{*}{$	4.8	\sigma	$} \\
&\small $\delta C_{10} =		-0.01	\pm	0.13						$ & & \\
\hline
\end{tabular}
}
{\color{black}
\begin{tabular}{|l|c|r|c|}
\hline 
\multicolumn{4}{|c|}{	\small All observables except $q^2 \in$ [6-8.68] GeV$^2$ bins 		} \\[-4pt]							
\multicolumn{4}{|c|}{\small{\bf 	50\% pc	} \; ($\chi^2_{\rm SM}=		304.0	$)} \\ \hline			
& b.f. value & $\chi^2_{\rm min}$ & ${\rm Pull}_{\rm SM}$  \\										
\hline \hline										
$\delta C_{9} $    	& $ 	-0.64	\pm	0.17	 $ & $ 	293.8	 $ & $	3.2	\sigma	 $  \\
\hline										
$\delta C_{10} $    	& $ 	-0.03	\pm	0.14	 $ & $ 	304.0	 $ & $	0.0	\sigma	 $  \\
\hline							          			
\hline										
$\delta C_{\rm LL}$	& $ 	-0.19	\pm	0.13	 $ & $ 	301.8	 $ & $	1.5	\sigma	 $  \\
\hline										
$\delta C_{\rm LR}$	& $ 	-0.23	\pm	0.11	 $ & $ 	300.2	 $ & $	1.9	\sigma	 $  \\
\hline										
$\delta C_{\rm RL}$	& $ 	0.00	\pm	0.08	 $ & $ 	302.7	 $ & $	1.1	\sigma	 $  \\
\hline										
$\delta C_{\rm RR}$	& $ 	0.07	\pm	0.13	 $ & $ 	303.7	 $ & $	0.5	\sigma	 $  \\
\hline										
\hline										
\multirow{2}{*}{$\{ \delta C_{9} ,\, \delta C_{10} \}$}	& \small $\delta C_{9} = 	-0.64	\pm	0.17	$ & \multirow{2}{*}{$	293.8	$} & \multirow{2}{*}{$	2.7	\sigma	$} \\
&\small $\delta C_{10} =		0.01	\pm	0.14						$ & & \\
\hline										
\end{tabular}
}
{\color{black}
\begin{tabular}{|l|c|r|c|}
\hline 
\multicolumn{4}{|c|}{	\small All observables except $q^2 \in$ [6-8.68] GeV$^2$ bins 		} \\[-4pt]							
\multicolumn{4}{|c|}{\small{\bf 	100\% pc	} \; ($\chi^2_{\rm SM}=		291.4	$)} \\ \hline			
& b.f. value & $\chi^2_{\rm min}$ & ${\rm Pull}_{\rm SM}$  \\										
\hline \hline										
$\delta C_{9} $    	& $ 	-0.56	\pm	0.20	 $ & $ 	286.0	 $ & $	2.3	\sigma	 $  \\
\hline										
$\delta C_{10} $    	& $ 	0.05	\pm	0.16	 $ & $ 	291.3	 $ & $	0.3	\sigma	 $  \\
\hline							          			
\hline										
$\delta C_{\rm LL}$	& $ 	-0.17	\pm	0.13	 $ & $ 	289.7	 $ & $	1.3	\sigma	 $  \\
\hline										
$\delta C_{\rm LR}$	& $ 	-0.14	\pm	0.14	 $ & $ 	290.5	 $ & $	0.9	\sigma	 $  \\
\hline										
$\delta C_{\rm RL}$	& $ 	0.00	\pm	0.08	 $ & $ 	290.4	 $ & $	1.0	\sigma	 $  \\
\hline										
$\delta C_{\rm RR}$	& $ 	0.02	\pm	0.15	 $ & $ 	291.4	 $ & $	0.0	\sigma	 $  \\
\hline										
\hline										
\multirow{2}{*}{$\{ \delta C_{9} ,\, \delta C_{10} \}$}	& \small $\delta C_{9} = 	-0.57	\pm	0.21	$ & \multirow{2}{*}{$	286.0	$} & \multirow{2}{*}{$	1.8	\sigma	$} \\
&\small $\delta C_{10} =		0.03	\pm	0.15						$ & & \\
\hline
\end{tabular}
}
}
\caption{NP fits to all $b\to s \ell\ell$ observables excluding $q^2 \in [6,8.68]$ GeV$^2$ bins when using GRvDV23 form factors~\cite{Gubernari:2023puw} for $B\to K^*, B_s\to \phi$ and GKvD18 form factors~\cite{Gubernari:2018wyi} for $B\to K$,  assuming 10\%, 50\% and 100\% power corrections from left to right, respectively.
\label{tab:allobs_except_608_GRvDV23_50_100_nonFactErr}} 
\end{center} 
\end{table}

%
Finally, we extend our analysis beyond the one- and two-dimensional fits presented in Table~\ref{tab:2025GRvDV_all_vs_allwo68_1D} by allowing NP contributions in all relevant Wilson coefficients. Since there is no theoretical justification for restricting NP to only a few operators, we perform multi-dimensional fits in which potentially  the coefficients $C_{7,8,9,10}$ and the scalar and pseudoscalar couplings $C_{Q_1}$ and $C_{Q_2}$ as well as their chirality-flipped counterparts can be varied simultaneously. We give the fit results (excluding the $q^2 \in [6, 8.68]~\text{GeV}^2$ bins) with four, six and also all universal Wilson coeeficients in Tables~\ref{tab:LHCb2025_wo68_4D_all},~\ref{tab:LHCb2025_wo68_6D_all} and~\ref{tab:LHCb2025_wo68_12D_all}.
It is noteworthy that the SM pull remains sizeable at $4.2\sigma$ even in the 12-dimensional fit of Table~\ref{tab:LHCb2025_wo68_12D_all}. Because all these NP fits are nested, the various NP  hypotheses can be compared by the Wilks' test~\cite{Wilks:1938dza} (for more details about Wilks' Theorem see Section~\ref{sec:data_driven}). Table~\ref{tab:LHCb2025_wo68_Wilks} provides a statistical comparison of the multidimensional fits with respect to the SM prediction and to each other. The ``Improvement'' column reports the gain in fit quality relative to NP scenario (of previous row) with smaller set of Wilson coefficients, evaluated using Wilks’ theorem.

\begin{table}[h!]
\begin{center}
\setlength\extrarowheight{3pt}
\scalebox{0.8}{
{\color{black}
\begin{tabular}{|c|c|c|c|}
\hline								
\multicolumn{4}{|c|}{\small All observables excluding $q^2 \in [6,8.68]$ GeV$^2$ bins with $\chi^2_{\rm SM}=		329.3			$} \\ 			
 \multicolumn{4}{|c|}{\small ($\chi^2_{\rm min}=	291.7	;\; {\rm Pull}_{\rm SM}=	5.3	\sigma$)} \\ 	
\hline \hline								
$\delta C_{7}$ & $\delta C_{8}$ & $\delta C_{9}$ & $\delta C_{10}$ \\								
$	0.06	\pm	0.02	$ & $	-0.74	\pm	0.22	$ &
$	-0.86	\pm	0.15	$ & $	0.06	\pm	0.14	$ \\
\hline								
\end{tabular}
} 
}
\caption{Four operator NP fit excluding the LHCb [6, 8] and CMS [6, 8.68] GeV$^2$ bins assuming 10\% power corrections. 
\label{tab:LHCb2025_wo68_4D_all}}
\end{center} 
\end{table}
%

\begin{table}[h!]
\begin{center}
\setlength\extrarowheight{3pt}
\scalebox{0.8}{
{\color{black}
\begin{tabular}{|c|c|c|c|c|c|}
\hline								
\multicolumn{6}{|c|}{\small All observables excluding $q^2 \in [6,8.68]$ bins with $\chi^2_{\rm SM}=		329.3			$} \\ 			
 \multicolumn{6}{|c|}{\small ($\chi^2_{\rm min}=	291.1	;\; {\rm Pull}_{\rm SM}=	4.9	\sigma$)} \\ 	
\hline \hline								
$\delta C_{7}$ & $\delta C_{8}$ & $\delta C_{9}$ & $\delta C_{10}$  & $\delta C_{Q_{1}}$  & $\delta C_{Q_{2}}$ \\								
$	0.06	\pm	0.02	$ & $	-0.74	\pm	0.22	$ &
$	-0.85	\pm	0.16	$ & $	0.04	\pm	0.19	$ &
$	-0.13	\pm	0.06	$ & $	0.06	\pm	0.08	$ \\
\hline								
\end{tabular}
}
}
\caption{Six operator NP fit excluding the LHCb [6, 8] and CMS [6, 8.68] GeV$^2$ bins assuming 10\% power corrections. 
\label{tab:LHCb2025_wo68_6D_all}}
\end{center} 
\end{table}
%

\begin{table}[h!]
\begin{center}
\setlength\extrarowheight{3pt}
\scalebox{0.7}{
{\color{black}
\begin{tabular}{|c|c|c|c|}
\hline																
\multicolumn{4}{|c|}{\small All observables excluding $q^2 \in [6,8.68]$ GeV$^2$ bins with $\chi^2_{\rm SM}=		329.3			$} \\											
 \multicolumn{4}{|c|}{\small  ($\chi^2_{\rm min}=	 	286.8	;\; {\rm Pull}_{\rm SM}=	4.2				\sigma$)} \\					
\hline \hline																
\multicolumn{2}{|c|}{$\delta C_7$} &  \multicolumn{2}{c|}{$\delta C_8$}\\																
\multicolumn{2}{|c|}{$	0.06	\pm	0.02	$} & \multicolumn{2}{c|}{$	-0.78	\pm	0.22	$}\\								
\hline																
\multicolumn{2}{|c|}{$\delta C_7^\prime$} &  \multicolumn{2}{c|}{$\delta C_8^\prime$}\\																
\multicolumn{2}{|c|}{$	0.00	\pm	0.01	$} & \multicolumn{2}{c|}{$	0.50	\pm	0.50	$}\\								
\hline																
$\delta C_{9}$ & $\delta C_{9}^{\prime}$ & $\delta C_{10}$ & $\delta C_{10}^{\prime}$ \\																
$	-0.88	\pm	0.16	$ & $	-0.26	\pm	0.20	$ & $	0.09	\pm	0.19	$ & $	-0.03	\pm	0.14	$ \\
\hline\hline																
$\delta C_{Q_{1}}$ & $\delta C_{Q_{1}}^{\prime}$ & $\delta C_{Q_{2}}$ & $\delta C_{Q_{2}}^{\prime}$ \\																
$	-0.12	\pm	0.09	$ & $	0.01	\pm	0.10	$ & $	0.40	\pm	0.09	$ & $	0.34	\pm	0.09	$ \\
\hline																
\end{tabular}
} 
}
\caption{Twelve operator NP fit excluding the LHCb [6, 8] and CMS [6, 8.68] GeV$^2$ bins assuming 10\% power corrections.
\label{tab:LHCb2025_wo68_12D_all}}
\end{center} 
\end{table}
%

\begin{table}[!h]
\begin{center}
\scalebox{0.8}{
{\color{black}
\begin{tabular}{|c|c|c|c|c|}\hline
\multicolumn{5}{|c|}{\small All observables excluding $q^2 \in [6,8.68]$ bins} \\				
\hline \hline									
Set of WC & param. & $\chi^2_{\rm min}$ & Pull$_{\rm SM}$ & Improvement\\ \hline									
SM                                          	&	0	&	329.3	& $	-	$ & $	-	$ \\
$C_9$                                 	&	1	&	302.6	& $	5.2	\sigma$ & $	5.2	\sigma$\\
$C_9, C_{10}$                   	&	2	&	302.6	& $	4.8	\sigma$ & $	0.0	\sigma$\\
$C_7,C_8,C_9,C_{10}$	&	4	&	291.7	& $	5.3	\sigma$ & $	2.9	\sigma$\\
$C_7,C_8,C_9,C_{10},C_{Q_1},C_{Q_2}$    	&	6	&	291.1	& $	4.9	\sigma$ & $	0.3	\sigma$\\
All WC (incl. primed)                       	&	12	&	286.8	& $	4.2	\sigma$ & $	0.5	\sigma$\\
\hline
\end{tabular}
}
}
\caption{\small Pull$_{\rm SM}$ of $1,2,4,6$ and 12 dimensional fit. 
The last row also includes the chirality-flipped counterparts of the Wilson coefficients.
In the last column the significance of improvement of the fit compared to  the scenario of the previous row is given. 
\label{tab:LHCb2025_wo68_Wilks}}
\vspace*{-0.6cm}
\end{center} 
\end{table}

\section{Data-driven analyses}\label{sec:data_driven}

In this section, we update our data-driven analyses presented in  
Refs.~\cite{Chobanova:2017ghn,Arbey:2018ics,Hurth:2020rzx,Hurth:2021nsi} in view of the new CMS and LHCb data. These analyses contributed to the question whether the large deviations from QCDf predictions in the angular observables of the decay $B \to K^*\ell\ell$   represent a sign for new physics beyond the Standard Model or a consequence of underestimated hadronic power corrections.  We present a direct comparison of two global fits to the data in the low-$q^2$ region based on the two different assumptions.
We already  know~\cite{Chobanova:2017ghn} that the new physics fit can be embedded into the general hadronic fit and therefore it is not and will not be possible to disprove the hadronic hypothesis in favour of the NP one within this analysis, with the set of observables considered in the present study. On the other hand, NP can hide within hadronic contributions and new observables like inclusive modes, confirming that the NP option can rule out the hadronic hypothesis.  

In this section, the fit includes the branching ratio of $B^{(+,0)} \to K^{*(+,0)} \gamma$ (to get reasonable bounds on $\delta C_7$), branching ratio and angular observables of $B^{(+,0)} \to K^{*(+,0)} \mu^+ \mu^-$, and angular observables of $B^{0} \to K^{*0} e^+ e^-$, all restricted to the low-$q^2$ region (either including or excluding the bin above 6 GeV$^2$). The full list of observables is provided as ancillary files.

Finally, we note that within the data-driven approach, we do not add any guesstimate of non-factorisable power corrections to the known QCDf prediction. The pull is calculated with respect to this {\it plain} QCDf prediction.

\subsection{Theoretical framework}
Already in Ref.~\cite{Ciuchini:2015qxb}, it was shown that the deviation in the low-$q^2$ region from the QCDf prediction can be fitted by hadronic power corrections. For this purpose, one has to derive the most general ansatz for the non-factorisable power corrections. 
We recall that the $b\to s\ell\ell$ transitions are described by an effective Hamiltonian that can be divided into hadronic and semileptonic components.
In case of the decay  $B\to K^* \mu^+ \mu^-$, the semileptonic part is the dominant contribution and is described by seven independent form factors 
$\tilde{S}, \tilde{V}_\lambda, \tilde{T}_\lambda$, with helicities $\lambda=\pm1,0$~\cite{Jager:2012uw,Jager:2014rwa}. 
Then the decay $B\to V \bar \ell \ell$, where $V$ is a vector meson, is described by the following eight helicity amplitudes:
\begin{align}\label{eq:HV}     
 H_V(\lambda) &=-i\, N^\prime \Big\{ C_9^{\rm eff} \tilde{V}_{\lambda} - C_{9}'  \tilde{V}_{-\lambda}
      + \frac{m_B^2}{q^2} \Big[\frac{2\,\hat m_b}{m_B} (C_{7}^{\rm eff} \tilde{T}_{\lambda} - C_{7}'  \tilde{T}_{-\lambda})
      - 16 \pi^2 {\cal N}_\lambda \Big] \Big\} ,  \\  
  H_A(\lambda) &= -i\, N^\prime (C_{10}  \tilde{V}_{\lambda} - C_{10}'\tilde{V}_{-\lambda}) , \\
  H_P &= i\, N^\prime \left\{  (C_{Q_2} - C_{Q_2}')         
            + \frac{2\,m_\ell \hat m_b}{q^2} \left(1+\frac{m_s}{m_b} \right)(C_{10}-C_{10}') \right\} \tilde{S}, \\
  H_S &= i\, N^\prime (C_{Q_1} - C_{Q_1}')\tilde{S}, 
\end{align}
At this level, it is manifest that the new physics contributions can be mimicked by hadronic ones~\cite{Jager:2012uw,Jager:2014rwa}. The effective part of $C_9^{{\rm eff}}$, i.e. $C_9 + Y(q^2)$, as well as the
non-factorisable contribution, ${\cal N}_\lambda(q^2)$, arise
from the hadronic part of the Hamiltonian, via the emission of a photon that subsequently turns into a lepton pair.
Due to the vectorial coupling of the photon to the lepton pair, the contributions from the hadronic effective Hamiltonian appear in the vectorial helicity amplitude ($H_V(\lambda)$).
A similar effect occurs with the short-distance contribution of $C_9$ (and $C_7$), resulting in an ambiguity when separating the NP effects of $\delta C_9$ (and $\delta C_7$) from the non-factorisable hadronic contributions.

Thus, ${\cal N}_\lambda(q^2)$ corresponds to the leading non-factorisable contribution calculable in QCDf, plus some unknown non-factorisable power corrections, which we denote as $h_\lambda$.
The most general ansatz for the unknown $h_\lambda$ terms respecting the analyticity of the amplitude
(up to higher-order terms in $q^2$) is given by (see Refs~\cite{Jager:2012uw,Jager:2014rwa})
\begin{align}\label{eq:hlambdapm}
  h_\pm(q^2)&= h_\pm^{(0)} + \frac{q^2}{1 \,{\rm GeV}^2}h_\pm^{(1)} + \frac{q^4}{1 \,{\rm GeV}^4}h_\pm^{(2)}\,,\\[-6pt]
\label{eq:hlambda0}
h_0(q^2)&= \sqrt{q^2}\times \left( h_0^{(0)} + \frac{q^2}{1\, {\rm GeV}^2}h_0^{(1)} + \frac{q^4}{1\, {\rm GeV}^4}h_0^{(2)}\right).
\end{align}
The extra $\sqrt{q^2}$ in the longitudinal amplitude assures that the longitudinal amplitude vanishes when the intermediate  $\gamma$ becomes on-shell.

\subsection{Hadronic 18 parameter fit to  helicity-dependent parametrisation}
\label{sec:HadFit_18param}
In the previous section, we have made guesstimates of the unknown non-factorisable power corrections. We assumed them to be $10\%$ of the leading order non-factorisable contribution, which is calculable in QCDf.
However, it is also possible to make just a fit to the data for these power corrections - using the most general parameterisation with 18 parameters~\cite{Ciuchini:2015qxb,Chobanova:2017ghn, Neshatpour:2017qvi,Ciuchini:2017mik,Arbey:2018ics,Hurth:2020rzx}. 
The results of this fit are summarised in Table~\ref{tab:HadronicFit_18param}, with a detailed illustration of its impact on some of the observables given in Appendix~\ref{sec:HadFit_Obs}. 

One finds that half of the fitted parameters in such a fit are still consistent with zero. However, this is a large change to our previous corresponding analysis~\cite{Hurth:2020rzx} where almost all parameters were compatible with zero within the $1\sigma$ range. Also, the pull compared with the QCDf prediction increases from $4.7\sigma$ to $9.2\sigma$ (in the case where bins up to 8.68~$\mbox{GeV}^2$ are included). When we leave out the problematic bins above 6~$\mbox{GeV}^2$, one finds a pull of $4.7\sigma$. And because there are still rather large uncertainties, one may expect that such a hadronic fit will improve further in the future. Indeed, we have shown in a previous analysis~\cite{Hurth:2020rzx} that assuming the present central value, the significance compared with the QCDf prediction will increase dramatically within the first and second LHCb upgrade. 
\begin{table}[h!]
\ra{0.90}
\rb{1.3mm}
\begin{center}
\setlength\extrarowheight{2pt}
\scalebox{0.85}{
{\color{black}
\begin{tabular}{|l||r|r|}
\hline
\multicolumn{3}{|c|}{\small 		$B\to K^*\, \gamma/\ell\ell$ observables - $q^2\leqslant 6$ GeV$^2$ bins				}\\
\multicolumn{3}{|c|}{ ($\chi^2_{\rm QCDf}=	158.8	\; \chi^2_{\rm min}=	94.9	;\; {\rm Pull}_{\rm QCDf}=	5.0	\sigma$)}   \\ 
\hline						
	&	\cen{Real}	&	\multicolumn{1}{c|}{Imaginary} 		\\
\hline						
$h_{+}^{(0)}$	&	$ ( 0.0 \pm 5.0 ) \times 10^{-5} $	&	$( 1.1 \pm 0.8 ) \times 10^{-4} $		\\
$h_{+}^{(1)}$	&	$( -2.0 \pm 8.0 ) \times 10^{-5} $	&	$( -7.0 \pm 12.0 ) \times 10^{-5} $		\\
$h_{+}^{(2)}$	&	$( 1.7 \pm 2.0 ) \times 10^{-5} $	&	$ ( 0.0 \pm 2.6 ) \times 10^{-5} $		\\
\hline						
$h_{-}^{(0)}$	&	$( -1.0 \pm 0.6 ) \times 10^{-4} $	&	$( -2.5 \pm 1.4 ) \times 10^{-4} $		\\
$h_{-}^{(1)}$	&	$( 5.0 \pm 6.0 ) \times 10^{-5} $	&	$( 3.7 \pm 1.4 ) \times 10^{-4} $		\\
$h_{-}^{(2)}$	&	$( 7.0 \pm 12.0 ) \times 10^{-6} $	&	$( -8.5 \pm 3.2 ) \times 10^{-5} $		\\
\hline						
$h_{0}^{(0)}$	&	$( 6.0 \pm 12.0 ) \times 10^{-5} $	&	$( 3.7 \pm 1.6 ) \times 10^{-4} $		\\
$h_{0}^{(1)}$	&	$( 8.0 \pm 10.0 ) \times 10^{-5} $	&	$( -1.1 \pm 1.5 ) \times 10^{-4} $		\\
$h_{0}^{(2)}$	&	$( -4.0 \pm 17.0 ) \times 10^{-6} $	&	$( -1.3 \pm 2.5 ) \times 10^{-5} $		\\
\hline
\end{tabular} 
}
{\color{black}
\begin{tabular}{|l||r|r|}
\hline
\multicolumn{3}{|c|}{\small 		$B\to K^*\, \gamma/\ell\ell$ observables - $q^2\leqslant 8.68$ GeV$^2$ bins				}\\
\multicolumn{3}{|c|}{ ($\chi^2_{\rm QCDf}=	269.8	\; \chi^2_{\rm min}=	133.6	;\; {\rm Pull}_{\rm QCDf}=	 9.2	\sigma$)}   \\ 
\hline						
	&	\cen{Real}	&	\multicolumn{1}{c|}{Imaginary} 		\\
\hline						
$h_{+}^{(0)}$	&	$( -2.0 \pm 4.0 ) \times 10^{-5} $	&	$( 6.0 \pm 7.0 ) \times 10^{-5} $		\\
$h_{+}^{(1)}$	&	$( 4.0 \pm 7.0 ) \times 10^{-5} $	&	$( 1.0 \pm 8.0 ) \times 10^{-5} $		\\
$h_{+}^{(2)}$	&	$( -6.0 \pm 13.0 ) \times 10^{-6} $	&	$( -1.3 \pm 1.6 ) \times 10^{-5} $		\\
\hline						
$h_{-}^{(0)}$	&	$( -7.0 \pm 5.0 ) \times 10^{-5} $	&	$( -7.0 \pm 13.0 ) \times 10^{-5} $		\\
$h_{-}^{(1)}$	&	$( 1.0 \pm 4.0 ) \times 10^{-5} $	&	$( 9.0 \pm 12.0 ) \times 10^{-5} $		\\
$h_{-}^{(2)}$	&	$( 1.5 \pm 0.5 ) \times 10^{-5} $	&	$( -9.0 \pm 26.0 ) \times 10^{-6} $		\\
\hline						
$h_{0}^{(0)}$	&	$( 9.0 \pm 10.0 ) \times 10^{-5} $	&	$( 3.8 \pm 1.3 ) \times 10^{-4} $		\\
$h_{0}^{(1)}$	&	$( 8.0 \pm 6.0 ) \times 10^{-5} $	&	$( -1.6 \pm 0.9 ) \times 10^{-4} $		\\
$h_{0}^{(2)}$	&	$( -9.0 \pm 7.0 ) \times 10^{-6} $	&	$( 1.0 \pm 1.1 ) \times 10^{-5} $		\\
\hline
\end{tabular} 
}
}
\caption{Hadronic power correction fit to $B\to K^*\, \gamma/\ell\ell$ observables for low-$q^2$ bins, with complex power corrections up to $q^4$ terms with 18 free parameters in total. 
\label{tab:HadronicFit_18param}}
\end{center} 
\end{table}

\subsection{New physics fits}
We can now fit the same experimental data from CMS and LHCb to the complex or real NP Wilson coefficients $\delta C_7$ and $\delta C_9$ or only to the complex or real NP Wilson coefficients $\delta C_9$ as given in Tables~\ref{tab:NewPhysicsFit_C7C9_4param},~\ref{tab:NewPhysicsFit_C9_2param} and~\ref{tab:NewPhysicsFit_C9_1param}. If we take the  low-$q^2$ bins beyond 6~$\mbox{\rm GeV}^2$ into account, we can compare the pulls with those of our earlier analysis in Ref.~\cite{Hurth:2020rzx}
{which shows} that the pulls have increased dramatically: In the case of the fit to the complex $\delta C_9$ from $5.8 \sigma$ to $9.3\sigma$, and in the case of the real coefficient 
from $6.0\sigma$ to $8.8\sigma$. If we omit the low-$q^2$ bins beyond 6~$ {\rm GeV}^2$ for the reasons discussed in section~\ref{sec:CMS_vs_LHCb}, we find $5.6\sigma$, which is compatible with the result of the one-parameter fit to $\delta C_9$ in section~\ref{sec:CMS_vs_LHCb}.

The imaginary parts in NP fits to a complex NP Wilson coefficient $\delta C_9$ are very large. But these results have to be taken with caution, because there are no strongly constraining CP-violating observables measured with high precision yet, therefore only CP-averaged observables are considered in this analysis.
\begin{table}[h!]
\ra{1.}
\rb{1.3mm}
\begin{center}
\setlength\extrarowheight{2pt}
\scalebox{0.85}{
{\color{black}
\begin{tabular}{|c|c|}
\hline
\multicolumn{2}{|c|}{\small 		$B\to K^*\, \gamma/\ell\ell$ observables - $q^2\leqslant 6$ GeV$^2$ bins				}\\
\multicolumn{2}{|c|}{ ($\chi^2_{\rm QCDf}=	158.8	\; \chi^2_{\rm min}=	113.4	;\; {\rm Pull}_{\rm QCDf}=	5.9	\sigma$)}   \\ 
\hline						
& \multicolumn{1}{c|}{best fit value}						\\
 \hline \hline						
$\delta C_7$ & $		(0.09  \pm 0.03)	+i(-0.20  \pm 0.03)		$	\\  [-6pt]
\footnotesize{\&} & 						\\  [-4pt]
$\delta C_9$ & $		(-1.65  \pm 0.26)	+i(1.90  \pm 0.50)		$	\\
\hline
\end{tabular}   
}
{\color{black}
\begin{tabular}{|c|c|}
\hline
\multicolumn{2}{|c|}{\small 		$B\to K^*\, \gamma/\ell\ell$ observables - $q^2\leqslant 8.68$ GeV$^2$ bins				}\\
\multicolumn{2}{|c|}{ ($\chi^2_{\rm QCDf}=	269.8	\; \chi^2_{\rm min}=	163.6	;\; {\rm Pull}_{\rm QCDf}=	 9.7	\sigma$)}   \\ 
\hline						
& \multicolumn{1}{c|}{best fit value}						\\
 \hline \hline						
$\delta C_7$ & $		(0.11  \pm 0.02)	+i(-0.22  \pm 0.02)		$	\\  [-6pt]
\footnotesize{\&} & 						\\  [-4pt]
$\delta C_9$ & $		(-1.87  \pm 0.19)	+i(2.10  \pm 0.40)		$	\\
\hline
\end{tabular} 
}
}
\caption{Two operator NP fits for complex $\delta C_7$ and $\delta C_{9}$  with 4 parameters considering $B\to K^*\, \gamma/\ell\ell$ observables for low-$q^2$ bins.
\label{tab:NewPhysicsFit_C7C9_4param}
}
\end{center} 
\end{table}  
\begin{table}[h!]
\ra{1.}
\rb{1.3mm}
\begin{center}
\setlength\extrarowheight{2pt}
\scalebox{0.85}{
{\color{black}
\begin{tabular}{|c|c|}
\hline
\multicolumn{2}{|c|}{\small 		$B\to K^*\, \gamma/\ell\ell$ observables - $q^2\leqslant 6$ GeV$^2$ bins				}\\
\multicolumn{2}{|c|}{ ($\chi^2_{\rm QCDf}=	158.8	\; \chi^2_{\rm min}=	122.6	;\; {\rm Pull}_{\rm QCDf}=	5.7	\sigma$)}   \\ 
\hline						
& \multicolumn{1}{c|}{best fit value}						\\
 \hline \hline						
$\delta C_9$ & $		(-0.69  \pm 0.16)	+i(-1.90  \pm 0.50)		$	\\
\hline
\end{tabular}    
}
{\color{black}
\begin{tabular}{|c|c|}
\hline
\multicolumn{2}{|c|}{\small 		$B\to K^*\, \gamma/\ell\ell$ observables - $q^2\leqslant 8.68$ GeV$^2$ bins				}\\
\multicolumn{2}{|c|}{ ($\chi^2_{\rm QCDf}=	269.8	\; \chi^2_{\rm min}=	177.9	;\; {\rm Pull}_{\rm QCDf}=	 9.3	\sigma$)}   \\ 
\hline						
& \multicolumn{1}{c|}{best fit value}						\\
 \hline \hline						
$\delta C_9$ & $		(-0.84  \pm 0.15)	+i(2.22  \pm 0.32)		$	\\
\hline
\end{tabular} 
}
}
\caption{One operator NP fits for complex $\delta C_{9}$  with 2 parameters considering $B\to K^*\, \gamma/\ell\ell$ observables for low-$q^2$ bins.
\label{tab:NewPhysicsFit_C9_2param}
}
\end{center} 
\end{table}  
\begin{table}[h!]
\ra{1.}
\rb{1.3mm}
\begin{center}
\setlength\extrarowheight{2pt}
\scalebox{0.85}{
{\color{black}
\begin{tabular}{|c|c|c|}
\hline
\multicolumn{2}{|c|}{\small 		$B\to K^*\, \gamma/\ell\ell$ observables - $q^2\leqslant 6$ GeV$^2$ bins				}\\
\multicolumn{2}{|c|}{ ($\chi^2_{\rm QCDf}=	158.8	\; \chi^2_{\rm min}=	127.7	;\; {\rm Pull}_{\rm QCDf}=	5.6	\sigma$)}   \\ 
\hline						
& \multicolumn{1}{c|}{best fit value}						\\
 \hline \hline						
$\delta C_9$ & $		-0.82  \pm 0.13			$	\\
\hline
\end{tabular}
}
{\color{black}
\begin{tabular}{|c|c|c|}
\hline
\multicolumn{2}{|c|}{\small 		$B\to K^*\, \gamma/\ell\ell$ observables - $q^2\leqslant 8.68$ GeV$^2$ bins				}\\
\multicolumn{2}{|c|}{ ($\chi^2_{\rm QCDf}=	269.8	\; \chi^2_{\rm min}=	193.1	;\; {\rm Pull}_{\rm QCDf}=	 8.8	\sigma$)}   \\ 
\hline						
& \multicolumn{1}{c|}{best fit value}						\\
 \hline \hline						
$\delta C_9$ & $		-1.13  \pm 0.10			$	\\
\hline
\end{tabular}
}
}
\caption{One operator NP fits for real $\delta C_{9}$ considering $B\to K^*\, \gamma/\ell\ell$ observables for low-$q^2$ bins.
\label{tab:NewPhysicsFit_C9_1param}
}
\end{center} 
\end{table}

\subsection{Test of helicity-independence of NP fit}

The independent NP fits with a complex $C_9$ in each  helicity amplitude separately represent a null test of the NP hypothesis. This can be done via ~\cite{Neshatpour:2017qvi,Arbey:2018ics,Hurth:2020rzx}:
\begin{align}
 h_\lambda (q^2)= -\frac{\tilde{V}_\lambda(q^2)}{16 \pi^2} \frac{q^2}{m_B^2}  \Delta C_9^{\lambda,\rm{PC}}\,,
\end{align}
with $\Delta C_9^{\lambda,\rm{PC}}$ being three complex (six real) $q^2$-independent parameters. This can be regarded as a minimalist description of hadronic power corrections, in which ruling out the NP hypothesis requires no extra $q^2$ term, and where it suffices if the fits to the three helicities are incompatible with each other.

However, if these three independent determinations of $\Delta C_9^\lambda$ are highly consistent with each other, then the NP description is strongly favoured. And while it is not possible in principle to rule out the hadronic explanation in that case, it is extremely unlikely that the power corrections for all different helicities would conspire to imitate the new physics description.

The results of the six-parameter hadronic fit are given in Table~\ref{tab:HadronicDeltaC9Fit_6param}. As can be seen, the results are not conclusive yet. The compatibility 
of the three complex numbers has not improved compared to our previous result in Ref.~\cite{Hurth:2020rzx}. The large uncertainties in the case of $\Delta C_9^{+,{\rm PC}}$ can be explained by the fact that 
the power corrections of $H_V(\lambda=+)$ are predicted to be smaller than those of $H_V(\lambda=-)$ by a factor of $m_B /\Lambda$~\cite{Jager:2012uw, Jager:2014rwa}.
And we have shown in Ref.~\cite{Hurth:2020rzx}, that with the two LHCb upgrades, we can expect a much clearer picture showing preference either for the NP or hadronic interpretation.
\begin{table}[h!]
\ra{1.}
\rb{1.3mm}
\begin{center}
\setlength\extrarowheight{2pt}
\scalebox{0.85}{
{\color{black}
\begin{tabular}{|c|c|}
\hline
\multicolumn{2}{|c|}{\small 		$B\to K^*\, \gamma/\ell\ell$ observables - $q^2\leqslant 6$ GeV$^2$ bins				}\\
\multicolumn{2}{|c|}{ ($\chi^2_{\rm QCDf}=	158.8	\; \chi^2_{\rm min}=	108.9	;\; {\rm Pull}_{\rm QCDf}=	5.8	\sigma$)}   \\ 
\hline						
& \multicolumn{1}{c|}{best fit value}						\\
 \hline \hline						
$\Delta C_9^{+,{\rm PC}}$ & $		(3.00  \pm 5.00)	+i(-10.00  \pm 5.00)		$	\\
\hline						
$\Delta C_9^{-,{\rm PC}}$ & $		(-0.79  \pm 0.18)	+i(-0.89  \pm 0.27)		$	\\
\hline						
$\Delta C_9^{0,{\rm PC}}$ & $		(-0.59  \pm 0.33)	+i(-0.96  \pm 0.34)		$	\\
\hline
\end{tabular}  
}
{\color{black}
\begin{tabular}{|c|c|}
\hline
\multicolumn{2}{|c|}{\small 		$B\to K^*\, \gamma/\ell\ell$ observables - $q^2\leqslant 8.68$ GeV$^2$ bins				}\\
\multicolumn{2}{|c|}{ ($\chi^2_{\rm QCDf}=	269.8	\; \chi^2_{\rm min}=	160.0	;\; {\rm Pull}_{\rm QCDf}=	 9.5	\sigma$)}   \\ 
\hline						
& \multicolumn{1}{c|}{best fit value}						\\
 \hline \hline						
$\Delta C_9^{+,{\rm PC}}$ & $		(1.00  \pm 4.00)	+i(-0.50  \pm 3.20)		$	\\
\hline						
$\Delta C_9^{-,{\rm PC}}$ & $		(-1.11  \pm 0.13)	+i(-1.21  \pm 0.28)		$	\\
\hline						
$\Delta C_9^{0,{\rm PC}}$ & $		(-0.72  \pm 0.28)	+i(-0.69  \pm 0.33)		$	\\
\hline
\end{tabular}    
}}
\caption{Hadronic power correction fits for the three helicities ($\lambda=\pm,0$) in the form of complex $\Delta C_9^{\lambda,{\rm PC}}$, considering $B\to K^* \gamma/\ell\ell$ observables for low-$q^2$ bins.
\label{tab:HadronicDeltaC9Fit_6param}
}
\end{center} 
\end{table}  

\subsection{Test of \texorpdfstring{$q^2$}{q2}-independence of NP fit}

Yet another test to differentiate between the hadronic and the NP interpretation is the analysis of the $q^2$ dependence. In Ref.~\cite{Ciuchini:2015qxb}, the authors claimed that a non-zero $q^4$ term in the hadronic fit ansatz is incompatible with the NP interpretation of the $B$ anomalies, as it could not be understood as a shift to the Wilson Coefficients. 
However, this argument misses the fact that such a $q^4$ term could also be produced by the $q^2$ dependence of the helicity form factors, which multiply the NP Wilson coefficients~\cite{Chobanova:2017ghn}. In Ref.~\cite{Alguero:2019ptt}, this problem is circumvented by the determination of the Wilson coefficients in a bin-by-bin fit (using a specific estimate of the non-local contribution from the $c\bar c$ loops based on Refs~\cite{Khodjamirian:2010vf,Khodjamirian:2012rm}).  Of course, this NP fit includes the assumption that possible NP contributions affect only single Wilson coefficients. The same authors have recently updated their analysis in Ref.~\cite{Alguero:2023jeh}. They found that the bin-by-bin analysis of the three channels $B \to K^{(*)}/ \phi \mu^+\mu^-$ {with} all independent $C_9$ fits are compatible with the best fit value of the global fit at the 2$\sigma$
level. It is worth noting that in their bin-by-bin fit combining all $b \to s \ell\ell$ modes, they find the largest deviation from the best fit point of the global fit in the  $6-8$~GeV$^2$ bin. More recently, another group has presented a similar analysis
including an estimate of the effects induced by the charmonium resonances via a dispersive method in order to subtract the dominant hadronic $q^2$-dependent contribution~\cite{Bordone:2024hui}. 

\begin{figure}[h!]
\begin{center}
\includegraphics[width=0.7\textwidth]{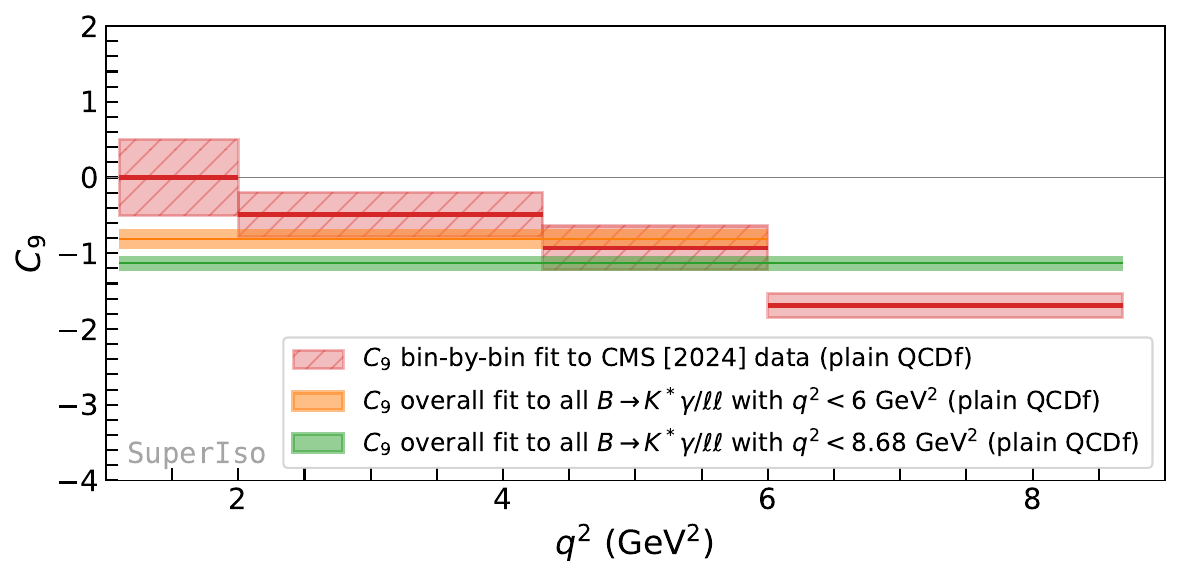}\\
\includegraphics[width=0.7\textwidth]{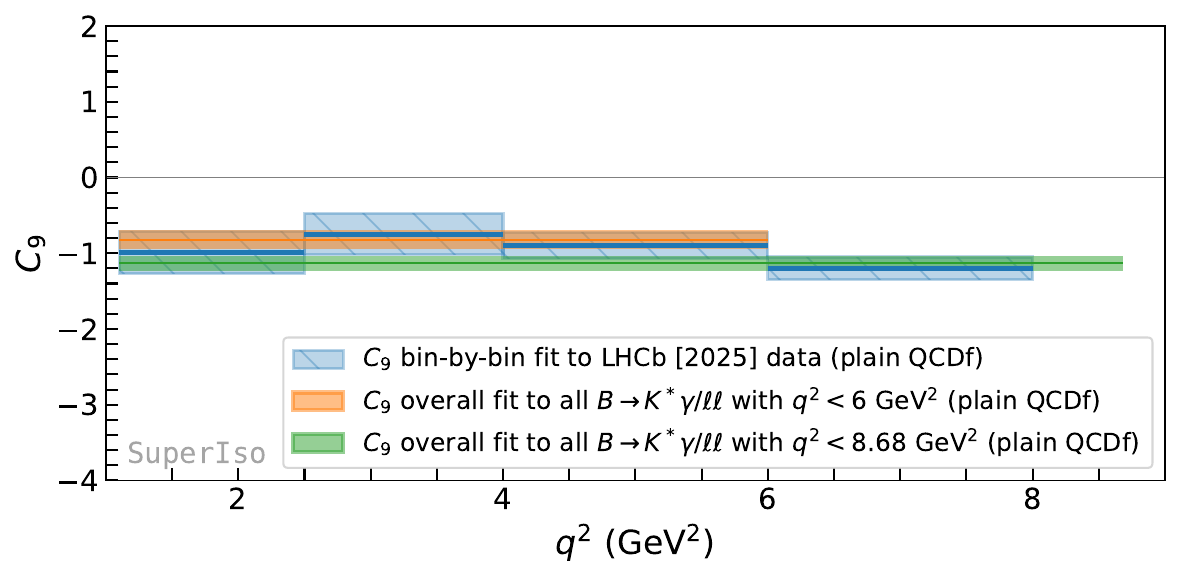}\\
\includegraphics[width=0.7\textwidth]{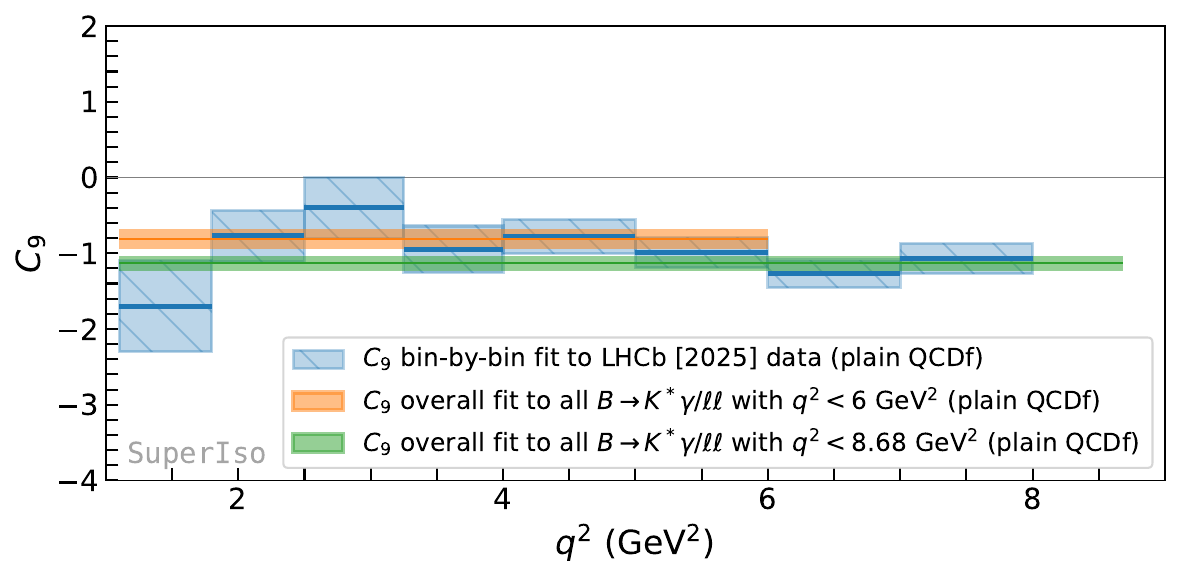}
\vspace{-0.2cm}
\caption{\small
Bin-by-bin fit to $C_9$ for CMS (top), LHCb config.~2 (middle), and LHCb config.~6 (bottom) results -- compared with combined fit up to 6~$\rm{GeV}^2$ (orange) or up to 8.68~$\rm{GeV}^2$ (green). 
For the LHCb data, config.~2 corresponds to $P_i^{(\prime)}$ observables with full-size bins, and config.~6 to $S_i$ observables with half-size bins. The results assume plain QCDf, without power corrections.
\label{fig:bin_by_bin_C9}}
\end{center}
\end{figure}

We have done this exercise using the LHCb and CMS data on $ B \to K^* \mu^+\mu^-$ in the low-$q^2$ region, without taking into account any hadronic contributions beyond the plain QCDf predictions, see Figure~\ref{fig:bin_by_bin_C9}. 
With the new LHCb data, we get a similar degree of consistency in the bin-by-bin determinations, although we have considered plain QCDf. This fact indicates that this test is far  from being conclusive yet:
Comparing with the overall fit to $C_9$ up to $6~\rm{GeV}^2$ (orange) or up to $8.68~\rm{GeV}^2$, where we used the combined LHCb and CMS data, we consider both config.~2 (full-size $P_i^{(\prime)}$ bins) and config.~6 (half-size $S_i$ bins) for the LHCb data. One finds excellent agreement for LHCb with the full-size bins, while the half-size bins yield somewhat larger fitted regions due to their higher experimental uncertainties; still, both configurations are compatible at about the 1$\sigma$ level, whereas the CMS data shows larger deviations (see Figure~\ref{fig:bin_by_bin_C9}).

Finally, we note that if these separate bin-by-bin determinations of the NP Wilson coefficients are highly inconsistent with each other, the NP interpretation can be ruled out, while strictly speaking, it is not possible to prove the NP interpretation in this way. However, at the point where one can prove the consistency of the individual determinations for smaller and smaller bins, it becomes rather unlikely that hadronic corrections can generate such a specific $q^2$-dependence.

\subsection{Wilks' test}  
Finally, we use Wilks' test to analyse the various interpretations of the deviation. This test includes all the various data-driven tests we discussed in this section and compares the various hypotheses with a precise statistical method~\cite{Chobanova:2017ghn}.
The Wilks' test allows for this direct comparison of the various hypotheses because the NP fits can be considered as scenarios embedded within the hadronic fits. Thus, the different hypotheses can be compared using likelihood ratio tests. As these are nested hypotheses, the $p$ values can be determined by applying Wilks' theorem~\cite{Wilks:1938dza}.
Here, the difference in $\chi^2$ between two hypotheses is itself a $\chi^2$ distribution with a number of degrees of freedom corresponding to the difference in the number of parameters.
The $p$ value therefore, indicates the significance of the newly added parameters. We then translate the $p$ values into Gaussian single parameter significances using the relation $\sigma = \sqrt{2}Erf^{-1}(1-p)$. We note that Wilks' theorem is only valid under the assumption of Gaussian distributed uncertainties. 

Therefore, as long as our various NP and hadronic fits are nested, we can compare them using Wilks' test. For example, the fit to real NP Wilson coefficients can be compared to the fit to the corresponding complex ones. A NP scenario can then be regarded as a nested scenario with respect to the hadronic fit. 

From the upper panel in Table~\ref{tab:Wilks_test}, when the largest low-$q^2$ bins are not included, we find significances of $5.0-5.9\sigma$ when comparing the various NP and hadronic descriptions of the data with the plain QCDf prediction. The number in the second row and the last column indicates that adding 17 hadronic parameters to the real NP $\delta C_9$ fit does improve the description of the data with a significance of $2.5\sigma$. The Table shows the comparative statistical preferences of various hypotheses: any preference among various models does not exceed $2.7\sigma$, which indicates that the situation with the present data is still not conclusive. There is no real change to our previous analysis in Ref.\cite{Hurth:2020rzx}. The high significance - in {the} cases where a complex Wilson coefficient $C_9$ is involved - is again a consequence of the lack of CP-violating observables measured with high precision.

\begin{table}[t!]
\ra{1.2}
\rb{1.3mm}
\begin{center}
\setlength\extrarowheight{2pt}
\scalebox{0.65}{
{\color{black}
\rowcolors{6}{}{light-gray}
\begin{tabular}{|l|c|c|c|c|c|c|c|c|c|c|c|}
\hline
\multicolumn{11}{|c|}{\small 		$B\to K^*\, \gamma/\ell\ell$ observables - $q^2\leqslant 6$ GeV$^2$ bins				}\\																
\hline																						
\multicolumn{1}{|c|}{\multirow{2}{*}{nr. of free}} & 1 & 2 & 2 & 4 & 3 & 6 & 6 & 9 & 12 & 18 \\ [-2pt]																						
\multicolumn{1}{|c|}{parameters} &																						
$\footnotesize \left(\!\!\begin{array}{c} {\rm Real} \\ \delta C_9 \end{array}\!\!\right)$ & $\footnotesize \left(\!\!\begin{array}{c} {\rm Real} \\ \delta C_7,\delta C_9 \end{array}\!\!\right)$ & $\footnotesize \left(\!\!\begin{array}{c} {\rm Comp.} \\ \delta C_9 \end{array}\!\!\right)$ & $\footnotesize \left(\!\!\begin{array}{c} {\rm Comp.} \\ \delta C_7,\delta C_9 \end{array}\!\!\right)$ & $\footnotesize \left(\!\!\begin{array}{c} {\rm Real} \\ \Delta C_9^{\lambda,{\rm PC}} \end{array}\!\!\right)$ & $\footnotesize \left(\!\!\begin{array}{c} {\rm Comp.} \\ \Delta C_9^{\lambda,{\rm PC}} \end{array}\!\!\right)$ & $\footnotesize \left(\!\!\begin{array}{c} {\rm Real} \\ h_{+,-,0}^{(0,1)} \end{array}\!\!\right)$ & $\footnotesize \left(\!\!\begin{array}{c} {\rm Real} \\ h_{+,-,0}^{(0,1,2)} \end{array}\!\!\right)$ & $\footnotesize \left(\!\!\begin{array}{c} {\rm Comp.} \\ h_{+,-,0}^{(0,1)} \end{array}\!\!\right)$ & $\footnotesize \left(\!\!\begin{array}{c} {\rm Comp.} \\ h_{+,-,0}^{(0,1,2)} \end{array}\!\!\right)$						\\ \hline																
0 (plain QCDf)	&$	5.6	$&$	5.4	$&$	5.7	$&$		5.9	$&$	5.0	$&$	5.8	$&$	5.2	$&$	5.0	$&$	5.4	$&$	5.0	$\\
1 {\small(Real $\delta C_9$)}	&$	\text{---}	$&$	1.3	$&$	2.3	$&$		3.0	$&$	0.5	$&$	3.1	$&$	2.0	$&$	2.0	$&$	2.7	$&$	2.5	$\\
2 {\small(Real $\delta C_7,\delta C_9$)}	&$	\text{---}	$&$	\text{---}	$&$	\text{---}	$&$		3.1	$&$	\text{---}	$&$	\text{---}	$&$	1.9	$&$	1.9	$&$	2.7	$&$	2.5	$\\
2 {\small(Comp. $\delta C_9$)}	&$	\text{---}	$&$	\text{---}	$&$	\text{---}	$&$		2.6	$&$	\text{---}	$&$	2.6	$&$	\text{---}	$&$	\text{---}	$&$	2.3	$&$	2.1	$\\
4 {\small(Comp. $\delta C_7,\delta C_9$)}	&$	\text{---}	$&$	\text{---}	$&$	\text{---}	$&$		\text{---}	$&$	\text{---}	$&$	\text{---}	$&$	\text{---}	$&$	\text{---}	$&$	\text{---}	$&$	1.3	$\\
3 {\small(Real $\Delta C_9^{\lambda,{\rm PC}}$)}	&$	\text{---}	$&$	\text{---}	$&$	\text{---}	$&$		\text{---}	$&$	\text{---}	$&$	3.5	$&$	2.4	$&$	2.2	$&$	3.0	$&$	2.7	$\\
6 {\small(Comp. $\Delta C_9^{\lambda,{\rm PC}}$)}	&$	\text{---}	$&$	\text{---}	$&$	\text{---}	$&$		\text{---}	$&$	\text{---}	$&$	\text{---}	$&$	\text{---}	$&$	\text{---}	$&$	1.0	$&$	1.0	$\\
6 {\small(Real $h_{+,-,0}^{(0,1)}$)}	&$	\text{---}	$&$	\text{---}	$&$	\text{---}	$&$		\text{---}	$&$	\text{---}	$&$	\text{---}	$&$	\text{---}	$&$	1.2	$&$	2.3	$&$	2.0	$\\
9 {\small(Real $h_{+,-,0}^{(0,1,2)}$)}	&$	\text{---}	$&$	\text{---}	$&$	\text{---}	$&$		\text{---}	$&$	\text{---}	$&$	\text{---}	$&$	\text{---}	$&$	\text{---}	$&$	\text{---}	$&$	2.0	$\\
12 {\small(Comp. $h_{+,-,0}^{(0,1)}$)}	&$	\text{---}	$&$	\text{---}	$&$	\text{---}	$&$		\text{---}	$&$	\text{---}	$&$	\text{---}	$&$	\text{---}	$&$	\text{---}	$&$	\text{---}	$&$	1.0	$\\
\hline
\end{tabular} 
}}
\\\vspace{0.3cm}
\scalebox{0.65}{
{\color{black}
\rowcolors{6}{}{light-gray}
\begin{tabular}{|l|c|c|c|c|c|c|c|c|c|c|c|}
\hline
\multicolumn{11}{|c|}{\small 		$B\to K^*\, \gamma/\ell\ell$ observables - $q^2\leqslant 8.68$ GeV$^2$ bins				}\\																
\hline																						
\multicolumn{1}{|c|}{\multirow{2}{*}{nr. of free}} & 1 & 2 & 2 & 4 & 3 & 6 & 6 & 9 & 12 & 18 \\ [-2pt]																						
\multicolumn{1}{|c|}{parameters} &																						
$\footnotesize \left(\!\!\begin{array}{c} {\rm Real} \\ \delta C_9 \end{array}\!\!\right)$ & $\footnotesize \left(\!\!\begin{array}{c} {\rm Real} \\ \delta C_7,\delta C_9 \end{array}\!\!\right)$ & $\footnotesize \left(\!\!\begin{array}{c} {\rm Comp.} \\ \delta C_9 \end{array}\!\!\right)$ & $\footnotesize \left(\!\!\begin{array}{c} {\rm Comp.} \\ \delta C_7,\delta C_9 \end{array}\!\!\right)$ & $\footnotesize \left(\!\!\begin{array}{c} {\rm Real} \\ \Delta C_9^{\lambda,{\rm PC}} \end{array}\!\!\right)$ & $\footnotesize \left(\!\!\begin{array}{c} {\rm Comp.} \\ \Delta C_9^{\lambda,{\rm PC}} \end{array}\!\!\right)$ & $\footnotesize \left(\!\!\begin{array}{c} {\rm Real} \\ h_{+,-,0}^{(0,1)} \end{array}\!\!\right)$ & $\footnotesize \left(\!\!\begin{array}{c} {\rm Real} \\ h_{+,-,0}^{(0,1,2)} \end{array}\!\!\right)$ & $\footnotesize \left(\!\!\begin{array}{c} {\rm Comp.} \\ h_{+,-,0}^{(0,1)} \end{array}\!\!\right)$ & $\footnotesize \left(\!\!\begin{array}{c} {\rm Comp.} \\ h_{+,-,0}^{(0,1,2)} \end{array}\!\!\right)$						\\ \hline																
0 (plain QCDf)	&$	 8.8	$&$	 8.6	$&$	 9.3	$&$		 9.7	$&$	 8.6	$&$	 9.5	$&$	 8.8	$&$	 9.0	$&$	 9.1	$&$	 9.2	$\\
1 {\small(Real $\delta C_9$)}	&$	\text{---}	$&$	1.4	$&$	3.9	$&$		4.8	$&$	1.9	$&$	4.6	$&$	3.1	$&$	3.9	$&$	4.3	$&$	4.8	$\\
2 {\small(Real $\delta C_7,\delta C_9$)}	&$	\text{---}	$&$	\text{---}	$&$	\text{---}	$&$		4.9	$&$	\text{---}	$&$	\text{---}	$&$	3.1	$&$	3.9	$&$	4.3	$&$	4.8	$\\
2 {\small(Comp. $\delta C_9$)}	&$	\text{---}	$&$	\text{---}	$&$	\text{---}	$&$		3.4	$&$	\text{---}	$&$	3.2	$&$	\text{---}	$&$	\text{---}	$&$	3.0	$&$	3.7	$\\
4 {\small(Comp. $\delta C_7,\delta C_9$)}	&$	\text{---}	$&$	\text{---}	$&$	\text{---}	$&$		\text{---}	$&$	\text{---}	$&$	\text{---}	$&$	\text{---}	$&$	\text{---}	$&$	\text{---}	$&$	2.7	$\\
3 {\small(Real $\Delta C_9^{\lambda,{\rm PC}}$)}	&$	\text{---}	$&$	\text{---}	$&$	\text{---}	$&$		\text{---}	$&$	\text{---}	$&$	4.6	$&$	2.9	$&$	3.7	$&$	4.1	$&$	4.7	$\\
6 {\small(Comp. $\Delta C_9^{\lambda,{\rm PC}}$)}	&$	\text{---}	$&$	\text{---}	$&$	\text{---}	$&$		\text{---}	$&$	\text{---}	$&$	\text{---}	$&$	\text{---}	$&$	\text{---}	$&$	1.3	$&$	2.6	$\\
6 {\small(Real $h_{+,-,0}^{(0,1)}$)}	&$	\text{---}	$&$	\text{---}	$&$	\text{---}	$&$		\text{---}	$&$	\text{---}	$&$	\text{---}	$&$	\text{---}	$&$	2.8	$&$	3.3	$&$	4.0	$\\
9 {\small(Real $h_{+,-,0}^{(0,1,2)}$)}	&$	\text{---}	$&$	\text{---}	$&$	\text{---}	$&$		\text{---}	$&$	\text{---}	$&$	\text{---}	$&$	\text{---}	$&$	\text{---}	$&$	\text{---}	$&$	3.3	$\\
12 {\small(Comp. $h_{+,-,0}^{(0,1)}$)}	&$	\text{---}	$&$	\text{---}	$&$	\text{---}	$&$		\text{---}	$&$	\text{---}	$&$	\text{---}	$&$	\text{---}	$&$	\text{---}	$&$	\text{---}	$&$	2.7	$\\
\hline
\end{tabular} 
}}
\caption{Improvement of the fits to $B\to K^* \gamma/\ell\ell$ observables for low-$q^2$ bins
for the hadronic fit and the scenarios with real and complex NP contributions to 
Wilson coefficients  $C_7$ and $C_9$ compared to the plain QCDf hypothesis and compared to each other.   \label{tab:Wilks_test}
}
\end{center} 
\end{table}  

In contrast, in the lower table (when the largest low-$q^2$ bins are included) 
we can read off from the first row that 
the various hadronic and NP scenarios offer a better description of the data than the plain QCDf prediction, with a significance larger than 
$8.6\sigma$. Compared to our previous results~\cite{Hurth:2020rzx}, this is a large increase of significance, where only $5-6\sigma$ were previously found. More importantly, one realises that the description of the data improves with a large significance of $3.9\sigma$ when going from the NP scenario of one real $\delta C_9$ to a hadronic fit with more than 9 parameters. This is an important change compared to our previous analysis in Ref~\cite{Hurth:2020rzx}, where one found only a change of $1.5\sigma$ in this case. {This is not unexpected, as the largest low-$q^2$ bins extend into a region where the QCDf approach is not considered very reliable. In the LHCb binning, the highest low-$q^2$ bin ([6, 8] GeV$^2$) was already included in our previous results. However, in the present analysis we also include the CMS result, whose largest low-$q^2$ bin ([6, 8.68] GeV$^2$) extends even further, approaching the tail of the $J/\psi$ resonance and moving deeper into the region where the QCDf treatment becomes increasingly questionable.} Therefore, it is difficult to establish a preference for the NP or the hadronic hypothesis based on these largest low-$q^2$ bins.

In any case, we have worked out projections of the present data for the two upgrade {stages} of the LHCb experiment in Ref.\cite{Hurth:2020rzx}. Assuming that future experimental data corresponds to the present best fit point of the NP fit to a real $\delta C_9$, one finds in both future upgrades scenarios - where we expect highly reduced uncertainties - large pulls which strongly favors the NP hypothesis - in spite of the fact that because the NP model is nested in the hadronic models we cannot disprove the hadronic option in favor of the NP one. 

In the other extreme future scenario, we assume the present best fit points of the 18-parameter hadronic fit, then of course the hadronic fit shows a large improvement in the description of the experimental data compared with the plain QCDf prediction or the NP fits. 
In that case, the Wilks' test shows a clear preference for the hadronic hypothesis.  
Thus, also in the case of this comprehensive Wilks' test, we can expect that the two LHCb upgrades will provide a clear  picture.

\section{Summary}\label{sec:conclusions}
In this manuscript, we have presented a critical assessment of the present $B$ anomalies based on the QCD factorisation (QCDf) approach.  

First, we have shown that the new CMS measurements of the angular observables 
of the $B \to K^* \mu^+\mu^-$ are compatible with the corresponding LHCb measurements at the 1$\sigma$ level. The updated 2025 LHCb data are in agreement with the previous LHCb results and also with the recent CMS data, but exhibit even larger tensions due to their smaller experimental uncertainties.

We also have shown in several examples that using different sets of local form factors based on LCSR and lattice calculations leads to notable changes in the NP significance. These large changes require further analysis. One explanation could be that the uncertainties in the form factor calculations have been underestimated so far - as already stated in Ref.~\cite{Chobanova:2017ghn}.

We find a NP significance of $5.2\sigma$ in the one-parameter fit to the Wilson coefficient $\delta C_9$ in our global fit to all $b \to s\ell\ell$ data assuming a 10\% guesstimate for the unknown power corrections to the QCDf prediction, and adopting the mentioned choice of form factors. 
This result corresponds to the global fit excluding the $q^2 \in [6, 8.68]~\text{GeV}^2$ region, where QCDf is not expected to be valid.
We have shown that a $100\%$ guesstimate for the unknown power corrections is needed to reduce the NP significance to $2.3\sigma$. This test demonstrates that there is still a significant deviation between data and QCDf predictions, which has to be understood. 

Within the data-driven approach, we have updated our earlier analyses in light of the new CMS and LHCb data, performing a comprehensive comparison between NP fits and general hadronic fits to the low-$q^2$ region. 
We have tested the helicity and $q^2$-independence of the NP description and found that present uncertainties are still too large for decisive conclusions. However, in the case of $q^2$-independence, the LHCb and CMS results indicate a slightly different picture, with the latter showing a greater inclination toward a $q^2$-dependence. A global comparison using Wilks’ test  (possible due to the nesting of the NP scenario into the hadronic one) confirms that, with present data, there is no strong statistical preference between NP and hadronic scenarios once the largest low-$q^2$ bins are excluded, whereas including them yields a preference for the hadronic fit—though this interpretation is complicated by the reduced reliability of QCDf in this region. Future LHCb and CMS upgrades should, however, provide much stronger discrimination between NP and hadronic scenarios.
 
\section*{Acknowledgements}
We thank A.~Khodjamirian and D.~Mishra for useful discussions, S.~Schmitt for valuable input on the $R_\phi$ results, and C.~Langenbruch for insightful discussions and help regarding the $B \to K^* \mu^+ \mu^-$ results. T.H. is supported by the Cluster of Excellence “Precision Physics, Fundamental Interactions, and Structure of Matter” (PRISMA+ EXC 2118/1) funded by the German Research Foundation (DFG) within the German Excellence Strategy (Project ID 390831469). T.H. also thanks the CERN theory group for its hospitality during his regular visits to CERN where part of the work was done.
This research is funded in part by the National Research Agency
(ANR) under project no. ANR-21-CE31-0002-01.

\appendix

\clearpage

\section{List of all observables}\label{sec:ObsList}
In this appendix, we provide the list of observables used in the global fit when including the $q^2 \in [6,8.68]$~GeV$^2$ bins, resulting in a total of 263 observables (see ancillary file \texttt{set\_AllConfig2.in}). For the set with 230 observables (\texttt{set\_AllConfig2\_wo68.in}), the $q^2 \in [6,8.68]$~GeV$^2$ bins are excluded. 

For the fits to angular observables only, discussed in Section~\ref{sec:CMS_vs_LHCb}, we use subsets of the observables listed in Tables~\ref{tab:Nob2025_obs_1_70}–\ref{tab:Nob2025_obs_181_263}, as given in  \texttt{set\_LHCbConfig2\_AngObs\_wo68.in}, \texttt{set\_CMS\_AngObs\_wo68.in},  and \texttt{set\_CMS\_LHCbConfig2\_AngObs\_wo\_68.in}. For the fit to the $S_i$ observables, we replace the $P_i^{(\prime)}$ observables of config.~2 (192--263 in Table~\ref{tab:Nob2025_obs_181_263}) with the corresponding $S_i$  observables of config.~1  (\texttt{set\_LHCbConfig1\_AngObs\_wo68.in}). 

For the hadronic fits presented in Section~\ref{sec:data_driven}, we employ the observables listed in \texttt{set\_HadronicFit\_below6.in} and \texttt{set\_HadronicFit\_below8.in}.

\begin{table}[bh!]
\begin{center}
\scalebox{0.65}{
\begin{tabular}{|l|l|l|}
\hline
nr. & observable & Ref. \\ \hline
1 & \texttt{ AI\_BKstargamma } & \cite{HeavyFlavorAveragingGroupHFLAV:2024ctg} \\ \hline
2 & \texttt{ BR\_BXsgamma } & \cite{HeavyFlavorAveragingGroupHFLAV:2024ctg} \\ \hline
3 & \texttt{ BRuntag\_Bsmumu } & \cite{ParticleDataGroup:2024cfk} \\ \hline
4 & \texttt{ BRuntag\_Bsee } & \cite{LHCb:2020pcv} \\ \hline
5 & \texttt{ BR\_BXsmumu\_1\_6 } & \cite{BaBar:2013qry} \\ \hline
6 & \texttt{ BR\_BXsmumu\_14.2\_22 } & \cite{BaBar:2013qry} \\ \hline
7 & \texttt{ BR\_BXsee\_1\_6 } & \cite{BaBar:2013qry} \\ \hline
8 & \texttt{ BR\_BXsee\_14.2\_22 } & \cite{BaBar:2013qry} \\ \hline
9 & \texttt{ BR\_B0Kstar0gamma } & \cite{ParticleDataGroup:2024cfk} \\ \hline
10 & \texttt{ BR\_BKstargamma } & \cite{ParticleDataGroup:2024cfk} \\ \hline

11 & \texttt{ dGamma/dq2\_BKstarmumu\_1.1\_6 } & \cite{LHCb:2014cxe} \\ \hline
12 & \texttt{ dGamma/dq2\_BKstarmumu\_15\_19 } & \cite{LHCb:2014cxe} \\ \hline

13 & \texttt{ R-1\_B0Kstar0ll\_0.1\_1.1 } & \cite{LHCb:2022vje} \\ \hline
14 & \texttt{ R-1\_B0Kstar0ll\_1.1\_6 } & \cite{LHCb:2022vje} \\ \hline
15 & \texttt{ R-1\_B0Kstar0ll\_0.045\_1.1\_Belle } & \cite{Belle:2019oag} \\ \hline
16 & \texttt{ R-1\_B0Kstar0ll\_1.1\_6\_Belle } & \cite{Belle:2019oag} \\ \hline
17 & \texttt{ R-1\_B0Kstar0ll\_15\_19\_Belle } & \cite{Belle:2019oag} \\ \hline
18 & \texttt{ dGamma/dq2\_B0K0mumu\_1.1\_6 } & \cite{LHCb:2014cxe} \\ \hline
19 & \texttt{ dGamma/dq2\_B0K0mumu\_15\_22 } & \cite{LHCb:2014cxe} \\ \hline
20 & \texttt{ dGamma/dq2\_BKmumu\_1.1\_6 } & \cite{LHCb:2014cxe} \\ \hline
21 & \texttt{ FH\_BKmumu\_1.1\_6 } & \cite{LHCb:2014auh} \\ \hline
22 & \texttt{ dGamma/dq2\_BKmumu\_15\_22 } & \cite{LHCb:2014cxe} \\ \hline
23 & \texttt{ FH\_BKmumu\_15\_22 } & \cite{LHCb:2014auh} \\ \hline
24 & \texttt{ R-1\_BKll\_0.1\_1.1 } & \cite{LHCb:2014cxe} \\ \hline
25 & \texttt{ R-1\_BKll\_1.1\_6 } & \cite{LHCb:2014cxe} \\ \hline
26 & \texttt{ dGamma/dq2\_Bsphimumu\_0.1\_0.98 } & \cite{LHCb:2021zwz} \\ \hline
27 & \texttt{ FL\_Bsphimumu\_0.1\_0.98 } & \cite{LHCb:2021xxq} \\ \hline
28 & \texttt{ S3\_Bsphimumu\_0.1\_0.98 } & \cite{LHCb:2021xxq} \\ \hline
29 & \texttt{ S4\_Bsphimumu\_0.1\_0.98 } & \cite{LHCb:2021xxq} \\ \hline
30 & \texttt{ S7\_Bsphimumu\_0.1\_0.98 } & \cite{LHCb:2021xxq} \\ \hline
31 & \texttt{ dGamma/dq2\_Bsphimumu\_1.1\_2.5 } & \cite{LHCb:2021zwz} \\ \hline
32 & \texttt{ dGamma/dq2\_Bsphimumu\_2.5\_4 } & \cite{LHCb:2021zwz} \\ \hline
33 & \texttt{ FL\_Bsphimumu\_1.1\_4 } & \cite{LHCb:2021xxq} \\ \hline
34 & \texttt{ S3\_Bsphimumu\_1.1\_4 } & \cite{LHCb:2021xxq} \\ \hline
35 & \texttt{ S4\_Bsphimumu\_1.1\_4 } & \cite{LHCb:2021xxq} \\ \hline
\end{tabular}
\quad
\begin{tabular}{|l|l|l|}\hline
nr. & observable & Ref. \\ \hline
36 & \texttt{ S7\_Bsphimumu\_1.1\_4 } & \cite{LHCb:2021xxq} \\ \hline
37 & \texttt{ dGamma/dq2\_Bsphimumu\_4\_6 } & \cite{LHCb:2021zwz} \\ \hline
38 & \texttt{ FL\_Bsphimumu\_4\_6 } & \cite{LHCb:2021xxq} \\ \hline
39 & \texttt{ S3\_Bsphimumu\_4\_6 } & \cite{LHCb:2021xxq} \\ \hline
40 & \texttt{ S4\_Bsphimumu\_4\_6 } & \cite{LHCb:2021xxq} \\ \hline
41 & \texttt{ S7\_Bsphimumu\_4\_6 } & \cite{LHCb:2021xxq} \\ \hline
42 & \texttt{ dGamma/dq2\_Bsphimumu\_6\_8 } & \cite{LHCb:2021zwz} \\ \hline
43 & \texttt{ FL\_Bsphimumu\_6\_8 } & \cite{LHCb:2021xxq} \\ \hline
44 & \texttt{ S3\_Bsphimumu\_6\_8 } & \cite{LHCb:2021xxq} \\ \hline
45 & \texttt{ S4\_Bsphimumu\_6\_8 } & \cite{LHCb:2021xxq} \\ \hline
46 & \texttt{ S7\_Bsphimumu\_6\_8 } & \cite{LHCb:2021xxq} \\ \hline
47 & \texttt{ dGamma/dq2\_Bsphimumu\_15\_19 } & \cite{LHCb:2021zwz} \\ \hline
48 & \texttt{ FL\_Bsphimumu\_15\_18.9 } & \cite{LHCb:2021xxq} \\ \hline
49 & \texttt{ S3\_Bsphimumu\_15\_18.9 } & \cite{LHCb:2021xxq} \\ \hline
50 & \texttt{ S4\_Bsphimumu\_15\_18.9 } & \cite{LHCb:2021xxq} \\ \hline
51 & \texttt{ S7\_Bsphimumu\_15\_18.9 } & \cite{LHCb:2021xxq} \\ \hline
52 & \texttt{ dGamma/dq2\_LambdabLambdamumu\_15\_20 } & \cite{LHCb:2015tgy} \\ \hline
53 & \texttt{ AlFB\_LambdabLambdamumu\_15\_20 } & \cite{LHCb:2018jna} \\ \hline
54 & \texttt{ AhFB\_LambdabLambdamumu\_15\_20 } & \cite{LHCb:2018jna} \\ \hline
55 & \texttt{ AlhFB\_LambdabLambdamumu\_15\_20 } & \cite{LHCb:2018jna} \\ \hline
56 & \texttt{ FL\_LambdabLambdamumu\_15\_20 } & \cite{LHCb:2015tgy} \\ \hline
57 & \texttt{ FL\_BKstarmumu\_0.1\_0.98 } & \cite{LHCb:2020lmf} \\ \hline
58 & \texttt{ AFB\_BKstarmumu\_0.1\_0.98 } & \cite{LHCb:2020lmf} \\ \hline
59 & \texttt{ S3\_BKstarmumu\_0.1\_0.98 } & \cite{LHCb:2020lmf} \\ \hline
60 & \texttt{ S4\_BKstarmumu\_0.1\_0.98 } & \cite{LHCb:2020lmf} \\ \hline
61 & \texttt{ S5\_BKstarmumu\_0.1\_0.98 } & \cite{LHCb:2020lmf} \\ \hline
62 & \texttt{ S7\_BKstarmumu\_0.1\_0.98 } & \cite{LHCb:2020lmf} \\ \hline
63 & \texttt{ S8\_BKstarmumu\_0.1\_0.98 } & \cite{LHCb:2020lmf} \\ \hline
64 & \texttt{ S9\_BKstarmumu\_0.1\_0.98 } & \cite{LHCb:2020lmf} \\ \hline
65 & \texttt{ FL\_BKstarmumu\_1.1\_2.5 } & \cite{LHCb:2020lmf} \\ \hline
66 & \texttt{ AFB\_BKstarmumu\_1.1\_2.5 } & \cite{LHCb:2020lmf} \\ \hline
67 & \texttt{ S3\_BKstarmumu\_1.1\_2.5 } & \cite{LHCb:2020lmf} \\ \hline
68 & \texttt{ S4\_BKstarmumu\_1.1\_2.5 } & \cite{LHCb:2020lmf} \\ \hline
69 & \texttt{ S5\_BKstarmumu\_1.1\_2.5 } & \cite{LHCb:2020lmf} \\ \hline
70 & \texttt{ S7\_BKstarmumu\_1.1\_2.5 } & \cite{LHCb:2020lmf} \\ \hline
\end{tabular}
}
\caption{Observables used in the global fit.}
\label{tab:Nob2025_obs_1_70}
\end{center}
\end{table}

\begin{table}[bh!]
\begin{center}
\scalebox{0.65}{
\begin{tabular}{|l|l|l|}
\hline
nr. & observable & Ref. \\ \hline
71 & \texttt{ S8\_BKstarmumu\_1.1\_2.5 } & \cite{LHCb:2020lmf} \\ \hline
72 & \texttt{ S9\_BKstarmumu\_1.1\_2.5 } & \cite{LHCb:2020lmf} \\ \hline
73 & \texttt{ FL\_BKstarmumu\_2.5\_4 } & \cite{LHCb:2020lmf} \\ \hline
74 & \texttt{ AFB\_BKstarmumu\_2.5\_4 } & \cite{LHCb:2020lmf} \\ \hline
75 & \texttt{ S3\_BKstarmumu\_2.5\_4 } & \cite{LHCb:2020lmf} \\ \hline
76 & \texttt{ S4\_BKstarmumu\_2.5\_4 } & \cite{LHCb:2020lmf} \\ \hline
77 & \texttt{ S5\_BKstarmumu\_2.5\_4 } & \cite{LHCb:2020lmf} \\ \hline
78 & \texttt{ S7\_BKstarmumu\_2.5\_4 } & \cite{LHCb:2020lmf} \\ \hline
79 & \texttt{ S8\_BKstarmumu\_2.5\_4 } & \cite{LHCb:2020lmf} \\ \hline
80 & \texttt{ S9\_BKstarmumu\_2.5\_4 } & \cite{LHCb:2020lmf} \\ \hline
81 & \texttt{ FL\_BKstarmumu\_4\_6 } & \cite{LHCb:2020lmf} \\ \hline
82 & \texttt{ AFB\_BKstarmumu\_4\_6 } & \cite{LHCb:2020lmf} \\ \hline
83 & \texttt{ S3\_BKstarmumu\_4\_6 } & \cite{LHCb:2020lmf} \\ \hline
84 & \texttt{ S4\_BKstarmumu\_4\_6 } & \cite{LHCb:2020lmf} \\ \hline
85 & \texttt{ S5\_BKstarmumu\_4\_6 } & \cite{LHCb:2020lmf} \\ \hline
86 & \texttt{ S7\_BKstarmumu\_4\_6 } & \cite{LHCb:2020lmf} \\ \hline
87 & \texttt{ S8\_BKstarmumu\_4\_6 } & \cite{LHCb:2020lmf} \\ \hline
88 & \texttt{ S9\_BKstarmumu\_4\_6 } & \cite{LHCb:2020lmf} \\ \hline
89 & \texttt{ FL\_BKstarmumu\_6\_8 } & \cite{LHCb:2020lmf} \\ \hline
90 & \texttt{ AFB\_BKstarmumu\_6\_8 } & \cite{LHCb:2020lmf} \\ \hline
91 & \texttt{ S3\_BKstarmumu\_6\_8 } & \cite{LHCb:2020lmf} \\ \hline
92 & \texttt{ S4\_BKstarmumu\_6\_8 } & \cite{LHCb:2020lmf} \\ \hline
93 & \texttt{ S5\_BKstarmumu\_6\_8 } & \cite{LHCb:2020lmf} \\ \hline
94 & \texttt{ S7\_BKstarmumu\_6\_8 } & \cite{LHCb:2020lmf} \\ \hline
95 & \texttt{ S8\_BKstarmumu\_6\_8 } & \cite{LHCb:2020lmf} \\ \hline
96 & \texttt{ S9\_BKstarmumu\_6\_8 } & \cite{LHCb:2020lmf} \\ \hline
97 & \texttt{ FL\_BKstarmumu\_15\_17 } & \cite{LHCb:2020lmf} \\ \hline
98 & \texttt{ AFB\_BKstarmumu\_15\_17 } & \cite{LHCb:2020lmf} \\ \hline
99 & \texttt{ S3\_BKstarmumu\_15\_17 } & \cite{LHCb:2020lmf} \\ \hline
100 & \texttt{ S4\_BKstarmumu\_15\_17 } & \cite{LHCb:2020lmf} \\ \hline
101 & \texttt{ S5\_BKstarmumu\_15\_17 } & \cite{LHCb:2020lmf} \\ \hline
102 & \texttt{ S7\_BKstarmumu\_15\_17 } & \cite{LHCb:2020lmf} \\ \hline
103 & \texttt{ S8\_BKstarmumu\_15\_17 } & \cite{LHCb:2020lmf} \\ \hline
104 & \texttt{ S9\_BKstarmumu\_15\_17 } & \cite{LHCb:2020lmf} \\ \hline
105 & \texttt{ FL\_BKstarmumu\_17\_19 } & \cite{LHCb:2020lmf} \\ \hline
106 & \texttt{ AFB\_BKstarmumu\_17\_19 } & \cite{LHCb:2020lmf} \\ \hline
107 & \texttt{ S3\_BKstarmumu\_17\_19 } & \cite{LHCb:2020lmf} \\ \hline
108 & \texttt{ S4\_BKstarmumu\_17\_19 } & \cite{LHCb:2020lmf} \\ \hline
109 & \texttt{ S5\_BKstarmumu\_17\_19 } & \cite{LHCb:2020lmf} \\ \hline
110 & \texttt{ S7\_BKstarmumu\_17\_19 } & \cite{LHCb:2020lmf} \\ \hline
111 & \texttt{ S8\_BKstarmumu\_17\_19 } & \cite{LHCb:2020lmf} \\ \hline
112 & \texttt{ S9\_BKstarmumu\_17\_19 } & \cite{LHCb:2020lmf} \\ \hline
113 & \texttt{ R-1\_BKstarll\_0.045\_6 } & \cite{LHCb:2021lvy} \\ \hline
114 & \texttt{ R-1\_B0K0ll\_1.1\_6 } & \cite{LHCb:2021lvy} \\ \hline
115 & \texttt{ R-1\_BKll\_1\_6\_Belle } & \cite{BELLE:2019xld} \\ \hline
116 & \texttt{ FH\_BKmumu\_1\_6\_CMS } & \cite{CMS:2018qih} \\ \hline
117 & \texttt{ dGamma/dq2\_B0Kstar0ee\_0.0009\_1 } & \cite{LHCb:2013pra} \\ \hline
118 & \texttt{ FL\_B0Kstar0ee\_0.0008\_0.257 } & \cite{LHCb:2020dof} \\ \hline
119 & \texttt{ ATRe\_B0Kstar0ee\_0.0008\_0.257 } & \cite{LHCb:2020dof} \\ \hline
120 & \texttt{ AT2\_B0Kstar0ee\_0.0008\_0.257 } & \cite{LHCb:2020dof} \\ \hline
121 & \texttt{ dGamma/dq2\_BKmumu\_0.1\_0.98\_CMS } & \cite{CMS:2024syx} \\ \hline
122 & \texttt{ dGamma/dq2\_BKmumu\_1.1\_2\_CMS } & \cite{CMS:2024syx} \\ \hline
123 & \texttt{ dGamma/dq2\_BKmumu\_2\_3\_CMS } & \cite{CMS:2024syx} \\ \hline
124 & \texttt{ dGamma/dq2\_BKmumu\_3\_4\_CMS } & \cite{CMS:2024syx} \\ \hline
125 & \texttt{ dGamma/dq2\_BKmumu\_4\_5\_CMS } & \cite{CMS:2024syx} \\ \hline
\end{tabular}
\quad
\begin{tabular}{|l|l|l|}\hline
nr. & observable & Ref. \\ \hline
126 & \texttt{ dGamma/dq2\_BKmumu\_5\_6\_CMS } & \cite{CMS:2024syx} \\ \hline
127 & \texttt{ dGamma/dq2\_BKmumu\_6\_7\_CMS } & \cite{CMS:2024syx} \\ \hline
128 & \texttt{ dGamma/dq2\_BKmumu\_7\_8\_CMS } & \cite{CMS:2024syx} \\ \hline
129 & \texttt{ dGamma/dq2\_BKmumu\_14.82\_16\_CMS } & \cite{CMS:2024syx} \\ \hline
130 & \texttt{ dGamma/dq2\_BKmumu\_16\_17\_CMS } & \cite{CMS:2024syx} \\ \hline
131 & \texttt{ dGamma/dq2\_BKmumu\_17\_18\_CMS } & \cite{CMS:2024syx} \\ \hline
132 & \texttt{ dGamma/dq2\_BKmumu\_18\_19.24\_CMS } & \cite{CMS:2024syx} \\ \hline
133 & \texttt{ dGamma/dq2\_BKmumu\_19.24\_22.9\_CMS } & \cite{CMS:2024syx} \\ \hline
134 & \texttt{ R-1\_BKll\_1.1\_6\_CMS } & \cite{CMS:2024syx} \\ \hline
135 & \texttt{ FL\_B0Kstar0mumu\_1.1\_2\_CMS } & \cite{CMS:2024atz} \\ \hline
136 & \texttt{ P1\_B0Kstar0mumu\_1.1\_2\_CMS } & \cite{CMS:2024atz} \\ \hline
137 & \texttt{ P2\_B0Kstar0mumu\_1.1\_2\_CMS } & \cite{CMS:2024atz} \\ \hline
138 & \texttt{ P3\_B0Kstar0mumu\_1.1\_2\_CMS } & \cite{CMS:2024atz} \\ \hline
139 & \texttt{ P4prime\_B0Kstar0mumu\_1.1\_2\_CMS } & \cite{CMS:2024atz} \\ \hline
140 & \texttt{ P5prime\_B0Kstar0mumu\_1.1\_2\_CMS } & \cite{CMS:2024atz} \\ \hline
141 & \texttt{ P6prime\_B0Kstar0mumu\_1.1\_2\_CMS } & \cite{CMS:2024atz} \\ \hline
142 & \texttt{ P8prime\_B0Kstar0mumu\_1.1\_2\_CMS } & \cite{CMS:2024atz} \\ \hline
143 & \texttt{ FL\_B0Kstar0mumu\_2\_4.3\_CMS } & \cite{CMS:2024atz} \\ \hline
144 & \texttt{ P1\_B0Kstar0mumu\_2\_4.3\_CMS } & \cite{CMS:2024atz} \\ \hline
145 & \texttt{ P2\_B0Kstar0mumu\_2\_4.3\_CMS } & \cite{CMS:2024atz} \\ \hline
146 & \texttt{ P3\_B0Kstar0mumu\_2\_4.3\_CMS } & \cite{CMS:2024atz} \\ \hline
147 & \texttt{ P4prime\_B0Kstar0mumu\_2\_4.3\_CMS } & \cite{CMS:2024atz} \\ \hline
148 & \texttt{ P5prime\_B0Kstar0mumu\_2\_4.3\_CMS } & \cite{CMS:2024atz} \\ \hline
149 & \texttt{ P6prime\_B0Kstar0mumu\_2\_4.3\_CMS } & \cite{CMS:2024atz} \\ \hline
150 & \texttt{ P8prime\_B0Kstar0mumu\_2\_4.3\_CMS } & \cite{CMS:2024atz} \\ \hline
151 & \texttt{ FL\_B0Kstar0mumu\_4.3\_6\_CMS } & \cite{CMS:2024atz} \\ \hline
152 & \texttt{ P1\_B0Kstar0mumu\_4.3\_6\_CMS } & \cite{CMS:2024atz} \\ \hline
153 & \texttt{ P2\_B0Kstar0mumu\_4.3\_6\_CMS } & \cite{CMS:2024atz} \\ \hline
154 & \texttt{ P3\_B0Kstar0mumu\_4.3\_6\_CMS } & \cite{CMS:2024atz} \\ \hline
155 & \texttt{ P4prime\_B0Kstar0mumu\_4.3\_6\_CMS } & \cite{CMS:2024atz} \\ \hline
156 & \texttt{ P5prime\_B0Kstar0mumu\_4.3\_6\_CMS } & \cite{CMS:2024atz} \\ \hline
157 & \texttt{ P6prime\_B0Kstar0mumu\_4.3\_6\_CMS } & \cite{CMS:2024atz} \\ \hline
158 & \texttt{ P8prime\_B0Kstar0mumu\_4.3\_6\_CMS } & \cite{CMS:2024atz} \\ \hline
159 & \texttt{ FL\_B0Kstar0mumu\_6\_8.68\_CMS } & \cite{CMS:2024atz} \\ \hline
160 & \texttt{ P1\_B0Kstar0mumu\_6\_8.68\_CMS } & \cite{CMS:2024atz} \\ \hline
161 & \texttt{ P2\_B0Kstar0mumu\_6\_8.68\_CMS } & \cite{CMS:2024atz} \\ \hline
162 & \texttt{ P3\_B0Kstar0mumu\_6\_8.68\_CMS } & \cite{CMS:2024atz} \\ \hline
163 & \texttt{ P4prime\_B0Kstar0mumu\_6\_8.68\_CMS } & \cite{CMS:2024atz} \\ \hline
164 & \texttt{ P5prime\_B0Kstar0mumu\_6\_8.68\_CMS } & \cite{CMS:2024atz} \\ \hline
165 & \texttt{ P6prime\_B0Kstar0mumu\_6\_8.68\_CMS } & \cite{CMS:2024atz} \\ \hline
166 & \texttt{ P8prime\_B0Kstar0mumu\_6\_8.68\_CMS } & \cite{CMS:2024atz} \\ \hline
167 & \texttt{ FL\_B0Kstar0mumu\_14.18\_16\_CMS } & \cite{CMS:2024atz} \\ \hline
168 & \texttt{ P1\_B0Kstar0mumu\_14.18\_16\_CMS } & \cite{CMS:2024atz} \\ \hline
169 & \texttt{ P2\_B0Kstar0mumu\_14.18\_16\_CMS } & \cite{CMS:2024atz} \\ \hline
170 & \texttt{ P3\_B0Kstar0mumu\_14.18\_16\_CMS } & \cite{CMS:2024atz} \\ \hline
171 & \texttt{ P4prime\_B0Kstar0mumu\_14.18\_16\_CMS } & \cite{CMS:2024atz} \\ \hline
172 & \texttt{ P5prime\_B0Kstar0mumu\_14.18\_16\_CMS } & \cite{CMS:2024atz} \\ \hline
173 & \texttt{ P6prime\_B0Kstar0mumu\_14.18\_16\_CMS } & \cite{CMS:2024atz} \\ \hline
174 & \texttt{ P8prime\_B0Kstar0mumu\_14.18\_16\_CMS } & \cite{CMS:2024atz} \\ \hline
175 & \texttt{ dGamma/dq2\_Bsphiee\_0.1\_1.1 } & \cite{LHCb:2024rto} \\ \hline
176 & \texttt{ dGamma/dq2\_Bsphiee\_1.1\_6 } & \cite{LHCb:2024rto} \\ \hline
177 & \texttt{ dGamma/dq2\_Bsphiee\_15\_19 } & \cite{LHCb:2024rto} \\ \hline
178 & \texttt{ R-1\_Bsphill\_0.1\_1.1 } & \cite{LHCb:2024rto} \\ \hline
179 & \texttt{ R-1\_Bsphill\_1.1\_6 } & \cite{LHCb:2024rto} \\ \hline
180 & \texttt{ R-1\_Bsphill\_15\_19 } & \cite{LHCb:2024rto} \\ \hline
\end{tabular}
}
\caption{Continued from Table~\ref{tab:Nob2025_obs_1_70}.}
\label{tab:Nob2025_obs_71_180}
\end{center}
\end{table}

\begin{table}[th!]
\begin{center}
\scalebox{0.65}{
\begin{tabular}{|l|l|l|}
\hline
nr. & observable & Ref. \\ \hline
181 & \texttt{ FL\_Bsphiee\_0.0009\_0.2615 } & \cite{LHCb:2024wnx} \\ \hline
182 & \texttt{ AT2\_Bsphiee\_0.0009\_0.2615 } & \cite{LHCb:2024wnx} \\ \hline
183 & \texttt{ AT2\_B0Kstar0ee\_0.0008\_1.12\_Belle } & \cite{Belle:2024mml} \\ \hline

184 & \texttt{ FL\_B0Kstar0ee\_1.1\_6 } & \cite{LHCb:2025pxz} \\ \hline
185 & \texttt{ P1\_B0Kstar0ee\_1.1\_6 } & \cite{LHCb:2025pxz} \\ \hline
186 & \texttt{ P2\_B0Kstar0ee\_1.1\_6 } & \cite{LHCb:2025pxz} \\ \hline
187 & \texttt{ P3\_B0Kstar0ee\_1.1\_6 } & \cite{LHCb:2025pxz} \\ \hline
188 & \texttt{ P4prime\_B0Kstar0ee\_1.1\_6 } & \cite{LHCb:2025pxz} \\ \hline
189 & \texttt{ P5prime\_B0Kstar0ee\_1.1\_6 } & \cite{LHCb:2025pxz} \\ \hline
190 & \texttt{ P6prime\_B0Kstar0ee\_1.1\_6 } & \cite{LHCb:2025pxz} \\ \hline
191 & \texttt{ P8prime\_B0Kstar0ee\_1.1\_6 } & \cite{LHCb:2025pxz} \\ \hline

192 & \texttt{ FL\_B0Kstar0mumu\_0.06\_0.98 } & \cite{LHCb:2025update} \\ \hline
193 & \texttt{ S2s\_B0Kstar0mumu\_0.06\_0.98 } & \cite{LHCb:2025update} \\ \hline
194 & \texttt{ S1c\_B0Kstar0mumu\_0.06\_0.98 } & \cite{LHCb:2025update} \\ \hline
195 & \texttt{ P1\_B0Kstar0mumu\_0.06\_0.98 } & \cite{LHCb:2025update} \\ \hline
196 & \texttt{ P2\_B0Kstar0mumu\_0.06\_0.98 } & \cite{LHCb:2025update} \\ \hline
197 & \texttt{ P3\_B0Kstar0mumu\_0.06\_0.98 } & \cite{LHCb:2025update} \\ \hline
198 & \texttt{ P4prime\_B0Kstar0mumu\_0.06\_0.98 } & \cite{LHCb:2025update} \\ \hline
199 & \texttt{ P5prime\_B0Kstar0mumu\_0.06\_0.98 } & \cite{LHCb:2025update} \\ \hline
200 & \texttt{ P6prime\_B0Kstar0mumu\_0.06\_0.98 } & \cite{LHCb:2025update} \\ \hline
201 & \texttt{ P8prime\_B0Kstar0mumu\_0.06\_0.98 } & \cite{LHCb:2025update} \\ \hline
202 & \texttt{ S6c\_B0Kstar0mumu\_0.06\_0.98 } & \cite{LHCb:2025update} \\ \hline
203 & \texttt{ dGamma/dq2\_B0Kstar0mumu\_0.06\_0.98 } & \cite{LHCb:2025update} \\ \hline

204 & \texttt{ FL\_B0Kstar0mumu\_1.1\_2.5\_LHCb2025c2 } & \cite{LHCb:2025update} \\ \hline
205 & \texttt{ S1c\_B0Kstar0mumu\_1.1\_2.5\_LHCb2025c2 } & \cite{LHCb:2025update} \\ \hline
206 & \texttt{ P1\_B0Kstar0mumu\_1.1\_2.5\_LHCb2025c2 } & \cite{LHCb:2025update} \\ \hline
207 & \texttt{ P2\_B0Kstar0mumu\_1.1\_2.5\_LHCb2025c2 } & \cite{LHCb:2025update} \\ \hline
208 & \texttt{ P3\_B0Kstar0mumu\_1.1\_2.5\_LHCb2025c2 } & \cite{LHCb:2025update} \\ \hline
209 & \texttt{ P4prime\_B0Kstar0mumu\_1.1\_2.5\_LHCb2025c2 } & \cite{LHCb:2025update} \\ \hline
210 & \texttt{ P5prime\_B0Kstar0mumu\_1.1\_2.5\_LHCb2025c2 } & \cite{LHCb:2025update} \\ \hline
211 & \texttt{ P6prime\_B0Kstar0mumu\_1.1\_2.5\_LHCb2025c2 } & \cite{LHCb:2025update} \\ \hline
212 & \texttt{ P8prime\_B0Kstar0mumu\_1.1\_2.5\_LHCb2025c2 } & \cite{LHCb:2025update} \\ \hline
213 & \texttt{ dGamma/dq2\_B0Kstar0mumu\_1.1\_2.5 } & \cite{LHCb:2025update} \\ \hline

214 & \texttt{ FL\_B0Kstar0mumu\_2.5\_4.0\_LHCb2025c2 } & \cite{LHCb:2025update} \\ \hline
215 & \texttt{ S1c\_B0Kstar0mumu\_2.5\_4.0\_LHCb2025c2 } & \cite{LHCb:2025update} \\ \hline
216 & \texttt{ P1\_B0Kstar0mumu\_2.5\_4.0\_LHCb2025c2 } & \cite{LHCb:2025update} \\ \hline
217 & \texttt{ P2\_B0Kstar0mumu\_2.5\_4.0\_LHCb2025c2 } & \cite{LHCb:2025update} \\ \hline
218 & \texttt{ P3\_B0Kstar0mumu\_2.5\_4.0\_LHCb2025c2 } & \cite{LHCb:2025update} \\ \hline
219 & \texttt{ P4prime\_B0Kstar0mumu\_2.5\_4.0\_LHCb2025c2 } & \cite{LHCb:2025update} \\ \hline
220 & \texttt{ P5prime\_B0Kstar0mumu\_2.5\_4.0\_LHCb2025c2 } & \cite{LHCb:2025update} \\ \hline
221 & \texttt{ P6prime\_B0Kstar0mumu\_2.5\_4.0\_LHCb2025c2 } & \cite{LHCb:2025update} \\ \hline
222 & \texttt{ P8prime\_B0Kstar0mumu\_2.5\_4.0\_LHCb2025c2 } & \cite{LHCb:2025update} \\ \hline
223 & \texttt{ dGamma/dq2\_B0Kstar0mumu\_2.5\_4.0\_LHCb2025c2 } & \cite{LHCb:2025update} \\ \hline
\end{tabular}
\quad
\begin{tabular}{|l|l|l|}\hline
nr. & observable & Ref. \\ \hline
224 & \texttt{ FL\_B0Kstar0mumu\_4.0\_6.0\_LHCb2025c2 } & \cite{LHCb:2025update} \\ \hline
225 & \texttt{ S1c\_B0Kstar0mumu\_4.0\_6.0\_LHCb2025c2 } & \cite{LHCb:2025update} \\ \hline
226 & \texttt{ P1\_B0Kstar0mumu\_4.0\_6.0\_LHCb2025c2 } & \cite{LHCb:2025update} \\ \hline
227 & \texttt{ P2\_B0Kstar0mumu\_4.0\_6.0\_LHCb2025c2 } & \cite{LHCb:2025update} \\ \hline
228 & \texttt{ P3\_B0Kstar0mumu\_4.0\_6.0\_LHCb2025c2 } & \cite{LHCb:2025update} \\ \hline
229 & \texttt{ P4prime\_B0Kstar0mumu\_4.0\_6.0\_LHCb2025c2 } & \cite{LHCb:2025update} \\ \hline
230 & \texttt{ P5prime\_B0Kstar0mumu\_4.0\_6.0\_LHCb2025c2 } & \cite{LHCb:2025update} \\ \hline
231 & \texttt{ P6prime\_B0Kstar0mumu\_4.0\_6.0\_LHCb2025c2 } & \cite{LHCb:2025update} \\ \hline
232 & \texttt{ P8prime\_B0Kstar0mumu\_4.0\_6.0\_LHCb2025c2 } & \cite{LHCb:2025update} \\ \hline
233 & \texttt{ dGamma/dq2\_B0Kstar0mumu\_4.0\_6.0\_LHCb2025c2 } & \cite{LHCb:2025update} \\ \hline

234 & \texttt{ FL\_B0Kstar0mumu\_6.0\_8.0\_LHCb2025c2 } & \cite{LHCb:2025update} \\ \hline
235 & \texttt{ S1c\_B0Kstar0mumu\_6.0\_8.0\_LHCb2025c2 } & \cite{LHCb:2025update} \\ \hline
236 & \texttt{ P1\_B0Kstar0mumu\_6.0\_8.0\_LHCb2025c2 } & \cite{LHCb:2025update} \\ \hline
237 & \texttt{ P2\_B0Kstar0mumu\_6.0\_8.0\_LHCb2025c2 } & \cite{LHCb:2025update} \\ \hline
238 & \texttt{ P3\_B0Kstar0mumu\_6.0\_8.0\_LHCb2025c2 } & \cite{LHCb:2025update} \\ \hline
239 & \texttt{ P4prime\_B0Kstar0mumu\_6.0\_8.0\_LHCb2025c2 } & \cite{LHCb:2025update} \\ \hline
240 & \texttt{ P5prime\_B0Kstar0mumu\_6.0\_8.0\_LHCb2025c2 } & \cite{LHCb:2025update} \\ \hline
241 & \texttt{ P6prime\_B0Kstar0mumu\_6.0\_8.0\_LHCb2025c2 } & \cite{LHCb:2025update} \\ \hline
242 & \texttt{ P8prime\_B0Kstar0mumu\_6.0\_8.0\_LHCb2025c2 } & \cite{LHCb:2025update} \\ \hline
243 & \texttt{ dGamma/dq2\_B0Kstar0mumu\_6.0\_8.0\_LHCb2025c2 } & \cite{LHCb:2025update} \\ \hline

244 & \texttt{ FL\_B0Kstar0mumu\_15.0\_17.0\_LHCb2025c2 } & \cite{LHCb:2025update} \\ \hline
245 & \texttt{ S1c\_B0Kstar0mumu\_15.0\_17.0\_LHCb2025c2 } & \cite{LHCb:2025update} \\ \hline
246 & \texttt{ P1\_B0Kstar0mumu\_15.0\_17.0\_LHCb2025c2 } & \cite{LHCb:2025update} \\ \hline
247 & \texttt{ P2\_B0Kstar0mumu\_15.0\_17.0\_LHCb2025c2 } & \cite{LHCb:2025update} \\ \hline
248 & \texttt{ P3\_B0Kstar0mumu\_15.0\_17.0\_LHCb2025c2 } & \cite{LHCb:2025update} \\ \hline
249 & \texttt{ P4prime\_B0Kstar0mumu\_15.0\_17.0\_LHCb2025c2 } & \cite{LHCb:2025update} \\ \hline
250 & \texttt{ P5prime\_B0Kstar0mumu\_15.0\_17.0\_LHCb2025c2 } & \cite{LHCb:2025update} \\ \hline
251 & \texttt{ P6prime\_B0Kstar0mumu\_15.0\_17.0\_LHCb2025c2 } & \cite{LHCb:2025update} \\ \hline
252 & \texttt{ P8prime\_B0Kstar0mumu\_15.0\_17.0\_LHCb2025c2 } & \cite{LHCb:2025update} \\ \hline
253 & \texttt{ dGamma/dq2\_B0Kstar0mumu\_15.0\_17.0\_LHCb2025c2 } & \cite{LHCb:2025update} \\ \hline

254 & \texttt{ FL\_B0Kstar0mumu\_17.0\_19.0\_LHCb2025c2 } & \cite{LHCb:2025update} \\ \hline
255 & \texttt{ S1c\_B0Kstar0mumu\_17.0\_19.0\_LHCb2025c2 } & \cite{LHCb:2025update} \\ \hline
256 & \texttt{ P1\_B0Kstar0mumu\_17.0\_19.0\_LHCb2025c2 } & \cite{LHCb:2025update} \\ \hline
257 & \texttt{ P2\_B0Kstar0mumu\_17.0\_19.0\_LHCb2025c2 } & \cite{LHCb:2025update} \\ \hline
258 & \texttt{ P3\_B0Kstar0mumu\_17.0\_19.0\_LHCb2025c2 } & \cite{LHCb:2025update} \\ \hline
259 & \texttt{ P4prime\_B0Kstar0mumu\_17.0\_19.0\_LHCb2025c2 } & \cite{LHCb:2025update} \\ \hline
260 & \texttt{ P5prime\_B0Kstar0mumu\_17.0\_19.0\_LHCb2025c2 } & \cite{LHCb:2025update} \\ \hline
261 & \texttt{ P6prime\_B0Kstar0mumu\_17.0\_19.0\_LHCb2025c2 } & \cite{LHCb:2025update} \\ \hline
262 & \texttt{ P8prime\_B0Kstar0mumu\_17.0\_19.0\_LHCb2025c2 } & \cite{LHCb:2025update} \\ \hline
263 & \texttt{ dGamma/dq2\_B0Kstar0mumu\_17.0\_19.0\_LHCb2025c2 } & \cite{LHCb:2025update} \\ \hline
 & & \\\hline
 & & \\\hline
 & & \\\hline
\end{tabular}
}
\caption{Continued from Table~\ref{tab:Nob2025_obs_71_180}.}
\label{tab:Nob2025_obs_181_263}
\end{center}
\end{table}

\clearpage

\section{Impact of the choice of form factors on tensions}
\subsection{\texorpdfstring{BR($B \to K \mu^+ \mu^-$)}{BR(B->Kmumu)} with different form factors}
\label{app:BR_BKmumu_FFcheck}

\begin{figure}[h!]
\begin{center}
\includegraphics[width=0.80\textwidth]{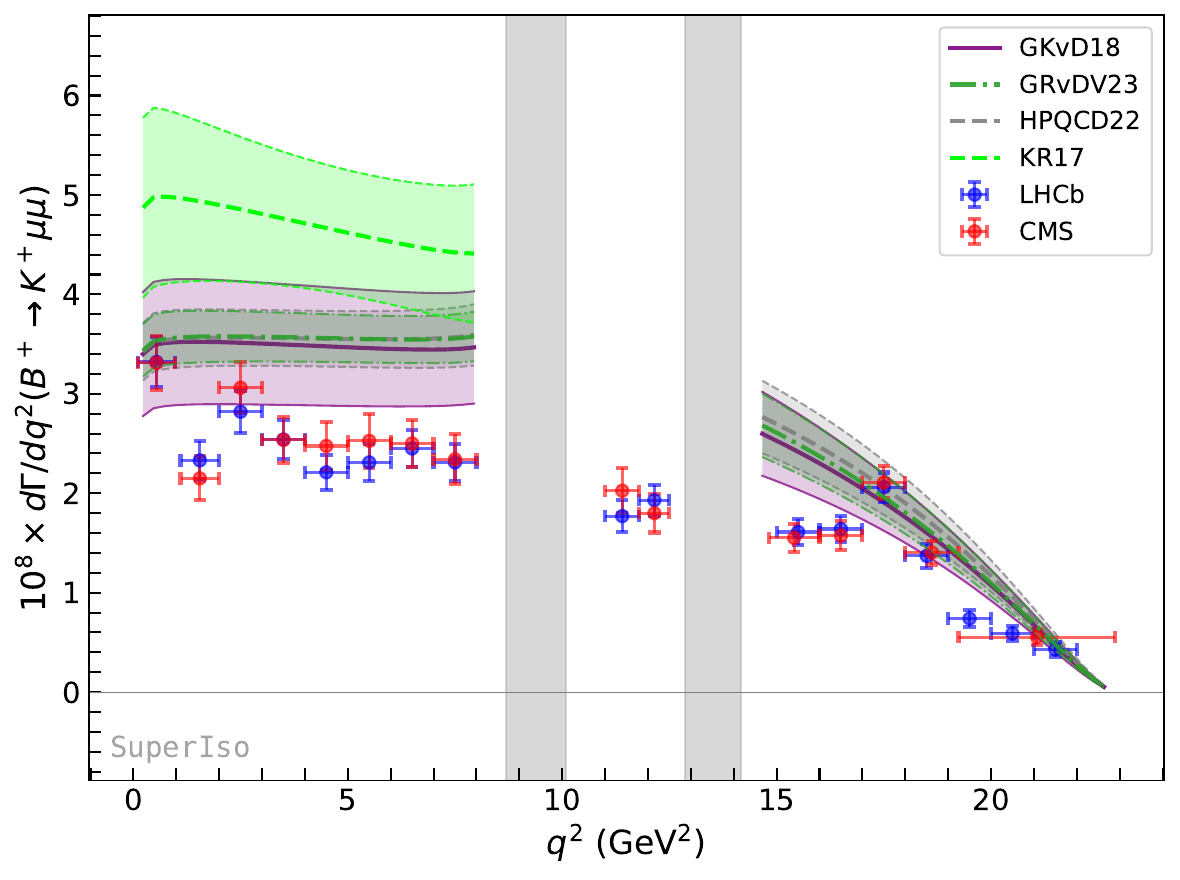}
\vspace{-0.4cm}
\caption{\small
Theoretical prediction for BR($B^+ \to K^+\mu^+\mu^-$) with different choices of $B\to K$ form factors. Notice that the dash-dotted dark green line indicating GRvDV23 almost coincides with the dashed gray line which corresponds to HPQCD22.
The experimental results are from LHCb~\cite{LHCb:2014auh} and CMS~\cite{CMS:2024syx}.
\label{fig:BR_BKmumu_FF}}
\end{center}
\end{figure}

In Figure~\ref{fig:BR_BKmumu_FF}, we illustrate the impact of different form factor choices on the Standard Model prediction for the branching ratio of $B \to K \mu^+ \mu^-$. We compare the results obtained using four different sets of form factors.

Using the weighted average of the LHCb~\cite{LHCb:2014auh} and CMS~\cite{CMS:2024syx} measurements in the $[1.1,6.0]$ GeV$^2$ bin only, we find that the level of tension with the SM prediction varies significantly depending on the form factor input -- ranging from $1.7\sigma$ to more than $4\sigma$ -- as noted in the list below and as illustrated in Figure~\ref{fig:BR_BKmumu_FF}.
\begin{itemize}
\item \textbf{GKvD18:} The combined LCSR (based on $B$-meson distribution amplitudes) and lattice QCD result from Ref.~\cite{Gubernari:2018wyi}, using lattice input from Ref.~\cite{Bouchard:2013eph} ($1.7\sigma$).

\item \textbf{GRvDV23:} The combined fit of Ref.~\cite{Gubernari:2023puw}, based on LCSR calculations from Ref.~\cite{Gubernari:2018wyi} and lattice QCD results from Refs.~\cite{Bouchard:2013eph,Bailey:2015dka,Parrott:2022rgu} ($4.1\sigma$).

\item \textbf{HPQCD22:} The lattice QCD result from Ref.~\cite{Parrott:2022rgu}, valid across the entire physical $q^2$ range ($3.6\sigma$).

\item \textbf{KR17:} The results of Ref.~\cite{Khodjamirian:2017fxg} calculated via LCSR with kaon distribution amplitude, applicable only for the low-$q^2$ region ($3.4\sigma$).

\end{itemize}

This comparison highlights the sensitivity of the branching ratio prediction to the choice of hadronic form factors. We emphasise again that we consider here one single bin in this comparison. The differences would be larger when all observables were taken into account. We also note that the results of Ref.~\cite{Parrott:2022rgu} denoted HPQCD22 in the list above were also used in Ref.~\cite{Gubernari:2023puw} in the low-$q^2$ region. One can see in Figure~\ref{fig:BR_BKmumu_FF} that this lattice result dominates the result of  Ref.~\cite{Gubernari:2023puw} denoted GRvDV23 in the low-$q^2$ region.

\subsection{\texorpdfstring{$P_2,P_5^\prime$($B \to K^* \mu^+ \mu^-$)}{P2,P5'(B->K*mumu)} with different form factors}
\label{app:P5prime_B0Kstar0mumu_FFcheck}

In this section, we compare the $P_2$ and $P_5^\prime$ observables of the $B \to K^* \mu^+ \mu^-$ decay using two different sets of $B \to K^*$ form factors: BSZ15 and GRvDV23, as introduced in Section~\ref{sec:CMS_vs_LHCb}.\footnote{In reproducing the $B\to K^*$ form factors from Ref.~\cite{Gubernari:2018wyi}, we find that the ancillary file provided for the LCSR+Lattice results (BKstar\_LCSR-Lattice.json) does not reproduce the uncertainties shown in Figure~6 of the paper. Specifically, the uncertainties at high $q^2$ are significantly larger, suggesting either an inconsistency in the ancillary files or that the figure was generated using inputs not reflected in the provided data.} We focus on the penultimate low-$q^2$ bin -- $[4,6]$ GeV$^2$ for LHCb and $[4.3,6]$ GeV$^2$ for CMS -- and report the local tensions between the SM predictions and experimental measurements (CMS~\cite{CMS:2024atz} and LHCb 2020~\cite{LHCb:2020lmf} and 2025~\cite{LHCb:2025update}) for each form factor choice. For the $P_5^\prime$ observable we find:

\begin{itemize}
\item \textbf{BSZ15:} $3.4\sigma$ for LHCb~2025, $2.6\sigma$ for LHCb~2020, and $3.0\sigma$ for CMS.
\item \textbf{GRvDV23:} $2.9\sigma$ for LHCb~2025, $2.3\sigma$ for LHCb~2020, and $2.7\sigma$ for CMS.
\end{itemize}
\vskip 10pt

\noindent
In the same bins, for the $P_2$ observable, we have:
\begin{itemize}
\item \textbf{BSZ15:} $4.0\sigma$ for LHCb~2025, $2.2\sigma$ for LHCb~2020, and $1.9\sigma$ for CMS.
\item \textbf{GRvDV23:} $3.3\sigma$ for LHCb~2025, $1.8\sigma$ for LHCb~2020, and $1.6\sigma$ for CMS.
\end{itemize}

\begin{figure}[h!]
\begin{center}
\includegraphics[width=0.6\textwidth]{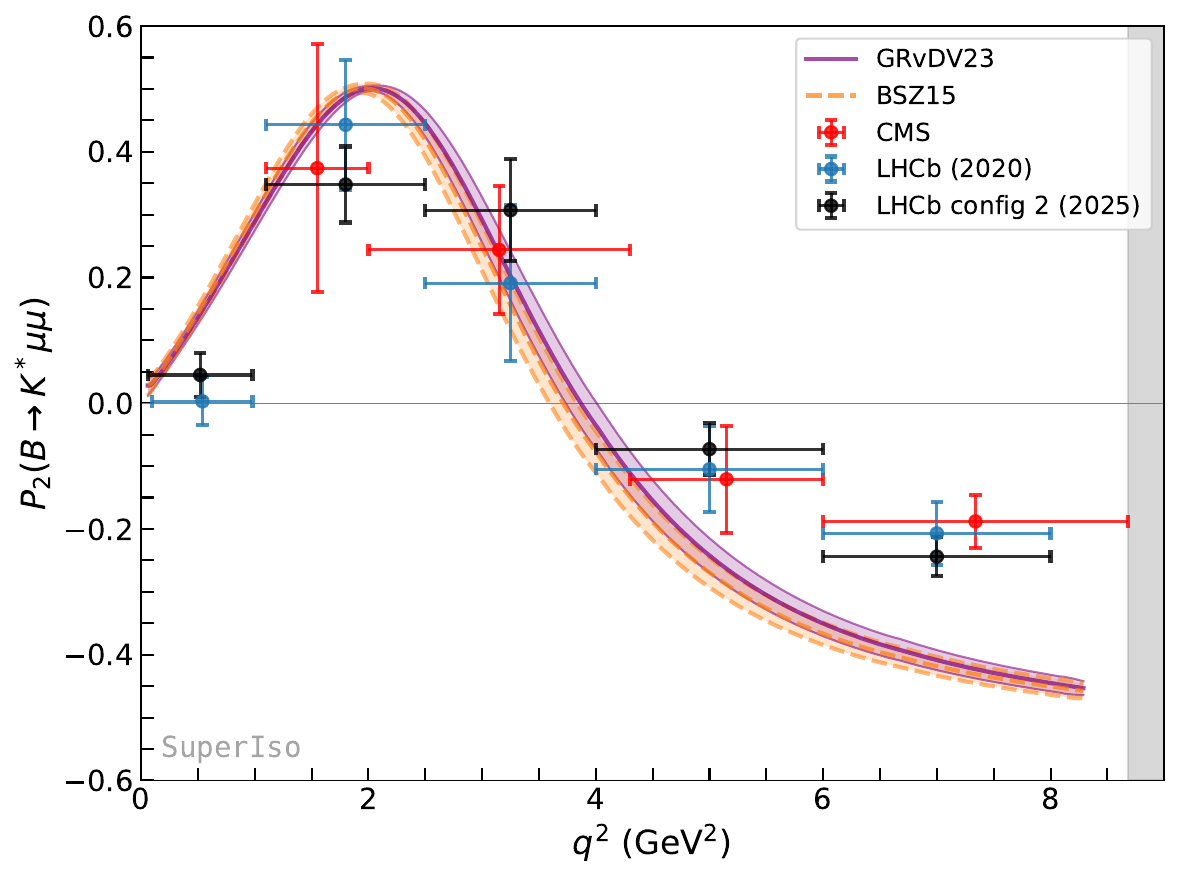}
\includegraphics[width=0.6\textwidth]{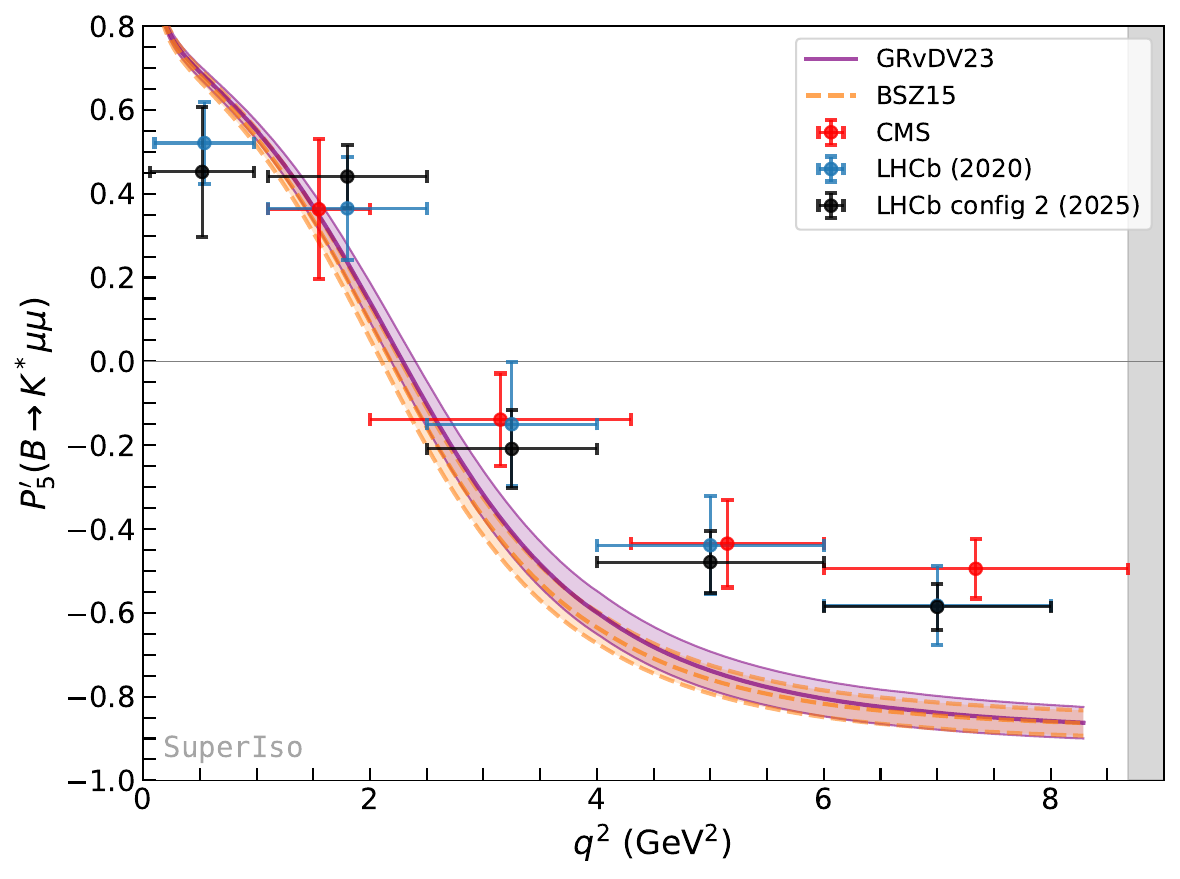}
\vspace{-0.4cm}
\caption{\small
Theoretical prediction for $P_2,P_5^\prime(B^0 \to K^{*0}\mu^+\mu^-)$ with two different choices of $B\to K^*$ form factors.
{The experimental results are from LHCb~\cite{LHCb:2020lmf} and CMS~\cite{CMS:2024atz}.}
The data points correspond to the CMS measurement~\cite{CMS:2024atz} and the LHCb results from 2020~\cite{LHCb:2020lmf} and 2025~\cite{LHCb:2025update}.
\label{fig:P5prime_B0Kstar0mumu_FF}}
\end{center}
\end{figure}

The results are also shown in Figure~\ref{fig:P5prime_B0Kstar0mumu_FF}.
This comparison demonstrates the sensitivity of the $B\to K^* \mu^+\mu^-$ observables, to the choice of hadronic form factors, with the BSZ15 resulting in larger tensions between theoretical prediction and measurement compared to when employing the GRvDV23 form factors. The cumulative impact of these two different choices of form factors on the $B\to K^* \mu^+\mu^-$ observables can be seen in the new physics fit of Figure~\ref{fig:angObs_LHCb_CMS_Pi_choice_1_2}. Note that each individual $P_i^{(\prime)}$ observable is constructed in such a way that its sensitivity to the local (soft) form factors is minimised.

\clearpage 
\subsection{BR($B_s\to \phi \mu^+\mu^-$) with different form factors}\label{app:BR_Bsphimumu_FFcheck}
\begin{figure}[h!]
\begin{center}
\includegraphics[width=0.6\textwidth]{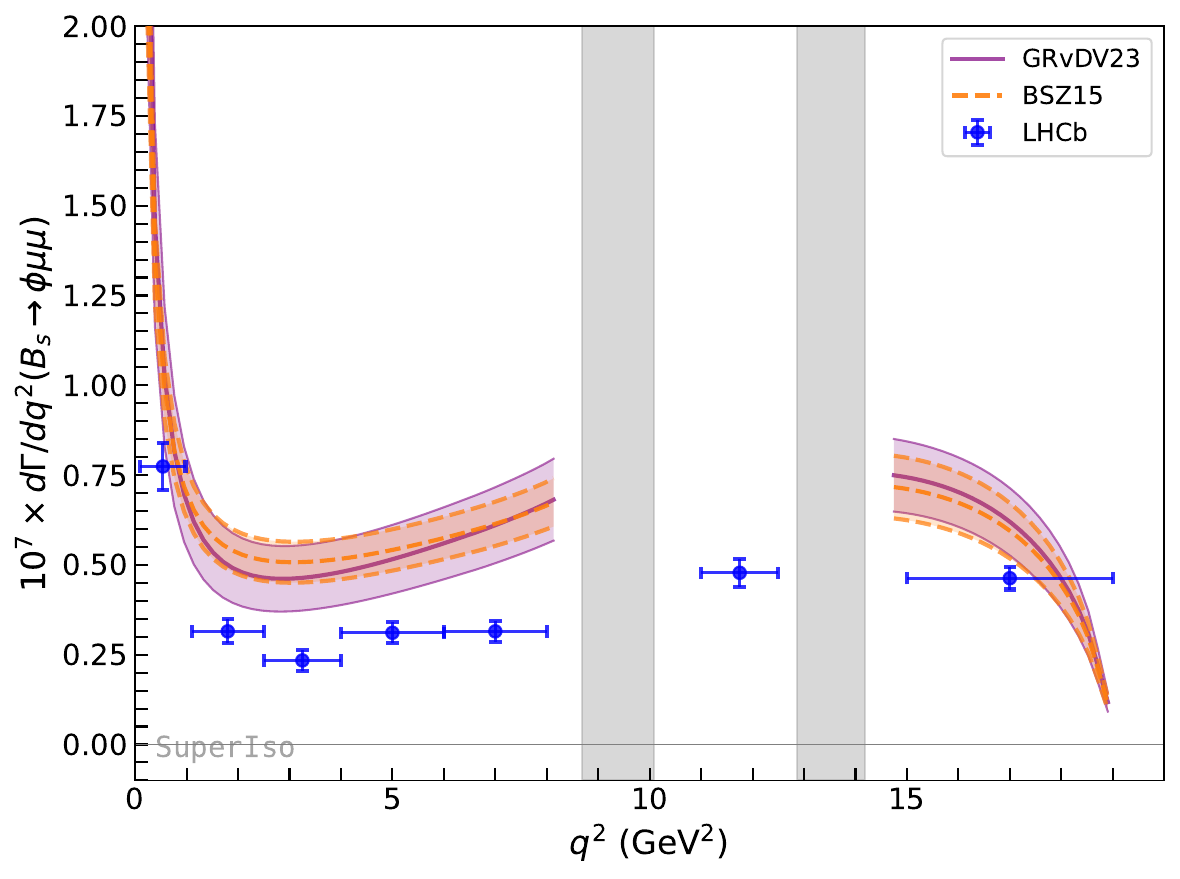}
\vspace{-0.4cm}
\caption{\small
Theoretical prediction for BR$(B_s \to \phi\mu^+\mu^-)$ with two different choices of $B_s \to \phi$ form factors. 
The experimental result is from LHCb~\cite{LHCb:2021zwz}.
}
\label{fig:BR_Bsphimumu_FF}
\end{center}
\end{figure}

In Fig.~\ref{fig:BR_Bsphimumu_FF}, we show the BR($B_s\to\phi \mu^+\mu^-$) using two different sets of $B_s \to \phi$ form factors: BSZ15 and GRvDV23, as described in section~\ref{sec:global_fits}, with the latter being the default set used in the global fit.  
Considering the LHCb measurement~\cite{LHCb:2021zwz} for the [1.1,6.0]~GeV$^2$ bin, we find the following tensions with the SM prediction for the two different form factor choices:
\begin{itemize}
    \item \textbf{BSZ15:} $4.1\sigma$
    \item \textbf{GRvDV23:} $2.2\sigma$
\end{itemize}

\clearpage

\section{\texorpdfstring{$\delta C_9$}{dC9} effects on key observables}\label{sec:dC9_effects}

In this section, we illustrate the effect of new physics in $\delta C_9$ on several $b \to s \ell\ell$ observables that exhibit the most pronounced tensions between SM predictions and experimental measurements.

\begin{figure}[h!]
\begin{center}
\includegraphics[width=0.48\textwidth]{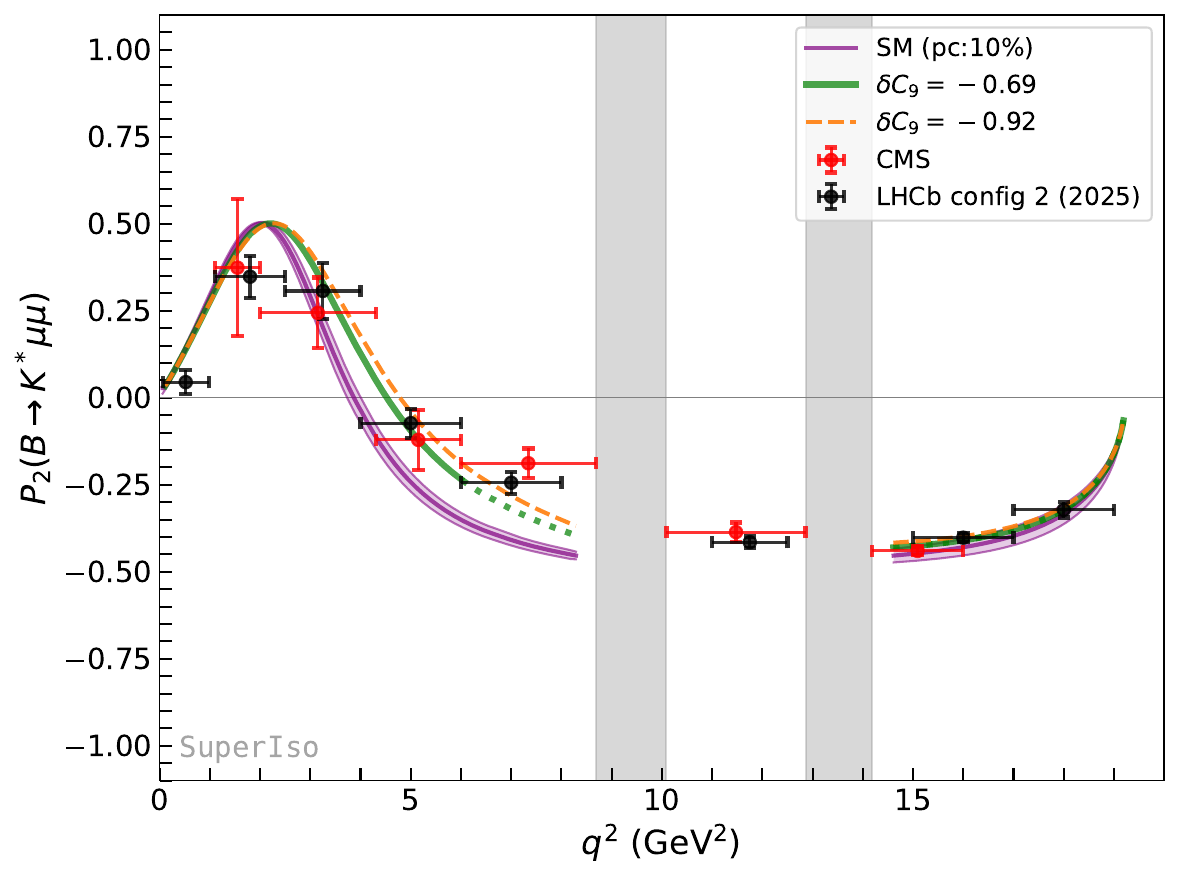}
\includegraphics[width=0.48\textwidth]{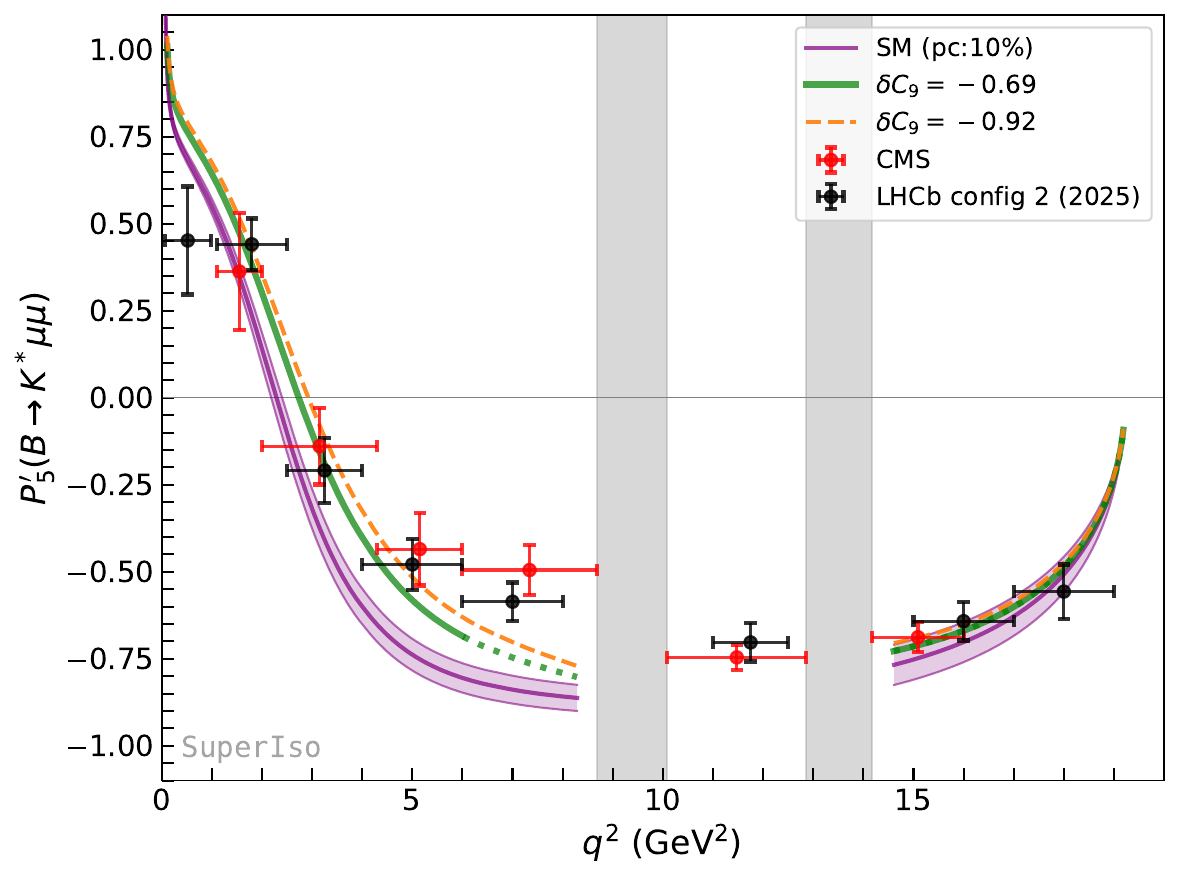}
\includegraphics[width=0.48\textwidth]{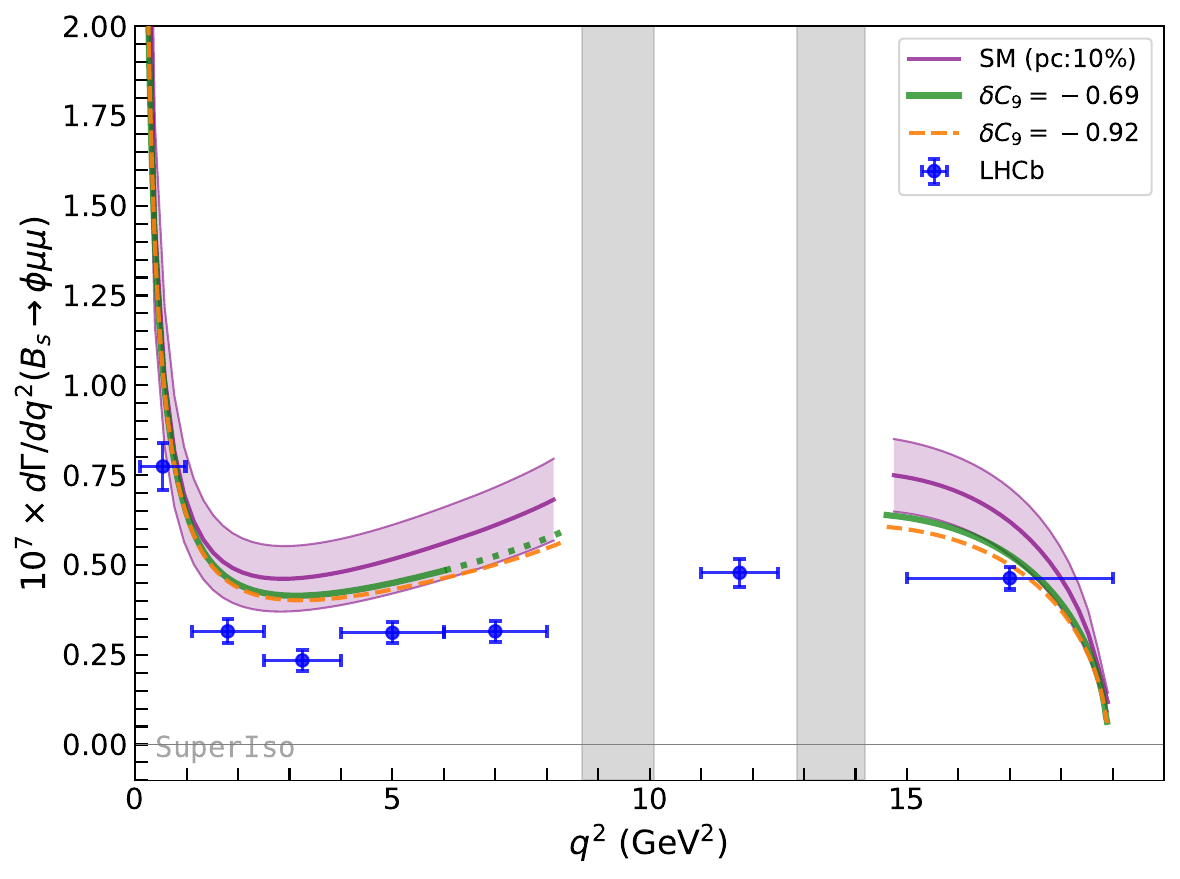}
\includegraphics[width=0.48\textwidth]{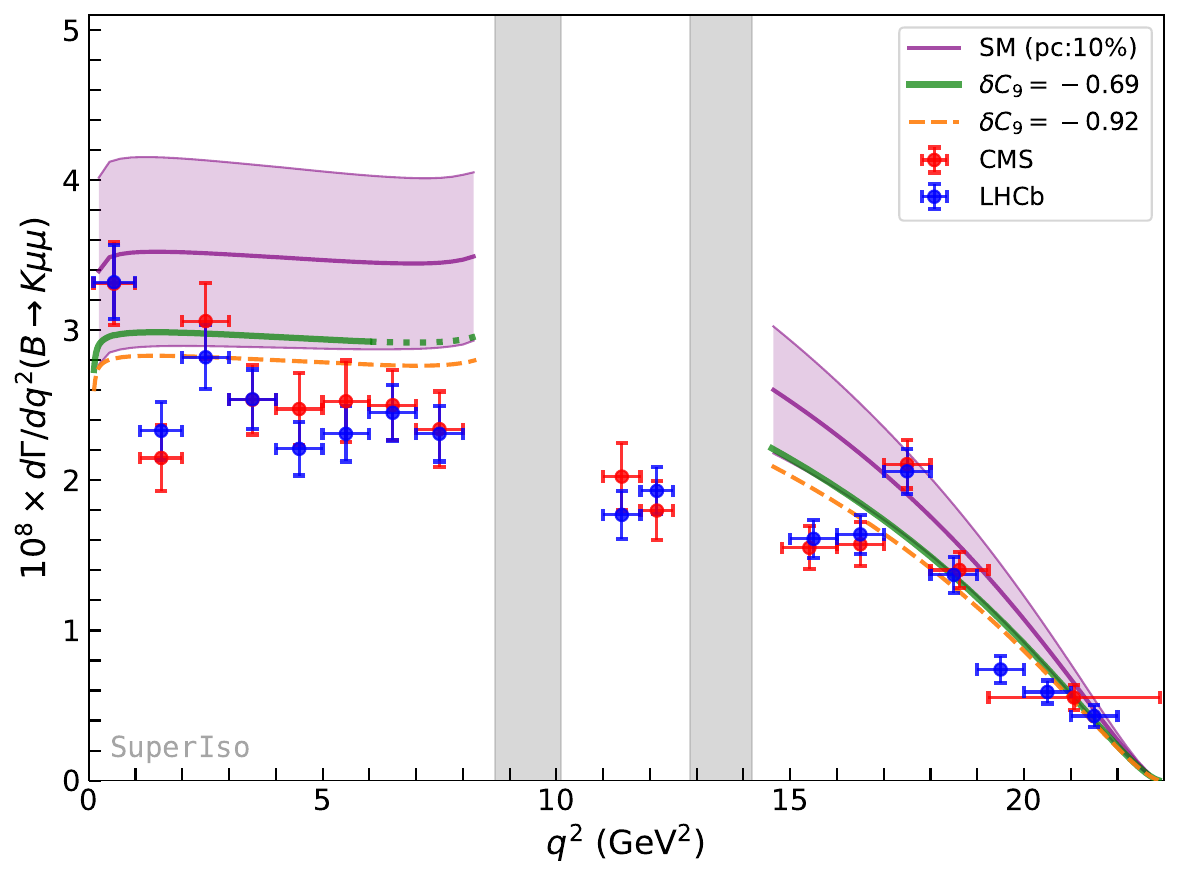}
\vspace{-0.2cm}
\caption{\small
Impact of $\delta C_9 = -0.69$ (green solid) and $\delta C_9 = -0.92$ (brown dashed), corresponding to the best-fit points from the left and right panels of Table~\ref{tab:2025GRvDV_all_vs_allwo68_1D}, respectively. The dotted green line represents the prediction of the fit with $\delta C_9 = -0.69$ in the $[6,8.68]$\,GeV$^2$ region, which was not included in that particular fit. 
\label{fig:C9_key_obs}}
\end{center}
\end{figure}

\clearpage
\section{Hadronic fit at the observable level}\label{sec:HadFit_Obs}

In this section, we illustrate the impact of the hadronic fit -- described in subsection~\ref{sec:HadFit_18param} and summarised in Table~\ref{tab:HadronicFit_18param} -- at the level of physical observables. 
\begin{figure}[!h]
\begin{center}
\includegraphics[width=0.48\textwidth]{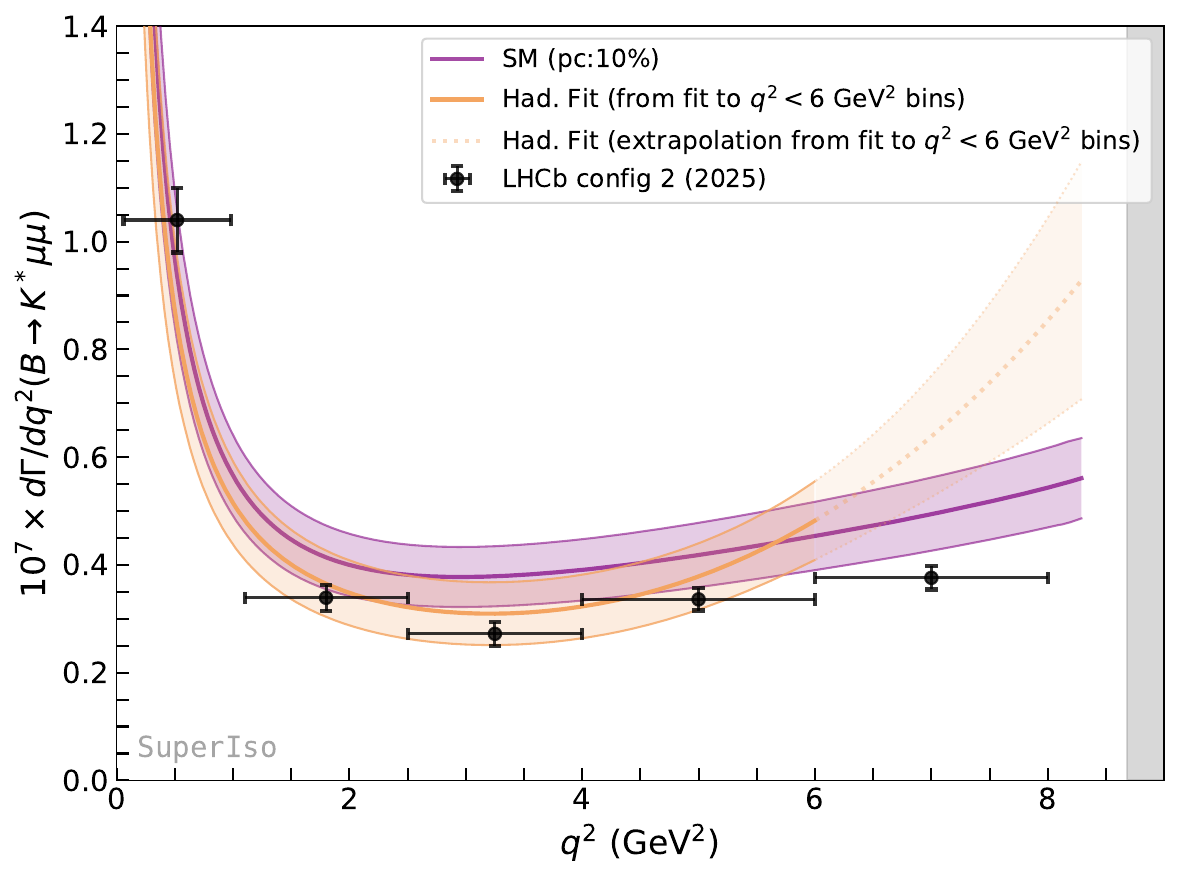}
\includegraphics[width=0.48\textwidth]{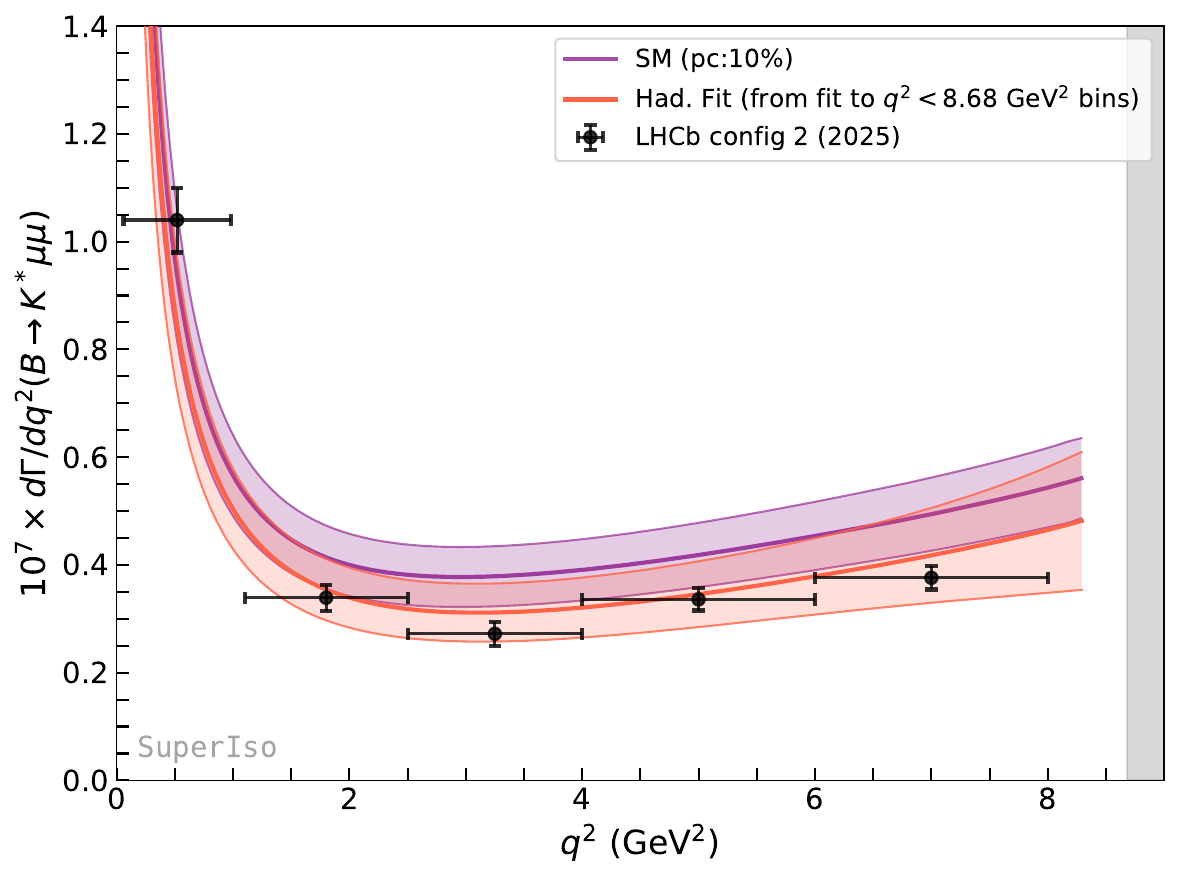}
\includegraphics[width=0.48\textwidth]{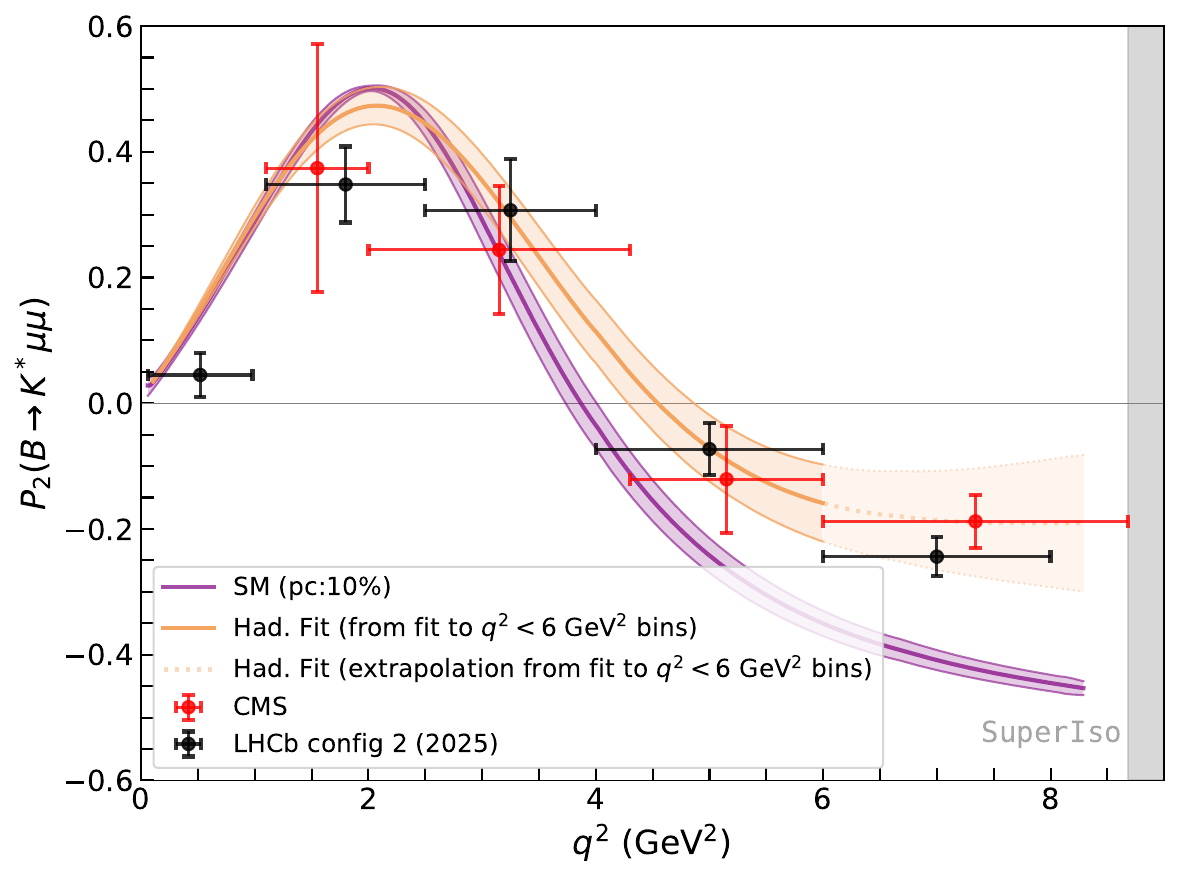}
\includegraphics[width=0.48\textwidth]{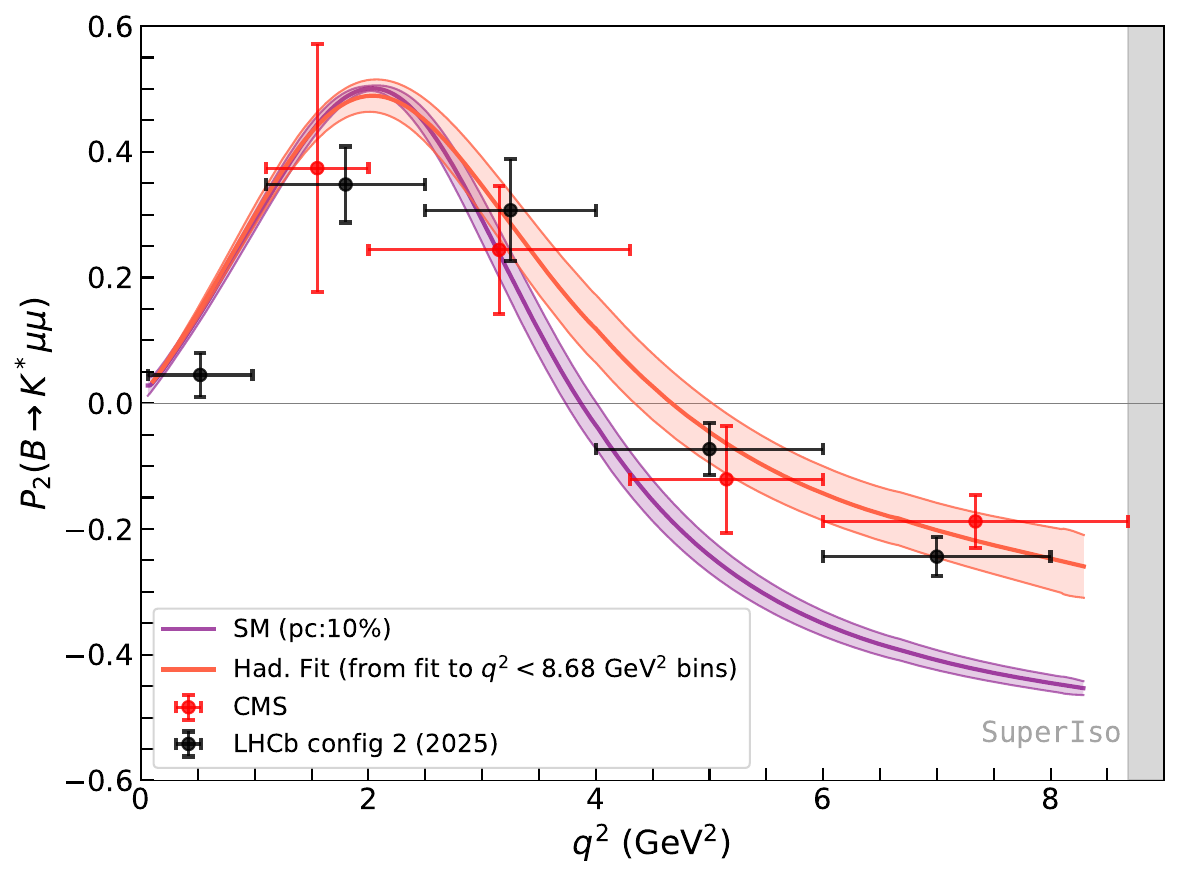}
\includegraphics[width=0.48\textwidth]{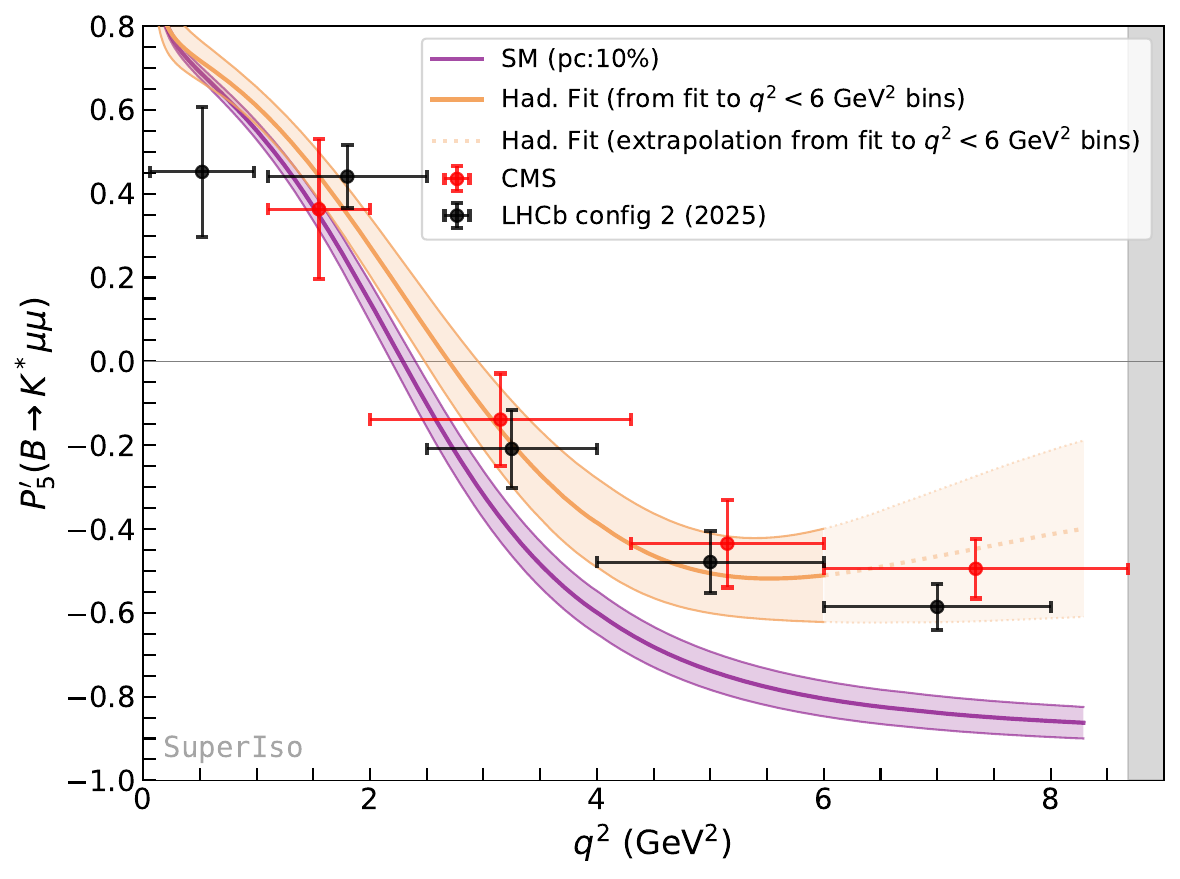}
\includegraphics[width=0.48\textwidth]{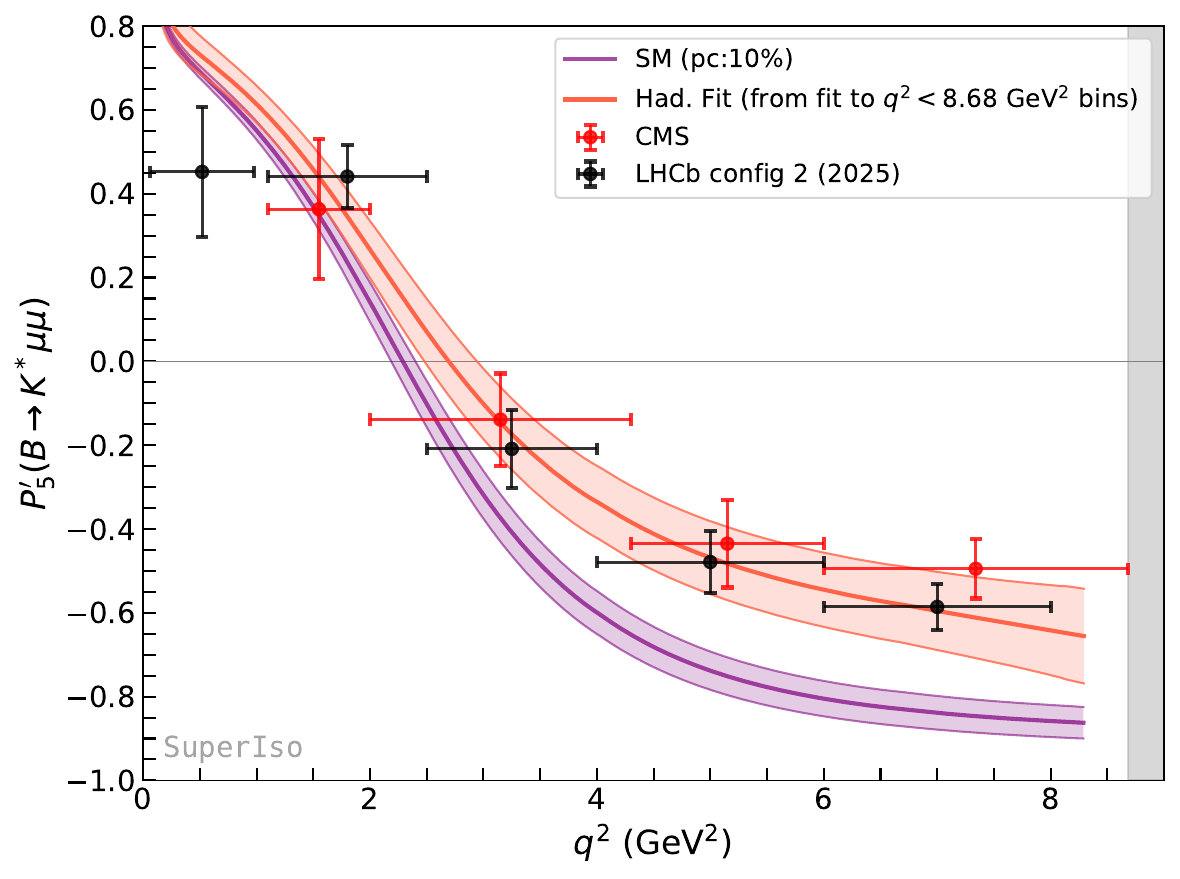}
\vspace{-0.2cm}
\caption{\small
Theoretical predictions of $B^0 \to K^{*0}\mu^+\mu^-$ observables BR, $P_2$ and $P_5^\prime$ for our SM prediction (assuming 10\% power corrections) as well as when considering the hadronic fit. On the left (right) the hadronic fit of to $B\to K^* \gamma/\ell\ell$ observables with $q^2\leq 6\;(8.68)$~GeV$^2$ bins have been considered as given in the left (right) panel of Table~\ref{tab:HadronicFit_18param}. In the left panels, the hadronic fits are extrapolated beyond its fitted range (6 GeV$^2$), with the extrapolated region marked by dotted lines.
\label{fig:B0Kstar0mumu_HadFit_18param_binned_BR_P2_P5prime_update}}
\end{center}
\end{figure}
Figure~\ref{fig:B0Kstar0mumu_HadFit_18param_binned_BR_P2_P5prime_update} displays the predictions for the observables BR, $P_2$, and $P_5^\prime$. The left panels correspond to fits performed up to 6~GeV$^2$, while the right panels show fits extended up to 8.68~GeV$^2$.

A comparison between the left and right plots highlights the effect of including the largest low-$q^2$ bins in the fit. This demonstrates how the inclusion of additional data points influences the hadronic fit in the low-$q^2$ region, leading to a reduction in uncertainties. Note that in the left panels, the fits are extrapolated beyond 6~GeV$^2$, as indicated by the dotted lines.

\bibliographystyle{JHEP} 
\bibliography{biblio_bsll2025}

\end{document}